\pdfoutput = 1

\documentclass[12pt,a4paper]{article}

\usepackage{amssymb,epsf,epsfig,amsmath,color}
\usepackage{multirow}
\usepackage{accents}
\usepackage{braket}
\usepackage{float}

\newcommand{\lsim}{
\mathrel{\hbox{\rlap{\hbox{\lower4pt\hbox{$\sim$}}}\hbox{$<$}}}}
\newcommand{\gsim}{
\mathrel{\hbox{\rlap{\hbox{\lower4pt\hbox{$\sim$}}}\hbox{$>$}}}}

\def\D0{D\O }

\begin{document}
\begin{titlepage}
\vspace*{-0.7truecm}
\begin{flushright}
Nikhef-2018-022
\end{flushright}

\vspace{1.6truecm}

\begin{center}
\boldmath
{\Large{\bf Decoding (Pseudo)-Scalar Operators in Leptonic and Semileptonic $B$ Decays}
}
\unboldmath
\end{center}

\vspace{0.8truecm}

\begin{center}
{\bf Giovanni Banelli\,${}^{a}$, Robert Fleischer\,${}^{a,b}$, Ruben Jaarsma\,${}^{a}$ and 
Gilberto~Tetlalmatzi-Xolocotzi\,${}^{a}$}

\vspace{0.5truecm}

${}^a${\sl Nikhef, Science Park 105, NL-1098 XG Amsterdam, Netherlands}

${}^b${\sl  Faculty of Science, Vrije Universiteit Amsterdam,\\
NL-1081 HV Amsterdam, Netherlands}

\end{center}

\vspace*{1.7cm}

\begin{abstract}
\noindent
We consider leptonic $B^-\to \ell^- \bar\nu_\ell$ and semileptonic $\bar B \to \pi \ell^- \bar\nu_\ell$, $\bar B \to \rho \ell^- \bar\nu_\ell$
decays and present a strategy to determine short-distance coefficients of New-Physics operators and the CKM element $|V_{ub}|$. 
As the leptonic channels play a central role, we illustrate this method for (pseudo)-scalar operators which may lift the helicity 
suppression of the corresponding transition amplitudes arising in the Standard Model. Utilising a new result by the Belle 
collaboration for the branching ratio of $B^-\to \mu^- \bar\nu_\mu$, we explore theoretically clean constraints and correlations 
between New Physics coefficients for leptonic final states with $\mu$ and $\tau$ leptons. In order to obtain stronger bounds
and to extract $|V_{ub}|$, we employ semileptonic $\bar B \to \pi \ell^- \bar\nu_\ell$ and $\bar B \to \rho \ell^- \bar\nu_\ell$ decays
as an additional ingredient, involving hadronic form factors which are determined through QCD sum rule and lattice calculations. 
In addition to a detailed analysis of the constraints on the New Physics contributions following from current data, we make predictions 
for yet unmeasured decay observables, compare them with experimental constraints and discuss the impact of CP-violating phases 
of the New-Physics coefficients. 
\end{abstract}

\vspace*{2.1truecm}

\vfill

\noindent
September 2018

\end{titlepage}

\thispagestyle{empty}
\vbox{}
\newpage

\setcounter{page}{1}

\section{Introduction}\label{sec:intro}
Leptonic transitions of $B$ mesons are the simplest weak decay class as the final-state particles do not have $SU(3)_{\rm C}$ colour quantum numbers. Consequently, the whole hadron dynamics is described by a single 
parameter, the $B$-meson decay constant
\begin{equation}
\langle 0|\overline{u}\gamma_\mu\gamma_5 b|B^-(p)\rangle =
i f_{B^-} p_\mu,
\end{equation}
which can be determined through lattice-QCD methods. 
While leptonic decays of neutral $B$ mesons are rare processes originating from flavour-changing neutral currents,
those of charged mesons, $B^-\to\ell^-\bar\nu_\ell$, are caused by charged-current interactions (with $\ell=e, \mu, \tau$). Within the Standard 
Model (SM), the branching ratio takes the following form:
\begin{equation}\label{SM-Br}
{\mathcal B}(B^-\to\ell^-\bar\nu_\ell)|_{\rm SM}=\frac{G_{\rm F}^2}{8\pi}
|V_{ub}|^2M_{B^-}m_\ell^2\left(1-\frac{m_\ell^2}{M_{B^-}^2}\right)^2f_{B^-}^2\tau_{B^-},
\end{equation}
where $G_{\rm F}$ is Fermi's constant, $\tau_{B^-}$ the lifetime of the $B^-$ meson, $M_{B^-}$ and $m_\ell$ are the 
$B^-$ and lepton masses, respectively, and the neutrino mass has been neglected. This branching ratio is 
suppressed by the Cabibbo--Kobayashi--Maskawa (CKM) element $V_{ub}$, and exhibits a helicity suppression, 
which is reflected by the proportionality to  $m_\ell^2$. Using $f_{B^-}=0.186 \pm 0.004 ~\hbox{GeV}$ \cite{Aoki:2016frl} and assuming the SM 
with \cite{Charles:2004jd}
\begin{equation} \label{eq:VubCKMFitter}
|V_{ub}|=(3.601\pm0.098)\times10^{-3},
\end{equation} 
we obtain 
\begin{eqnarray}
{\mathcal B}(B^-\to\tau^-\bar\nu_\tau)|_{\rm SM} &=& (7.92\pm0.55)\times10^{-5} \\
{\mathcal B}(B^-\to\mu^-\bar\nu_\mu)|_{\rm SM} &=& (3.56\pm0.25)\times10^{-7} \label{eq:SMleptBrmu} \\
{\mathcal B}(B^-\to e^-\bar\nu_e)|_{\rm SM} &=& (8.33\pm0.58)\times10^{-12}.\label{eq:SMleptBr}
\end{eqnarray}
In the case of $\ell=\tau$, the helicity suppression is very ineffective due to the large $\tau$ mass. Consequently, 
despite the challenging $\tau$ reconstruction, the $B^-\to\tau^-\bar\nu_\tau$ mode could already be observed by 
the BaBar and Belle collaborations about a decade ago, with the current average by the Particle Data Group 
(PDG) given as follows \cite{PDG}:
\begin{equation} \label{br-btaunu-avg}
{\mathcal B}(B^-\to\tau^-\bar\nu_\tau)= (1.09\pm0.24)\times10^{-4}.
\end{equation}
On the other hand, $B^-\to\mu^-\bar\nu_\mu$ and $B^-\to e^-\bar\nu_e$ with their SM branching ratios in 
regimes of  $10^{-7}$ and $10^{-11}$, respectively, appear 
much more challenging to measure. Nevertheless, the Belle collaboration has recently performed a new
search for the former channel, finding a $2.4\,\sigma$ excess over the background \cite{Belle-munu}:
\begin{equation}\label{Belle-res}
{\mathcal B}(B^-\to\mu^- \bar\nu_\mu)=(6.46\pm2.22|_{\rm stat}\pm1.60_{\rm syst})\times10^{-7} 
= (6.46 \pm 2.74)\times10^{-7}.
\end{equation}
For the electronic channel, only an upper bound is available, which was obtained by the Belle collaboration in 
2007 \cite{Belle-enu}:
\begin{equation}\label{Belle-enu-limit}
{\mathcal B}(B^- \to e^- \bar\nu_e) < 9.8 \times 10^{-7} \, \mbox{(90\% C.L.)}.
\end{equation}

Decays of $B$ mesons into final states with $\tau$ leptons are receiving a lot of interest in view of experimental results 
which indicate possible signals of New Physics (NP), where the ratios $R_{D^{(*)}}$ of the branching ratios
of $\bar B \to D^{(*)} \tau^- \bar\nu_\tau$ and $\bar B \to D^{(*)} \mu^- \bar\nu_\mu$ decays are in the focus
(see, for instance, Refs.~\cite{Lees:2012xj}--\cite{ABDS} and references therein). 
The exciting feature is an indication of a violation of the universality between $\tau$ and $\mu$. These results
are complemented by measurements of the ratios of branching ratios of the rare decays $B\to K^{(*)} \mu^+\mu^-$ and
$B\to K^{(*)} e^+e^-$ which may signal a violation of the universality of muons and electrons in these processes
(for recent overviews, see Refs.~\cite{WNSS,HN}). 

In this paper, we propose a new strategy to probe NP effects by utilising leptonic $B$ decays and the interplay with
their semileptonic counterparts. Both decay classes are actually caused by the same low-energy effective
Hamiltonian. We will obtain constraints on short-distance coefficients using the Belle result in Eq.~(\ref{Belle-res})
and address the question of how large the branching ratio of  $B^-\to e^-\bar\nu_e$ could be due to a lift of the 
helicity suppression through NP effects. We have addressed a similar question for the leptonic rare decays 
$B^0_{s,d}\to e^+e^-$, which could be enhanced to the level of the $B^0_{s,d}\to \mu^+\mu^-$ channels through 
new (pseudo)-scalar interactions \cite{FJTX-1}. We shall also make predictions for various semileptonic decay 
ratios which will allow us to fully reveal the underlying decay dynamics, and extract $|V_{ub}|$ while also allowing
for NP contributions.

A subtle point is given by CP-violating phases which may be present in the short-distance coefficients of NP operators. 
As is well known from discussions of non-leptonic meson decays, CP-violating asymmetries arising directly at the 
decay amplitude level,
\begin{equation}\label{aCP-dir}
a_{\rm CP}\equiv \frac{{\mathcal B}(\bar B \to \bar f)-{\mathcal B}(B \to f)}{{\mathcal B}(\bar B \to \bar f)
+{\mathcal B}(B \to f)},
\end{equation}
are induced by the interference between decay amplitudes with both non-trivial CP-violating and non-trivial CP-conserving 
phase differences \cite{RF-rev}. While the former originate from phases of CKM matrix elements in the SM or possible 
CP-violating NP phases, the latter could be generated through strong interactions or absorptive parts of loop diagrams. 
In the SM, the direct CP asymmetries vanish hence in (semi)-leptonic decays at leading order in the weak 
interactions while higher-order-effects can only generate negligible effects \cite{BSEGR}--
\cite{FV}. Due to the 
lack of sizeable CP-conserving phase differences, direct CP asymmetries (\ref{aCP-dir}) of $B^-\to \ell^-\bar\nu_\ell$, $\bar B\to \pi \ell^-\bar\nu_\ell$, and $\bar B\to \rho \ell^-\bar\nu_\ell$ decays can also not take sizeable values in the presence of NP contributions. 
Consequently, we cannot get empirical evidence for such phases through possible direct CP asymmetries in such
modes, in contrast to non-leptonic $B$ decays where strong interactions are at work to generate strong
phase differences. On the other hand, in the leptonic rare $\bar B^0_q\to\ell^+\ell^-$ decays of neutral 
$\bar B^0_q$ mesons ($q=d,s$), the impact of $B^0_q$--$\bar B^0_q$ mixing may induce CP-violating 
asymmetries, thereby indicating possible CP-violating NP phases \cite{CPV-paper}.

Contributions of NP to $B^-\to \ell^-\bar\nu_\ell$, $\bar B\to \pi \ell^-\bar\nu_\ell$ and $\bar B\to \rho \ell^-\bar\nu_\ell$ processes have also been addressed in, e.g., Refs.~\cite{Khodjamirian:2011}--
\cite{Ivanov:2017hun}.

The outline of this paper is as follows: after introducing briefly the theoretical framework in Section~\ref{sec:theo}, we discuss the leptonic $B^-\to \ell^-\bar\nu_\ell$ decays in Section~\ref{sec:lept}. The semileptonic $\bar B\to \rho \ell^-\bar\nu_\ell$ and $\bar B\to \pi \ell^-\bar\nu_\ell$ modes are analysed in Section~\ref{sec:semilept}, where we will also combine them with the leptonic constraints to obtain regions for short-distance coefficients. The hadronic form factors, which are required for the study of experimental data, are discussed in Appendix~\ref{sec:HFF}. In both Sections~\ref{sec:lept}~and~\ref{sec:semilept} we will also address the impact of CP-violating phases on the regions for the short-distance coefficients. Then, in Section~\ref{sec:Vub} we determine $|V_{ub}|$ in the presence of NP contributions. Finally, we give predictions for the not yet measured branching ratios $\mathcal{B}(B^-\rightarrow \mu^- \bar{\nu}_{\mu})$, $\mathcal{B}(\bar{B}\rightarrow \rho \tau^- \bar{\nu}_{\tau})$, and $\mathcal{B}(\bar{B}\rightarrow \pi \tau^- \bar{\nu}_{\tau})$ in Section~\ref{sec:predictions}. The conclusions are summarised in Section~\ref{sec:concl}.

\boldmath
\section{Theoretical Framework}\label{sec:theo}
\unboldmath
In the SM, the leptonic decays
\begin{eqnarray}
B^-\rightarrow \ell^- \bar{\nu}_{\ell},
\end{eqnarray}
(with $\ell= e,~\mu,~\tau $) originate from charged-current interactions due to the 
$W^-$ exchange between quark and lepton currents, which are effectively described by the four-fermion operator
\begin{equation}\label{eq:vectorleft}
{\cal O}^{\ell}_{V_L}=(\bar q \gamma^\mu P_L b)(\bar \ell \gamma_\mu P_L  \nu_{\ell}).
\end{equation}
The ${\cal O}^{\ell}_{V_L}$ operator also contributes to semileptonic transitions. For instance, for $q=u$, we have
\begin{eqnarray} \label{eq:semlepchannels}
B^-\rightarrow \rho^0 \ell^- \bar{\nu}_{\ell},&\quad\quad&\bar{B}^0\rightarrow \rho^+ \ell^- \bar{\nu}_{\ell},\nonumber\\
B^-\rightarrow \pi^0 \ell^- \bar{\nu}_{\ell},&\quad\quad&\bar{B}^0\rightarrow \pi^+ \ell^- \bar{\nu}_{\ell}.
\end{eqnarray}

In extensions of the SM, interactions with NP particles may lead to
\begin{eqnarray}
{\cal O}^{\ell}_{S}= (\bar q b)(\bar \ell   P_L \nu_{\ell}),&\quad& 
{\cal O}^{\ell}_{P}= (\bar q \gamma_5  b)(\bar \ell P_L  \nu_{\ell}),\nonumber\\
\quad 
\mathcal{O}^{\ell}_{V_R}=(\bar q \gamma^\mu P_R b)(\bar \ell \gamma_\mu P_L  \nu_{\ell}),&\quad&
{\cal O}^{\ell}_{T}=(\bar q \sigma^{\mu \nu}P_L b)(\bar \ell \sigma_{\mu \nu} P_L \nu_{\ell}),
\end{eqnarray}
where ${\cal O}^{\ell}_{S}$, ${\cal O}^{\ell}_{P}$, $\mathcal{O}^{\ell}_{V_R}$ and ${\cal O}^{\ell}_{T}$ correspond to a scalar, pseudoscalar, (an extra) vector, and a tensor operator, respectively. Notice that we are assuming the neutrinos to be left-handed and to have the same flavour as the lepton in each one of the operators.

We utilize the recent result by the Belle collaboration for ${\mathcal B}(B^- \to \mu^- \bar{\nu}_\mu)$ \cite{Belle-munu} given in Eq.~(\ref{Belle-res}). Combining this observable with experimental data for $B^- \to \tau^- \bar{\nu}_\tau$ and the semileptonic channels in Eq.~(\ref{eq:semlepchannels}), we are in a position to probe lepton flavour universality in decays mediated by a $b \to u \ell \bar{\nu}_\ell$ transition. This is complimentary to the $R_{D^{(\ast)}}$ observables, which involve $b \to c \ell \bar{\nu}_\ell$ transitions. New vector currents are often considered to explain the experimental measurements of these observables. Here we concentrate on the study of the effects of ${\cal O}^{\ell}_{S}$ and ${\cal O}^{\ell}_{P}$. Due to the structure of the formulae, they lift the helicity suppression of the leptonic decays. Consequently, these channels put strong constraints on the corresponding short-distance coefficients, while still allowing for interesting phenomenological predictions.
We then obtain the following low-energy effective Hamiltonian:
\begin{equation}\label{Heff}
{\cal H}_{\rm eff}= \frac{4 G_{\rm F}}{\sqrt{2}}V_{qb}\left[C_{V_L} {\cal O}_{V_L}^\ell  +
C_{S}^\ell{\cal O}_{S}^\ell   
+ 
C_{P}^\ell{\cal O}_{P}^\ell  \right] + h.c.
\end{equation}
In our analysis the vector coefficient takes its SM value $C_{V_L}=1$.

A prominent example of such NP contributions is the effect of charged Higgs bosons which arise in the context of type II Two-Higgs-Doublet-Models (2HDM) \cite{Hou:1992sy}, where 
\begin{eqnarray}\label{eq:C2HDM}
C^{\ell}_{P}&=& C_S^\ell = -\tan^2\beta \Bigl(\frac{m_b m_\ell}{M^2_{H^{\pm}}}\Bigl).
\end{eqnarray}
Here, $\tan\beta$ is the ratio of the vacuum expectation values and $M_{H^{\pm}}$ denotes the mass of the charged Higgs boson. A more recent discussion using the Georgi--Machacek model was given in 
Ref.~\cite{BGPS}, together with scenarios having leptoquarks.

\boldmath
\section{Leptonic $B^-\to\ell^-\bar{\nu}_\ell$ Decays}\label{sec:lept}
\unboldmath
Using the Hamiltonian in Eq.~(\ref{Heff}) we obtain the following branching ratio for the leptonic decays \cite{Kamenik:2008tj}:
\begin{equation}\label{eq:BrNP}
{\mathcal B}(B^-\to\ell^-\bar\nu_\ell)={\mathcal B}(B^-\to\ell^-\bar\nu_\ell)|_{\rm SM}
\left|1+ \frac{M_{B^-}^2}{m_\ell (m_b+m_u)} C_P^\ell \right|^2,
\end{equation}
where the prefactor ${\mathcal B}(B^-\to\ell^-\bar\nu_\ell)|_{\rm SM}$ is the SM branching ratio, which is given in Eq.~(\ref{SM-Br}). Here $m_b$ and $m_u$ are quark masses which enter through 
the use of the equations of motion of the quark fields. To get a better
understanding of the effect of the NP contributions to Eq.~(\ref{eq:BrNP}), we rewrite this expression as
\begin{eqnarray}\label{eq:Blept_split}
{\mathcal B}(B^-\to\ell^-\bar\nu_\ell)&=& 
\frac{G_{\rm F}^2}{8\pi}|V_{ub}|^2 M^3_{B^-} f_{B^-}^2\tau_{B^-}
\Bigl(1-\frac{m^2_{\ell}}{M^2_{B^-}} \Bigl)^2 \nonumber\\
&&\hspace*{-1.9truecm}\times \Biggl[\Bigl(\frac{m_\ell}{M_{B^-}}\Bigl)^2 
+ 2 \frac{m_\ell}{M_{B^-}}\left(\frac{M_{B^-}}{m_b + m_u}\right)  \Re\Bigl(C_P^\ell\Bigl) + \left( \frac{M_{B^-}}{m_b + m_u}\right)^2 |C_P^\ell|^2 \Biggl],
\end{eqnarray}
where we can see how the term proportional to $|C_P^\ell|^2$ may potentially play a dominant role since it is not suppressed by powers of $m_{\ell}/M_{B^-}$.

There is a subtlety related with Eqs.~(\ref{eq:BrNP})~and~(\ref{eq:Blept_split}) when allowing for 
physics beyond the SM: the point is that the value of $|V_{ub}|$ extracted from sophisticated analyses of semileptonic $B$ decays (for an overview, see Ref.~\cite{PDG}) may include NP contributions, thereby precluding us from calculating the SM branching ratio. In order to deal with this issue, our analysis will be based on the study of ratios of branching fractions.

\subsection{Constraints on pseudoscalar NP coefficients from leptonic decay observables}\label{Sec:Leptonicdecay}
We start our analysis by determining bounds for the pseudoscalar Wilson coefficient $C^{\ell}_P$. To this end, we consider the ratio of two leptonic decays to obtain
\begin{equation}\label{eq:Ru}
R^{\ell_1}_{\ell_2}\equiv 
\frac{m^2_{\ell_2}}{m^2_{\ell_1}}\left(\frac{M_{B^-}^2-m_{\ell_2}^2}{M_{B^-}^2-m_{\ell_1}^2}\right)^2
\frac{{\mathcal B}(B^-\to\ell_1^-\bar\nu_{\ell_1})}{{\mathcal B}(B^-\to \ell_2^-\bar\nu_{\ell_2})} =
\left| \frac{1+{\cal C}_{\ell_1; P}}{1+{\cal C}_{\ell_2; P}}  \right|^2,
\end{equation}
with
\begin{equation}\label{eq:CScomplex}
{\cal C}_{\ell; P}\equiv |{\cal C}_{\ell; P}|e^{i\phi_\ell} = \left[\frac{M_{B^-}^2}{m_\ell(m_b+m_q)}\right] C_P^\ell,
\end{equation}
where we have allowed for a generic CP-violating phase $\phi^q_\ell$. We would like to highlight some interesting features of $R^{\ell_1}_{\ell_2}$: unlike the leptonic branching ratios themselves, this quantity has the advantage that it does not depend on $|V_{ub}|$. Moreover, it is theoretically clean as the decay constants cancel. Finally, in the absence of NP contributions, corresponding to $C^{\ell_1}_P=C^{\ell_2}_P=0$, we have by definition $R^{\ell_1}_{\ell_2}|_{\rm SM}=1$.

Let us first assume that the NP short-distance coefficients $C^{\ell}_P$ in Eq.~(\ref{eq:CScomplex}) are real, and study the constraints we obtain from the leptonic decay ratios defined in Eq.~(\ref{eq:Ru}). For the specific determination of $R^{\ell_1}_{\ell_2}$, we consider the tau--muon and the electron--muon pairs, i.e.\ $R^{\tau}_{\mu}$  and $R^{e}_{\mu}$, respectively.

\begin{center}
\begin{figure}[H]
\begin{center}
\includegraphics[width=0.55\textwidth]{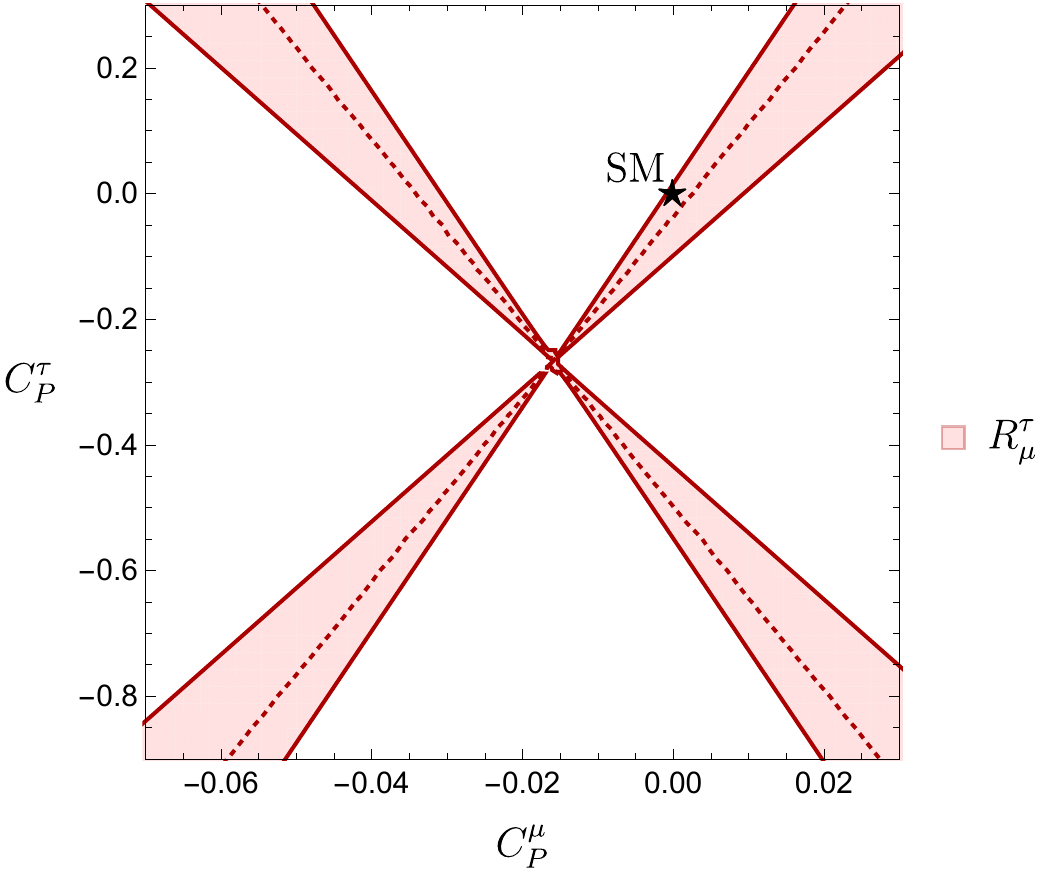} \qquad
\caption{Allowed regions in the  $C^{\mu}_P$--$C^{\tau}_P$ plane following from the leptonic ratio $R^{\tau}_{\mu}$.}
\label{fig:a}
\end{center}
\end{figure}
\end{center}

We obtain the experimental value of $R^{\tau}_{\mu}$ using as numerical inputs Eqs.~(\ref{br-btaunu-avg}) and (\ref{Belle-res}), yielding
\begin{eqnarray}\label{eq:RTaumu}
R^{\tau}_{\mu}=0.76\pm 0.36.
\end{eqnarray}
By comparing this experimental result with the corresponding theoretical expression, we determine the allowed regions in the $C^{\mu}_P$--$C^{\tau}_P$ plane, resulting in the cross-shaped area shown in Fig.~\ref{fig:a}. Here and throughout the rest of this work, the dotted lines define the central value of the corresponding observable, and the $1\sigma$ allowed regions are bounded by the solid lines. It is interesting to note that the constraints obtained are in agreement with the SM point $(C_P^\mu = C_P^\tau = 0)$, which is indicated by the black star in Fig.~\ref{fig:a}.

\begin{center}
\begin{figure}[H]
\begin{center}
\includegraphics[width=0.55\textwidth]{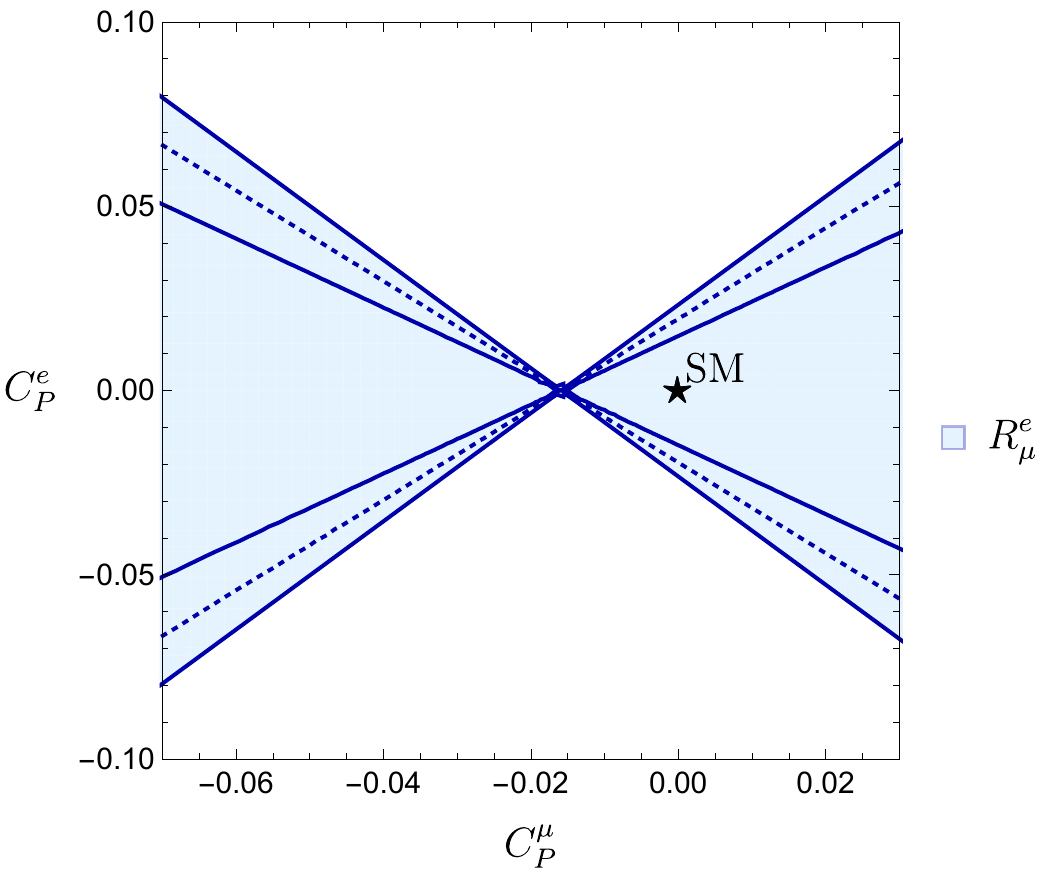} \qquad
\caption{Allowed regions in the $C^{\mu}_P$--$C^{e}_P$ plane following from the leptonic ratio $R^{e}_{\mu}$. The dotted line defines where the bound in Eq.~(\ref{eq:Remubound}) is saturated, with an error indicated by the solid lines due to the uncertainty of the measurement of $\mathcal{B}(B^-\rightarrow \mu^- \bar{\nu}_{\mu})$.}
\label{fig:b}
\end{center}
\end{figure}
\end{center}

To calculate $R^{e}_{\mu}$, we require $\mathcal{B}(B^-\rightarrow e^- \bar{\nu}_e)$ and $\mathcal{B}(B^-\rightarrow \mu^- \bar{\nu}_{\mu})$. For the former only the upper bound in Eq.~(\ref{Belle-enu-limit}) is available. Since this quantity defines the numerator in $R^{e}_{\mu}$, we obtain the experimental bound 
\begin{eqnarray}\label{eq:Remubound}
R^{e}_{\mu}<6.48 \times 10^4,
\end{eqnarray}
with an overall error of $\pm 2.75\times 10^{4}$ induced by the uncertainty associated with $\mathcal{B}(B^-\rightarrow \mu^- \bar{\nu}_{\mu})$. Let us now determine the allowed regions in the $C^{\mu}_P$--$C^{e}_P$ plane derived from the experimental bound in Eq.~(\ref{eq:Remubound}). The result is given by the wedge-shaped regions in Fig.~\ref{fig:b}, which contain the SM point $C_P^\mu = C_P^e = 0$. A future measurement of $\mathcal{B}(B^-\rightarrow e^- \bar{\nu}_e)$ will allow us to determine stringent constraints from $R^{e}_{\mu}$. In that case, we expect a cross-shaped region analogous to the one found for $R^{\tau}_{\mu}$ in Fig.~\ref{fig:a}.

As discussed in Section~\ref{sec:theo}, an important NP scenario that leads to new pseudo-scalar effects in semileptonic decays is the 2HDM. It is instructive to have a closer look at the impact of this scenario on the ratio $R^{\ell_1}_{\ell_2}$ defined in Eq.~(\ref{eq:Ru}), yielding
\begin{eqnarray} \label{eq:2HDMnocons1}
\mathcal{C}_{\ell;P}|_{\rm 2HDM}=-\tan^2\beta \Bigl(\frac{M_{B^-}}{M_{H^{\pm}}}\Bigl)^2\frac{m_b}{m_b + m_u}.
\end{eqnarray}
The right-hand side does actually not depend on the lepton flavour $\ell$, i.e.
\begin{equation}
\mathcal{C}_{\mu;P}|_{\rm 2HDM}=\mathcal{C}_{e;P}|_{\rm 2HDM}=\mathcal{C}_{\tau;P}|_{\rm 2HDM},
\end{equation}
leading to the pattern
\begin{eqnarray} \label{eq:2HDMnocons2}
R^{\tau}_{\mu}|_{\rm 2HDM}=R^{e}_{\mu}|_{\rm 2HDM}=1,
\end{eqnarray}
as in the SM.

\subsection{Implications of CP-violating phases}\label{sec:CPlept}
Let us now explore the impact of CP-violating phases of the NP coefficients $C^{\ell}_P$. To this end, we write
\begin{equation}
C^{\ell}_P=|C^{\ell}_P|e^{i\phi^{\ell}_P},
\end{equation}
where $\phi^{\ell}_P$ coincides with the CP-violating phase of $\mathcal{C}^{u}_{\ell;P}$ in Eq.~(\ref{eq:CScomplex}), and obtain
\begin{equation}\label{eq:complexRu}
R^{\ell_1}_{\ell_2}=\frac{1+2|{\cal C}_{\ell_1;P}|\cos\phi_{\ell_1;P} + |{\cal C}_{\ell_1;P}|^2}{1+
2|{\cal C}_{\ell_2;P}|\cos\phi_{\ell_2}+|{\cal C}_{\ell_2;P}|^2 }.
\end{equation}
We may now convert the experimental value for $R^{\ell_1}_{\ell_2}$ into a correlation between $|{\cal C}_{\ell_1;P}|$ and $|{\cal C}_{\ell_2;P}|$ for given combinations of the CP-violating phases $\phi_{\ell_1}^u$ and $\phi_{\ell_2}^u$:
\begin{equation}\label{eq:Cl1_of_Cl2}
|{\cal C}_{\ell_1;P}| = -\cos\phi_{\ell_1}\pm\sqrt{R^{\ell_1}_{\ell_2}\left[1+2 |{\cal C}_{\ell_2;P}|\cos\phi_{\ell_2} +
|{\cal C}_{\ell_2}|^2 \right]-
\sin^2\phi_{\ell_1}}.
\end{equation}
Assuming real coefficients, i.e.\ $\phi_\mu^u, \phi_\tau^u\in\{0^\circ,180^\circ\}$, yields
\begin{equation}\label{eq:Cul1l2}
|{\cal C}_{\ell_1;P}| = \mp 1 \pm \sqrt{R^{\ell_1}_{\ell_2}}\bigl|1\pm|{\cal C}_{\ell_2;P}|\bigr|,
\end{equation}
which results in a linear correlation between $|{\cal C}_{\ell_1;P}|$ and $|{\cal C}_{\ell_2;P}|$.

\begin{center}
\begin{figure}[H]
\begin{center}
\includegraphics[width=0.6\textwidth]{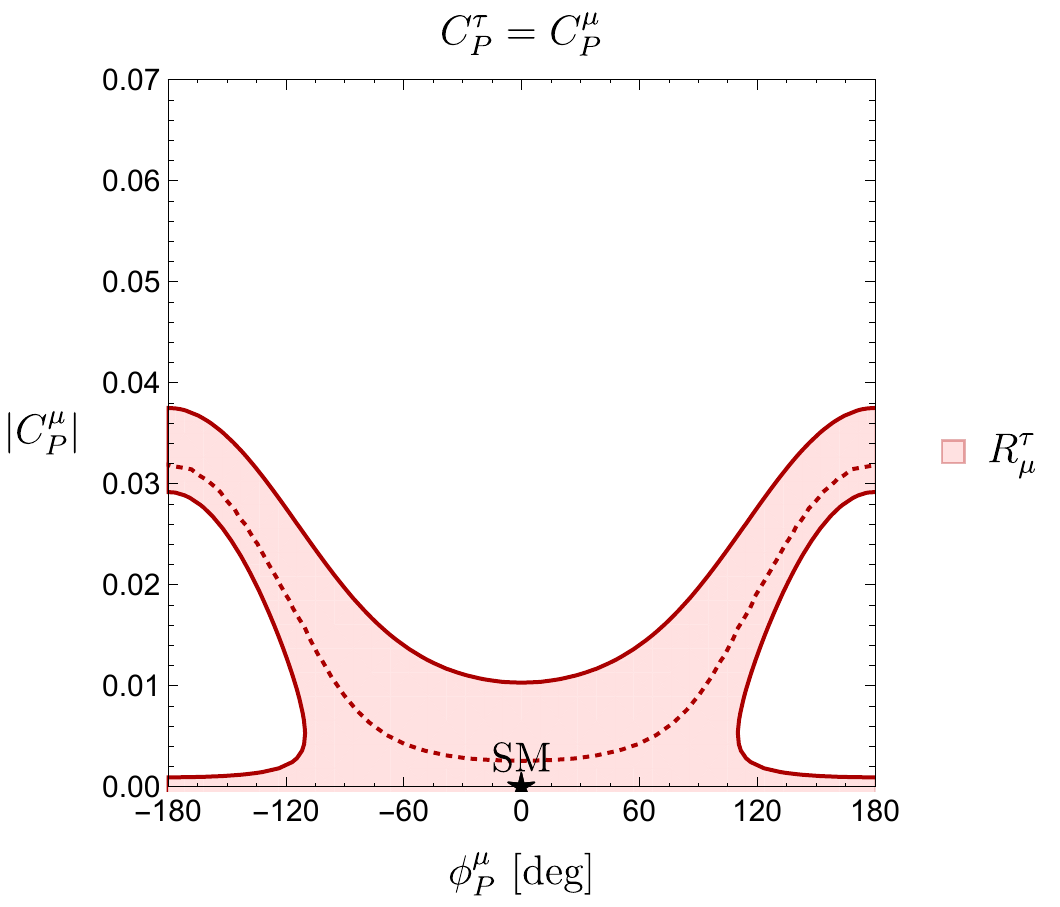} \qquad
\caption{New physics regions in the $\phi_P^\mu$--$|C_P^\mu|$ plane, assuming flavour universality for the pseudoscalar Wilson coefficients of
$\mu$ and $\tau$.}
\label{fig:c}
\end{center}
\end{figure}
\end{center}

Mapping Eq.~(\ref{eq:complexRu}) to the observable $R^{\tau}_{\mu}$ leads to four unknown parameters: $|C^{\mu}_P|$, $\phi^{\mu}_P$, $|C^{\tau}_P|$ and $\phi^{\tau}_P$. Therefore, in order to study the correlation between $|C_P^\mu|$ and its complex phase $\phi_P^\mu$, we have to make an assumption
for $|C_P^\tau|$ and $\phi_P^\tau$. In the case of universal Wilson coefficients for muons and taus, satisfying the relation 
\begin{equation}
C_P^\mu=C_P^\tau,  
\end{equation}
we find 
\begin{equation}
\left(\frac{M_{B^-}}{m_b+m_u}\right)|C^{\mu}_P|=-a\pm\sqrt{a^2-b} \ ,
\end{equation}
with
\begin{equation}
a\equiv\left[\frac{\frac{M_{B^-}}{m_\tau}-R^{\tau}_{\mu}\frac{M_{B^-}}{m_\mu}}{\left(\frac{M_{B^-}}{m_\tau}\right)^2-
R^{\tau}_{\mu}\left(\frac{M_{B^-}}{m_\mu}\right)^2}\right]\cos\phi^{\mu}_P, \quad
b=\frac{1 - R^{\tau}_{\mu}}{\left(\frac{M_{B^-}}{m_\tau}\right)^2-R^{\tau}_{\mu}\left(\frac{M_{B^-}}{m_\mu}\right)^2}.
\end{equation}
The resulting correlation is shown in Fig.~\ref{fig:c}.

\begin{center}
\begin{figure}[H]
\begin{center}
\includegraphics[width=0.6\textwidth]{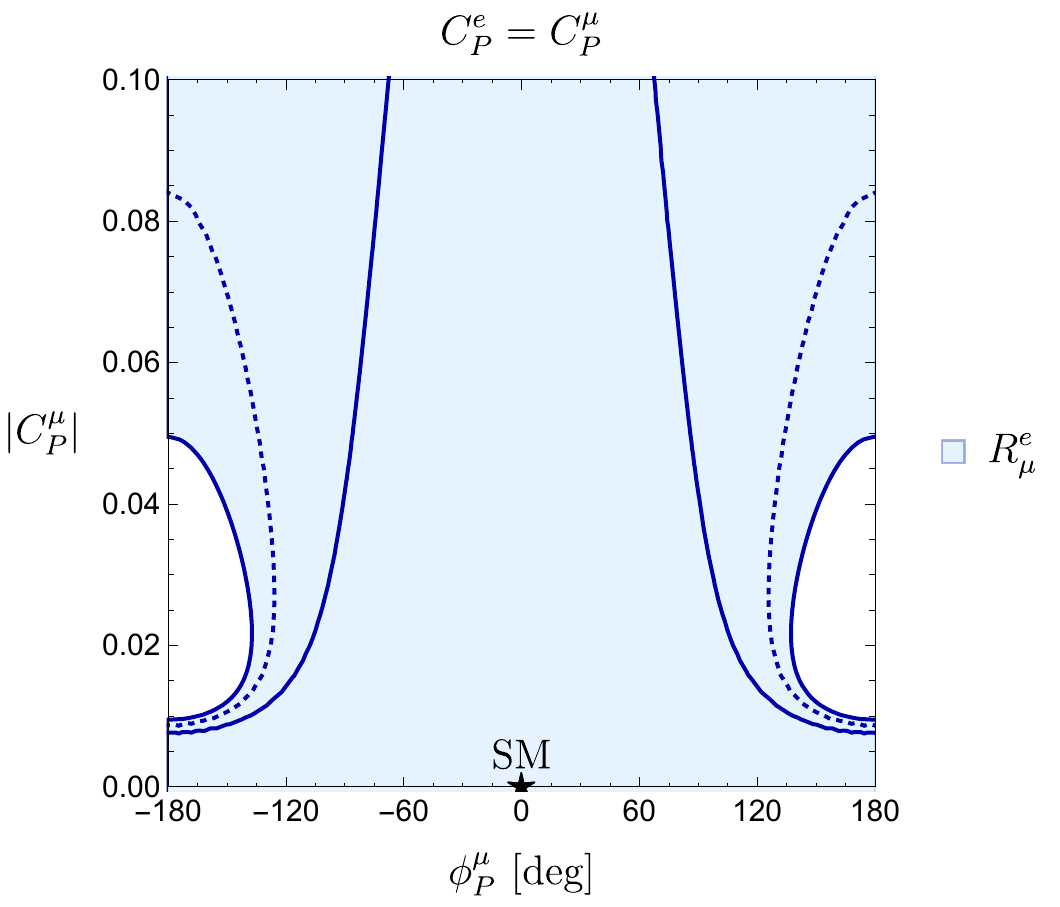} \qquad
\caption{New physics regions in the $\phi_P^\mu$--$|C_P^\mu|$ plane, assuming flavour universality for the pseudoscalar Wilson coefficients of
$\mu$ and $e$.}
\label{fig:d}
\end{center}
\end{figure}
\end{center}

On the other hand, assuming
\begin{equation}
C_P^\mu=C_P^e,
\end{equation}
yields the constraint from $R^{e}_{\mu}$ shown in Fig.~\ref{fig:d}. By looking at Figs.~\ref{fig:c} and \ref{fig:d}, we can see how the SM point $|C^{\mu}_P| = \phi^{\mu}_P = 0$ is consistent with the constraints derived from the current data.

\begin{center}
\begin{figure}[H]
\begin{center}
\includegraphics[width=0.45\textwidth]{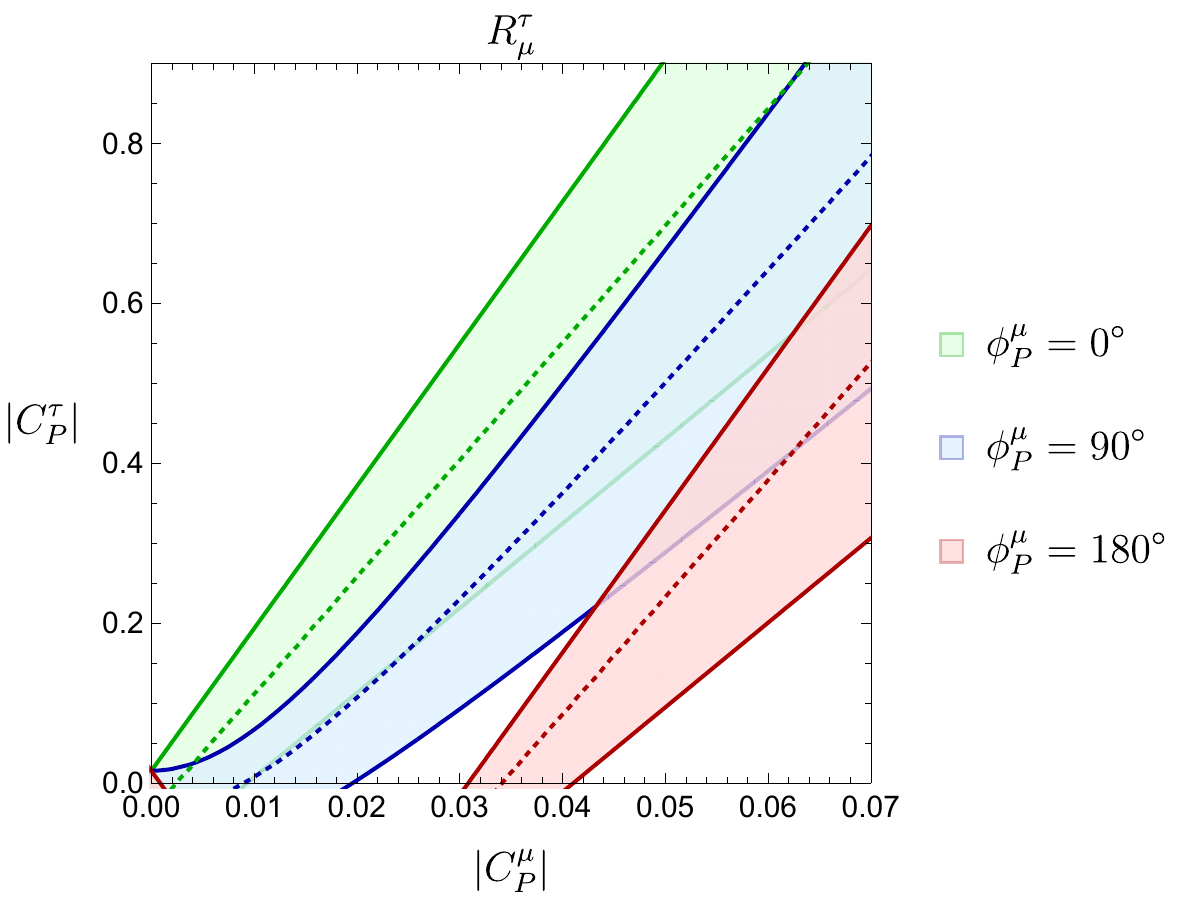}
\includegraphics[width=0.45\textwidth]{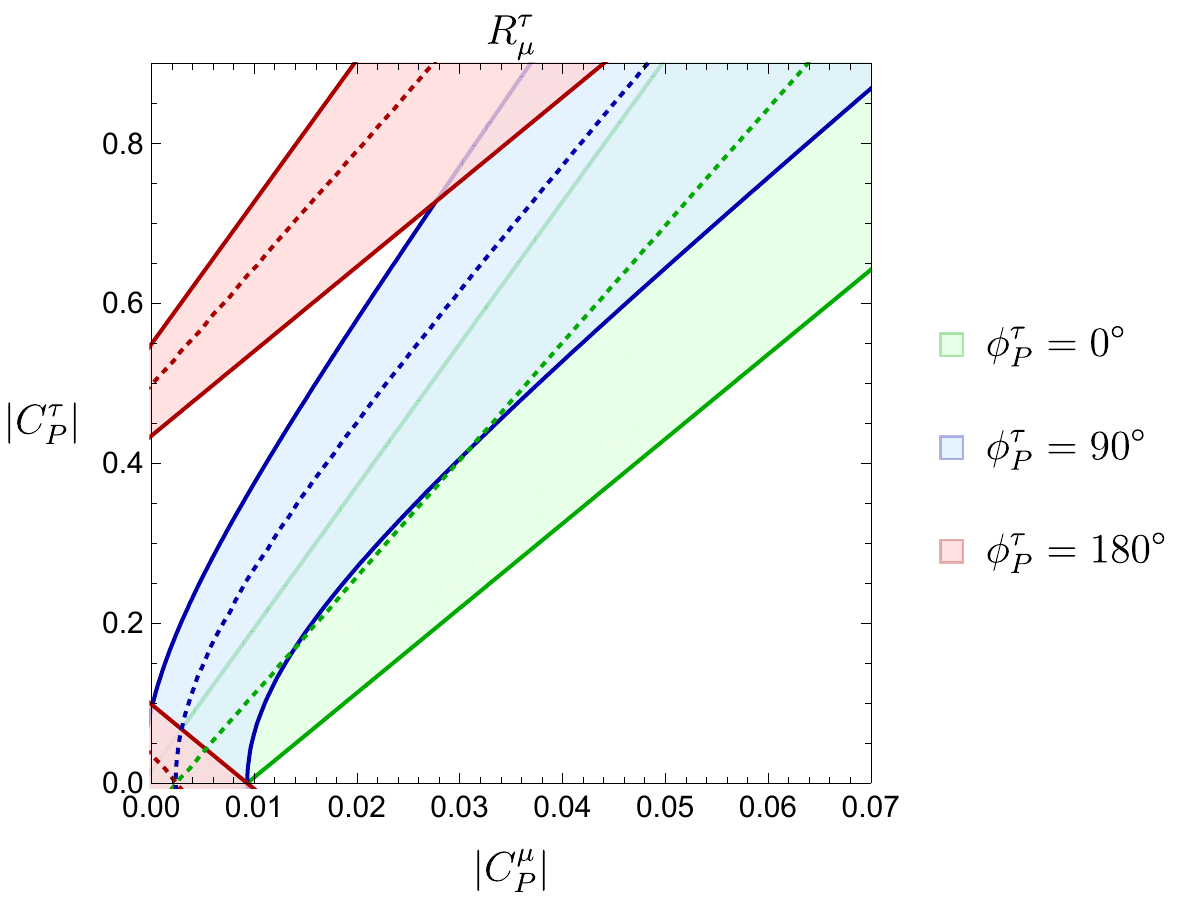}
\caption{Constraints in the $|C^{\mu}_P|$--$|C^{\tau}_P|$ plane from $R^{\tau}_{\mu}$ for different values of $\phi_P^\mu$ and $\phi_P^\tau$. In the left panel, we have $\phi_P^\tau = 0^\circ$ while $\phi_P^\mu$ is varied, whereas in the right panel the roles of $\phi_P^\mu$ and $\phi_P^\tau$ are reversed.}
\label{fig:aa}
\end{center}
\end{figure}
\end{center}

It is also interesting to explore the correlations in the  $|C_P^\mu|$--$|C_P^\tau|$ plane that arise for different combinations of the CP-violating phases $\phi^{\mu}_P$ and $\phi^{\tau}_P$. For example, assuming $\phi^{\tau}_P=0^\circ$ and considering different values of $\phi_P^\mu$, we obtain the patterns shown in the left panel of Fig.~\ref{fig:aa}. On the other hand, the right panel shows how the contours are affected when we keep $\phi_P^\mu = 0^\circ$ and vary $\phi_P^\tau$. For $\phi^{\mu,\tau}_P=0^{\circ}, 180^{\circ}$ the central values of the observables in Fig.~\ref{fig:aa} obey the linear correlation indicated in Eq.~(\ref{eq:Cul1l2}).

\newpage

\boldmath
\section{Semileptonic $\bar{B} \to \rho \ell^- \bar{\nu}_\ell$ and $\bar{B} \to \pi \ell^- \bar{\nu}_\ell$ Decays}\label{sec:semilept}
\unboldmath
We may improve the constraints on the NP short-distance contributions if in addition to the leptonic processes described in Sec.~\ref{sec:lept} we also include semileptonic decays caused by the transition $b\rightarrow u\ell^-\bar{\nu}_{\ell}$. The relevant decays for our analysis are $\bar B\to \rho \ell^- \bar\nu_\ell$ and $\bar B\to \pi \ell^- \bar\nu_\ell$. The first mode depends only on $C^{\ell}_P$ and therefore can be considered the counterpart of the leptonic channels. On the other hand, the process $\bar B\to \pi \ell^- \bar\nu_\ell$ is sensitive to the short distance contribution $C^{\ell}_S$.

The expressions for semileptonic decays have a more complicated structure than those for the leptonic modes due to the hadronic form factors used to calculate the transitions $\bar B\to \rho$ and $\bar B\to \pi$. The kinematical regimes for the semileptonic decays are described in terms of $q^2 \equiv (p_B - p_{\rho,\pi})^2$, where $p_B$ and $p_{\rho,\pi}$ are the four-momenta of the $B$-meson and the $\rho$ or $\pi$, respectively. For low momentum transfer, i.e. $q^2 \in [0,q^2_{\rm{max}}]$ where $12~{\rm{GeV}}^2\leq q^2_{\rm{max}}\leq 16~{\rm{GeV}}^2$, the non-perturbative hadronic form factors are estimated using QCD sum rules. For higher $q^2$ values, lattice determinations are available. Quark models were also used for the determination of the hadronic form factors, as discussed in Ref.~\cite{Ivanov:2017hun}.

The calculation of the non-perturbative contributions to $\bar{B}\rightarrow \pi$ decays
is well developed. As a matter of fact, the corresponding form factors are currently known with good precision. Here we use the parameterization that extrapolates from high to low $q^2$  values introduced originally in Ref.~\cite{Lattice:2015tia} and discussed in more detail in Appendix~\ref{sec:HFF}. In contrast, the form factors for the $\bar{B}\rightarrow \rho$ transitions are less precisely known, and only determinations referring independently to either the low  $q^2$ or high $q^2$ regimes are available in the literature. Moreover, high $q^2$ calculations are more than one decade old \cite{Bowler:2004zb} and have large uncertainties. Later in this section, we will argue on the importance of improving these results.

Semileptonic decays are used for the exclusive determination of $|V_{ub}|$, which is typically done using SM expressions. Therefore, a value of $|V_{ub}|$  based on this approach may already be affected by NP contributions. Consequently, using this parameter as an input in other NP
studies may lead to wrong conclusions.
 To avoid this problem, we propose a different method for the determination of $|V_{ub}|$, which is described in more detail in Sec.~\ref{sec:Vub}. Our strategy is based on two key steps: we first obtain the NP short-distance contributions $C^{\ell}_S$ and $C^{\ell}_P$ using only ratios of branching fractions of leptonic and semileptonic processes. Then, we substitute these results in the individual expressions for the branching fractions in order to extract the value of $|V_{ub}|$.

\boldmath
\subsection{ $\bar B\to \rho \ell^- \bar\nu_\ell$}
\unboldmath

Let us start our study of semileptonic decays  by analyzing the processes  $\bar{B}_d^0 \to \rho^+ \ell^- \bar\nu_\ell$ and $B^- \to \rho^0 \ell^- \bar\nu_\ell$. To simplify the notation, we will refer to both of 
them as $\bar B\to \rho \ell^- \bar\nu_\ell$ when writing expressions that hold  for both cases. Whenever a distinction is required, we will make the charges of the $B$ and $\rho$ mesons explicit. The expression for the decay width of the process $\bar B\to \rho \ell^- \bar\nu_\ell$ in the presence of pseudoscalar NP particles reads \cite{STTW}
\begin{eqnarray}\label{eq:difrho}
&&\frac{d\mathcal{B}(\bar{B}\rightarrow \rho \ell^- \bar{\nu}_{\ell})}{dq^2}=
\frac{G^2_F \tau_B |V_{ub}|^2}{24 \pi^3 m^2_{B}}
\Biggl\{
\Biggl[\frac{1}{4}\Bigl(1 + \frac{m^2_{\ell}}{2q^2}\Bigl)
\Bigl(H^{\rho~2}_{V,+} + H^{\rho~2}_{V,-} + H^{\rho~2}_{V,0}\Bigl) + \frac{3}{8}\frac{m^2_{\ell}}{q^2} H^{\rho~2}_{V, t}\Biggl]   \nonumber \\
&&
+\frac{3}{8}|C^{\ell}_P|^2 H^{\rho~2}_S
+ \frac{3}{4}\Re\Bigl[C^{\ell*}_P\Bigl]\frac{m_{\ell}}{\sqrt{q^2}}
H^{\rho}_S H^{\rho}_{V, t}
\Biggl\}\frac{(q^2-m^2_{\ell})^2}{q^2} |\vec{p}_{\rho}|,\end{eqnarray}
where  $q^2$ is the four-momentum transfer to the leptonic system composed by the $\ell$ and the $\bar{\nu}_{\ell}$, which satisfies
\begin{equation} \label{eq:qsqbounds}
m_\ell^2 \leq q^2 \leq (M_{B} - M_\rho)^2.
\end{equation}
The hadronic form factors in the helicity basis are given by $H^{\rho}_{V,\pm}, H^{\rho}_{V, 0}, H^{\rho}_{V, t}, H^{\rho}_{S}$; more details about these quantities are provided in Appendix \ref{sec:HFF}.
The norm of the  three-momentum of the $\rho$ meson in the rest frame of the $\bar B$ meson is given by
\begin{equation}\label{eq:vecprho}
|\vec{p}_\rho|=\frac{\sqrt{\Bigl[(M_{B}-M_\rho)^2-q^2\Bigl]\Bigl[(M_{B}+M_\rho)^2-q^2\Bigl]}}{2M_{B}}.
\end{equation}
In addition, the angular distribution contains more observables that are sensitive to (pseudo)-scalar operators. In particular, the coefficient $J_{6c}$, which enters the forward-backward asymmetry, only takes a non-vanishing value when there are new scalar contributions \cite{Altmannshofer:2008dz}. For a discussion on the angular analysis of $\bar{B}\rightarrow \rho \ell^- \bar{\nu}_{\ell}$ see Ref.~\cite{Bernlochner:2014ova}.

To constrain the Wilson coefficients $C^{\ell}_P$  (for $\ell=e, \mu, \tau$), we introduce the following $|V_{ub}|$-independent ratios:
\begin{eqnarray}\label{eq:idealratiosrho}
\mathcal{R}^e_{e;\rho} \equiv \frac{\mathcal{B}(B^-\rightarrow e \bar{\nu}_e)}{\mathcal{B}(\bar{B}\rightarrow \rho e^- \bar{\nu}_e)},
\quad\quad
\mathcal{R}^{\mu}_{\mu;\rho} \equiv \frac{\mathcal{B}(B^-\rightarrow \mu^- \bar{\nu}_{\mu})}{\mathcal{B}(\bar{B}\rightarrow \rho \mu^- \bar{\nu}_{\mu})},
\quad
\quad
\mathcal{R}^{\tau}_{\tau;\rho} \equiv \frac{\mathcal{B}(B^-\rightarrow \tau^- \bar{\nu}_{\mu})}{\mathcal{B}(\bar{B}\rightarrow \rho \tau^- \bar{\nu}_{\tau})}.
\end{eqnarray}
Unfortunately, there is not enough experimental information available to evaluate these observables. To the best of our knowledge, in the case of the $B\rightarrow \rho$ semileptonic transitions, there are only measurements for the decay probabilities of the combined channels $\bar{B} \rightarrow \rho e^- \bar{\nu}_e$ and $\bar{B} \rightarrow \rho \mu^- \bar{\nu}_{\mu}$ available for different $q^2$ bins. 
In view of the recent results on lepton flavour universality violations, we urge to have independent experimental determinations for each leptonic flavour, and then assess the effects of potential NP contributions for $e^-$, $\mu^-$ and $\tau^-$ independently. In our study, we consider therefore the leptonic average
\begin{eqnarray}\label{eq:average2}
\Braket{{\mathcal B}(\bar B^0\rightarrow \rho^+ \ell^- \bar{\nu}_{\ell})}_{[\ell=~ e, \mu]}&=&
\frac{1}{2}\Bigl({\mathcal B}(\bar B^0\rightarrow \rho^+ e^- \bar{\nu}_{e}) + {\mathcal B}(\bar B^0\rightarrow \rho^+ \mu^- \bar{\nu}_{\mu})\Bigl),
\end{eqnarray}
and correspondingly for the $B^-$ meson. In addition, we use the isospin symmetry to introduce a second average
\begin{eqnarray}\label{eq:average1}
\Braket{{\mathcal B}(\bar B \rightarrow \rho  \ell^- \bar{\nu}_{\ell})}
&=&\frac{1}{2} \Bigl(\Braket{{\mathcal B}(\bar B^0\rightarrow \rho^+ \ell^- \bar{\nu}_{\ell})}
+ 2  \Braket{{\mathcal B}( B^-\rightarrow \rho^0 \ell^- \bar{\nu}_{\ell})}
\Bigl).
\end{eqnarray}

We start by studying the behaviour of the semileptonic decay $\bar{B} \rightarrow \rho \ell^- \bar{\nu}_{\ell}$ for values of $q^2$ within the low-$q^2$ range $0\leq q^2\leq 12~\rm{GeV}^2$, since this is the range for which QCD sum rule calculations of the form factors are available. The experimental information provided by Belle \cite{Sibidanov:2013rkk} in this region leads to
\begin{eqnarray}\label{eq:Exprho}
\Braket{{\mathcal B}(\bar B^0\rightarrow \rho^+ \ell^- \bar{\nu}_{\ell})}_{[\ell=~ e, \mu],~q^2\leq 12~\rm{GeV}^2}&=&(1.90 \pm 0.20)\times 10^{-4},\nonumber\\
2\Braket{{\mathcal B}( B^-\rightarrow \rho^0 \ell^- \bar{\nu}_{\ell})}_{[\ell=~ e, \mu],~q^2\leq 12~\rm{GeV}^2}&=&(2.03 \pm 0.16) \times 10^{-4}.
\end{eqnarray}
We combine the previous measurements through a weighted average \cite{PDG} using the isospin symmetry as indicated in Eq.~(\ref{eq:average1}), yielding
\begin{eqnarray}\label{eq:Brhoaverage}
\Braket{{\mathcal B}(\bar B\rightarrow \rho \ell^- \bar{\nu}_{\ell})}_{[\ell=~ e, \mu],~q^2\leq 12~\rm{GeV}^2}&=&(1.98 \pm 0.12) \times 10^{-4}.
\end{eqnarray}
This allows us to introduce the following ratio as an alternative to the observables in Eq.~(\ref{eq:idealratiosrho}):
\begin{eqnarray}\label{eq:leptonicoversemileptonicrho-theo}
\mathcal{R}^{\mu}_{\Braket{e, \mu}; \rho ~ [q^2\leq 12]~\rm{GeV}^2}&\equiv&\mathcal{B}(B^-\rightarrow \mu^- \bar{\nu})/
\Braket{{\mathcal B}(\bar B \rightarrow \rho \ell^- \bar{\nu}_{\ell})}_{[\ell=~ e, \mu],~q^2\leq 12~\rm{GeV}^2}.
\end{eqnarray}
Using the experimental information in Eqs.~(\ref{eq:SMleptBrmu}) and (\ref{eq:Brhoaverage}), we then obtain
\begin{eqnarray}\label{eq:leptonicoversemileptonicrho}
\mathcal{R}^{\mu}_{\Braket{e, \mu}; \rho ~ [q^2\leq 12]~\rm{GeV}^2}&=&(3.3 \pm 1.4)\times 10^{-3}.
\end{eqnarray}

We proceed with the evaluation of the SM value. Applying the formulae in Eqs.~(\ref{eq:difrho}), (\ref{eq:average2}), (\ref{eq:average1}), (\ref{eq:leptonicoversemileptonicrho-theo}) and evaluating the corresponding form factors as indicated in Appendix~\ref{sec:HFF}, we get
\begin{eqnarray}
\mathcal{R}^{\mu}_{\Braket{e, \mu}; \rho ~ [q^2\leq 12]~\rm{GeV}^2}\Bigl |_{\rm{SM}}&=& (1.52 \pm 0.29)\times 10^{-3}.
\end{eqnarray}
We note that this value agrees with the experimental information at the $(1$--$2)\sigma$ level.

To conclude this section, we would like to obtain better insights into the structure of Eq.~(\ref{eq:difrho}). For the purpose of the discussion in the remainder of this section, it is convenient to define
\begin{eqnarray}\label{eq:newparam}
s\equiv \sqrt{q^2}, \quad \quad \xi_{\ell}\equiv \frac{m_{\ell}}{s},
\end{eqnarray}
and write Eq.~(\ref{eq:difrho}) in terms of these parameters as follows:
\begin{eqnarray}\label{eq:rhosparameter}
&&\frac{1}{s^2}\frac{d\mathcal{B}(\bar{B}\rightarrow \rho \ell^- \bar{\nu}_{\ell})}{ds^2}=
\frac{G^2_F \tau_B |V_{ub}|^2}{24 \pi^3 m^2_{B}}
\Biggl\{
\Biggl[\frac{1}{4}\Bigl(1 + \frac{1}{2} \xi_{\ell}^2\Bigl)
\Bigl(H^{\rho~2}_{V,+} + H^{\rho~2}_{V,-} + H^{\rho~2}_{V,0}\Bigl) + \frac{3}{8}\xi_{\ell}^2 H^{\rho~2}_{V, t}\Biggl] \nonumber\\
&&+\frac{3}{8}|C_P|^2 H^{\rho~2}_S + \frac{3}{4}\Re\Bigl[C^{*}_P\Bigl]\xi_{\ell}
H^{\rho}_S H^{\rho}_{V, t}
\Biggl\} \Bigl(1-\xi^2_{\ell}\Bigl)^2 |\vec{p}_{\rho}|.
\end{eqnarray}
When $s$ is sufficiently large, we have $\xi_\ell \ll 1$ and we may neglect the terms proportional to $\xi_\ell$.
We see that in this case and within the SM, only the term proportional to  
$H^{\rho~2}_{V,+} + H^{\rho~2}_{V,-} + H^{\rho~2}_{V,0}$ contributes 
to Eq.~(\ref{eq:rhosparameter}). It should be noted that this term is flavour universal,  i.e. it does not depend on $m_{\ell}$.

One has to be careful when neglecting $\xi_\ell$ terms in Eq.~(\ref{eq:rhosparameter}) as the bounds on $q^2$ in Eq.~(\ref{eq:qsqbounds}) yield
\begin{equation}\label{eq:slimits}
m^2_\ell \leq s^2 \leq (M_{B} - M_\rho)^2.
\end{equation}
Consequently, at low momentum transfer, $\xi_\ell$ is $\mathcal{O}(1)$ and cannot be neglected. It is a priori not obvious whether Eq.~(\ref{eq:rhosparameter}) gives accurate results for $\xi_\ell = 0$ when integrating over the range $0 \leq s^2\leq 12~\rm{GeV}^2$. In order to shed more light on this issue, we compare $\Braket{\mathcal{B}(\bar B\rightarrow \rho \ell^- \bar{\nu_{\ell}})}$
with the full rate where $\xi_\ell \neq 0$. To that end, we introduce
\begin{equation}
\delta^{\rho;\ \ell}_{\rm SL} \equiv \frac{\Braket{\mathcal{B}(\bar B\rightarrow \rho \ell^- \bar{\nu}_{\ell})}_{\xi_\ell=0}-\Braket{\mathcal{B}(\bar B\rightarrow \rho \ell^- \bar{\nu}_{\ell})}}{\Braket{\mathcal{B}(\bar B\rightarrow \rho \ell^- \bar{\nu}_{\ell})}},
\end{equation}
where we integrate over the given kinematic range and take the isospin average. Assuming the SM, the numerical evaluation gives
\begin{equation}
\delta^{\rho;\ e}_{\rm SL}=4.1\times 10^{-8}, \qquad \delta_{\rm SL}^{\rho; \ \mu} = 3.0\times10^{-3},
\end{equation}
thereby demonstrating that integrating Eq.~(\ref{eq:rhosparameter}) for $\xi_\ell = 0$ provides a good approximation of the branching ratio for the light lepton flavours. Consequently, the assumption of flavour universality works particularly well within the SM in the case of electrons and muons. This justifies  the usual approach  followed for the extraction of $|V_{ub}|$ of averaging over light leptons with the aim of improving the precision by increasing the statistics. On the other hand, for $\ell=\tau$ we find
\begin{equation}
\delta^{\rho;\ \tau}_{\rm SL} = 1.62,
\end{equation}
showing that in this case the leptonic mass cannot be neglected. This result is not surprising since the range in Eq.~(\ref{eq:slimits}) yields
\begin{equation}
0.16 \leq \xi_\tau^2 \leq 1,
\end{equation}
showing how the relatively large mass of the $\tau$ has a non-negligible impact on the phase space of the integral to calculate the semileptonic branching fraction.

\subsubsection{Constraints on pseudoscalar NP coefficients from \boldmath $\bar{B}\rightarrow \rho \ell^- \bar{\nu}_{\ell}$ \unboldmath}

\begin{center}
\begin{figure}[H]
\begin{center}
\includegraphics[width=0.6\textwidth]{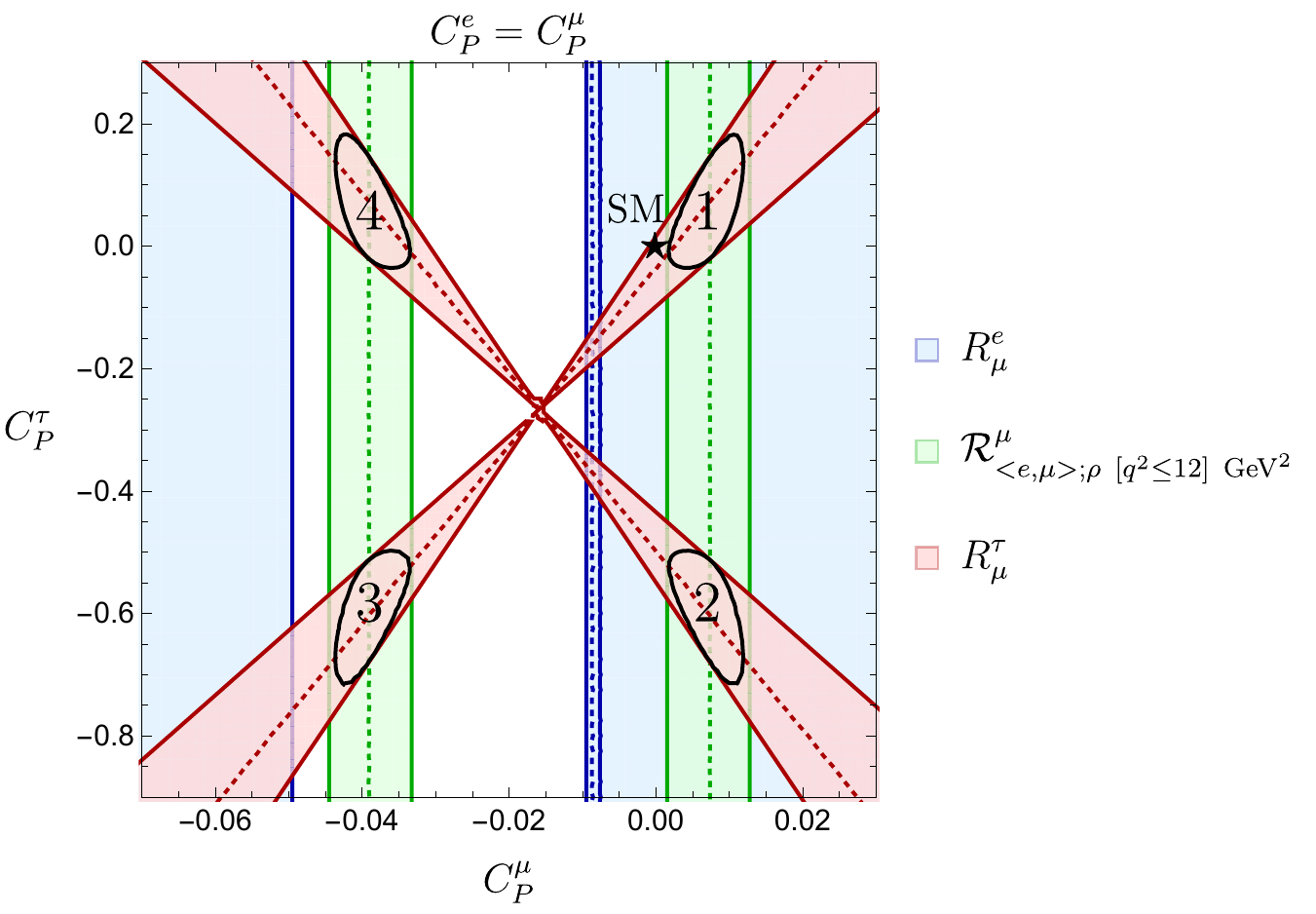} \qquad
\caption{Allowed regions in the  $C^{\mu}_P$--$C^{\tau}_P$ plane utilizing the ratios $R^{e}_{\mu}$, $R^{\tau}_{\mu}$ and $\mathcal{R}^{\mu}_{\Braket{e, \mu}; \rho~[q^2\leq 12]~\rm{GeV}^2}$ under the assumption $C^e_P=C^{\mu}_P$.}
\label{fig:lepsemmutau}
\end{center}
\end{figure}
\end{center}

Using the observable in Eq.~(\ref{eq:leptonicoversemileptonicrho-theo}) and making the assumption $C^{e}_P=C^{\mu}_P$, we can derive further constraints on the regions shown in Fig.~\ref{fig:a}. In particular, the range for $C_P^\mu$ following from $\mathcal{R}^{\mu}_{\Braket{e, \mu};\rho~[q^2\leq 12]~\rm{GeV}^2}$ yields the green vertical bands shown in Fig.~\ref{fig:lepsemmutau}. The combination with $R_\mu^\tau$ gives us then four allowed regions. Performing a $\chi^2$ fit to these two observables yields the $1\sigma$ allowed regions given by the black contours. Since the Wilson coefficients $C_P^\mu$ and $C_P^e$ are correlated, we may in addition include the ratio $R^{e}_{\mu}$ to obtain even stronger constraints. For $C_P^e=C_P^\mu$, this observable yields the blue region in Fig.~\ref{fig:lepsemmutau}, selecting the right green band and excluding solutions 3 and 4 satisfying $C_P^\mu < 0$.

\begin{center}
\begin{figure}[H]
\begin{center}
\includegraphics[width=0.6\textwidth]{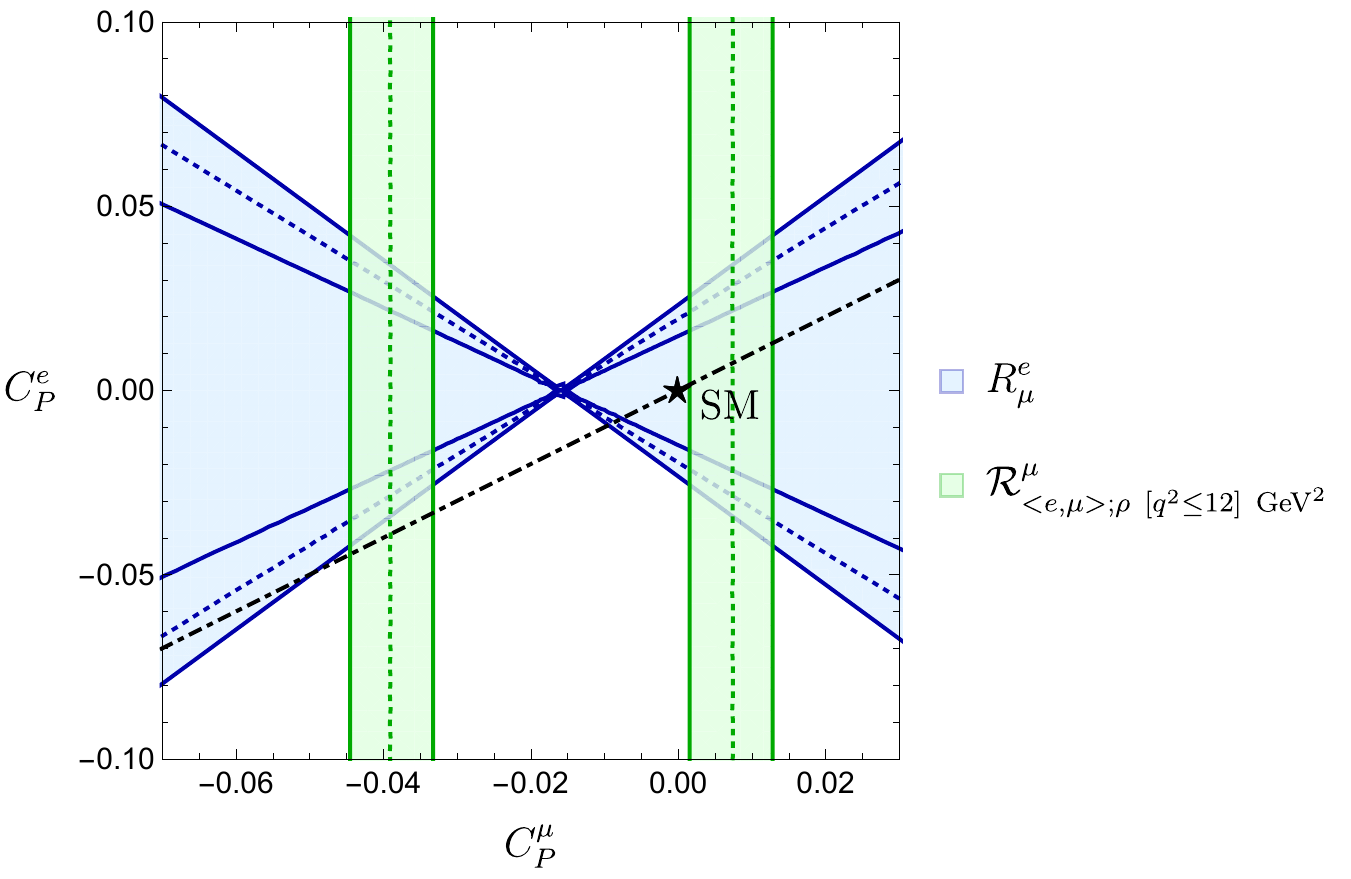} \qquad
\caption{Allowed regions in the  $C^{\mu}_P$--$C^{e}_P$ plane utilizing the ratios $R^{e}_{\mu}$ and $\mathcal{R}^{\mu}_{\Braket{e, \mu};\rho~[q^2\leq 12]~\rm{GeV}^2}$. The dashed-dotted line indicates the correlation arising for $C_P^e = C_P^\mu$.}
\label{fig:lepsemmue}
\end{center}
\end{figure}
\end{center}

Giving up on the condition $C^e_P=C^{\mu}_P$, we can constrain these coefficients independently and refine the bounds in Fig.~\ref{fig:b}. We then obtain the results shown in Fig.~\ref{fig:lepsemmue}, where the dashed-dotted line corresponds to $C_P^e = C_P^\mu$.  We see how in both Figs.~\ref{fig:lepsemmutau}~and~\ref{fig:lepsemmue} the SM point is about $1 \sigma$ away from the allowed regions given by the  intersection of our different constraints.

\boldmath\label{eq:RatrhoP}
\subsection{ $\bar B\to \pi \ell^- \bar\nu_\ell$}
\unboldmath

Until now we have studied different leptonic and semileptonic constraints on the Wilson coefficient $C^{\ell}_P$ only. To obtain sensitivity for  $C^{\ell}_S$, we include the processes $\bar B^0 \to \pi^+ \ell^- \bar\nu_\ell$ and $B^- \to \pi^0 \ell^- \bar\nu_\ell$ with $\ell=e,\mu,\tau$, to be denoted generically as $\bar B \to \pi \ell^- \bar\nu_\ell$. The corresponding differential branching ratio in the presence of scalar NP contributions takes the following form \cite{STTW}:
\begin{eqnarray} \label{eq:dBrpi}
&&\frac{d\mathcal{B}(\bar B\rightarrow \pi \ell^- \bar{\nu_{\ell}})}{dq^2}=\frac{G^2_F \tau_{B}|V_{ub}|^2}
{24 \pi^3 M^2_{B}}\Biggl\{ \Biggl[ 
\Bigl( 1+\frac{m^2_\ell}{2q^2} \Bigl)\frac{H^{\pi~2}_{V, 0}}{4} + \frac{3}{8}\frac{m^2_{\ell}}{q^2}H^{\pi~2}_{V, t}
\Biggl]\nonumber\\
&&+ \frac{3}{8}|C^{\ell}_S|^2 H^{\pi~2}_S + \frac{3}{4}\Re\Bigl[ C^{\ell~*}_S\Bigl]\frac{m_{\ell}}{\sqrt{q^2}}
H^{\pi}_{S} H^{\pi}_{V,t}
\Biggl\}\frac{(q^2-m^2_{\ell})^2}{q^2} |\vec{p}_{\pi}|.
\end{eqnarray}
The full kinematical range for $q^2$ is
\begin{equation} \label{eq:qsqboundspi}
m_\ell^2 \leq q^2 \leq (M_{B} - M_\pi)^2.
\end{equation}
The hadronic form factors in the helicity basis are denoted as $H^{\pi}_{V, 0}$, $H^{\pi}_{V, t}$, $H^{\pi}_S$ and are described in more detail in Appendix~\ref{sec:HFF}. Moreover, the three momentum of the pion is given by
\begin{equation}
|\vec{p}_\pi|=\frac{\sqrt{\Bigl[(M_{B}-M_\pi)^2-q^2\Bigl]\Bigl[(M_{B}+M_\pi)^2-q^2\Bigl]}}{2M_{B}}.
\end{equation}

In analogy with Eq.~(\ref{eq:idealratiosrho}), we introduce the following observables:
\begin{eqnarray}\label{eq:idealratiospion}
\mathcal{R}^e_{e;\pi} \equiv \frac{\mathcal{B}(B^-\rightarrow e \bar{\nu}_e)}{\mathcal{B}(\bar{B}\rightarrow \pi e^- \bar{\nu}_e)},
\quad
\mathcal{R}^{\mu}_{\mu;\pi} \equiv \frac{\mathcal{B}(B^-\rightarrow \mu^- \bar{\nu}_{\mu})}{\mathcal{B}(\bar{B}\rightarrow \pi \mu^- \bar{\nu}_{\mu})},
\quad
\mathcal{R}^{\tau}_{\tau;\pi} \equiv \frac{\mathcal{B}(B^-\rightarrow \tau^- \bar{\nu}_{\tau})}{\mathcal{B}(\bar{B}\rightarrow \pi \tau^- \bar{\nu}_{\tau})}.
\end{eqnarray}
This set of ratios is sensitive to  $C^{\ell}_P$ and $C^{\ell}_S$. Just as for the $\bar{B}\rightarrow \rho$ 
processes, we do not have independent determinations of the $\bar{B}\rightarrow \pi e^- \bar{\nu}_e$ and $\bar{B}\rightarrow \pi \mu^- \bar{\nu}_\mu$ branching ratios. Instead, the following leptonic averages are available experimentally \cite{PDG}:
\begin{eqnarray}\label{eq:Exppi}
\Braket{{\mathcal B}(\bar B^0\rightarrow \pi^+ \ell^- \bar{\nu}_{\ell})}_{[\ell=~ e, \mu]}&=&(1.50 \pm 0.06)\times 10^{-4},\nonumber\\
2\Braket{{\mathcal B}( B^-\rightarrow \pi^0 \ell^- \bar{\nu}_{\ell})}_{[\ell=~ e, \mu]}&=&(1.56 \pm 0.05) \times 10^{-4}.
\end{eqnarray}
We combine these determinations using again the isospin symmetry to obtain
\begin{eqnarray}\label{eq:piexp}
\Braket{{\mathcal B}(\bar{B}\rightarrow \pi \ell^- \bar{\nu}_{\ell})}_{[\ell=~ e, \mu]}&=&(1.53 \pm 0.04)\times 10^{-4},
\end{eqnarray}
and introduce the observable
\begin{eqnarray}\label{eq:leptonicoversemileptonicpi}
\mathcal{R}^{\mu}_{\Braket{e, \mu}; \pi}&\equiv&\mathcal{B}(B^-\rightarrow \mu^- \bar{\nu})/
\Braket{{\mathcal B}(\bar B \rightarrow \pi \ell^- \bar{\nu}_{\ell})}_{[\ell=~ e, \mu]},
\end{eqnarray}
which takes the current experimental value
\begin{eqnarray}
\mathcal{R}^{\mu}_{\Braket{e, \mu}; \pi}&=&(4.2 \pm 1.8)\times 10^{-3}.
\end{eqnarray}
This may be compared with the SM value, for which we obtain
\begin{equation}
\mathcal{R}^{\mu}_{\Braket{e, \mu}; \pi}|_{\rm SM} = (3.18 \pm 0.96) \times 10^{-3},
\end{equation}
which is in good agreement with the experimental value.

We may rewrite Eq.~(\ref{eq:dBrpi}) using the parameterization introduced in Eq.~(\ref{eq:newparam}), yielding
\begin{eqnarray} \label{eq:dBrpinewpar}
&&\frac{1}{s^2}\frac{d\mathcal{B}(\bar B\rightarrow \pi \ell^- \bar{\nu}_{\ell})}{ds^2}=\frac{G^2_F \tau_{B}|V_{ub}|^2}
{24 \pi^3 M^2_{B}}\Biggl\{ \Biggl[ 
\Bigl( 1+\frac{1}{2}\xi^2_{\ell} \Bigl)\frac{H^{\pi~2}_{V, 0}}{4} + \frac{3}{8} \xi^2_{\ell} H^{\pi~2}_{V, t}
\Biggl]\nonumber\\
&&+ \frac{3}{8}|C^{\ell}_S|^2 H^{\pi~2}_S + \frac{3}{4}\Re\Bigl[ C^{\ell~*}_S\Bigl]
\xi_{\ell}
H^{\pi}_{S} H^{\pi}_{V,t}
\Biggl\} \Bigl(1-\xi^2_{\ell}\Bigl)^2 |\vec{p}_{\pi}|.
\end{eqnarray}
As for the $\bar{B} \rightarrow \rho$ transitions, we assess the validity of Eq.~(\ref{eq:dBrpinewpar}) through the difference
\begin{equation}
\delta_{\rm SL}^{\pi; \ \ell} \equiv \frac{\Braket{\mathcal{B}(\bar B\rightarrow \pi \ell^- \bar{\nu}_{\ell})}_{\xi_\ell=0}-\Braket{\mathcal{B}(\bar B\rightarrow \pi \ell^- \bar{\nu}_{\ell})}}{\Braket{\mathcal{B}(\bar B\rightarrow \pi \ell^- \bar{\nu}_{\ell})}},
\end{equation}
where we consider again the isospin average. For $\delta_{\rm SL}^{\pi;e}$ and $\delta_{\rm SL}^{\pi;\mu}$, we find tiny values at the $10^{-8}$ and $10^{-3}$ levels, respectively, when considering the SM. This shows that taking $\xi_\ell = 0$ in  Eq.~(\ref{eq:dBrpinewpar}) provides a good approximation of the branching ratio. On the other hand, for $\ell = \tau$, the correction factor due to the mass of the $\tau$ lepton is $\delta_{\rm SL}^{\pi; \ \tau} = 42 \%$.

\subsubsection{Constraints on (pseudo)-scalar NP coefficients from \boldmath $\bar{B}\rightarrow \pi \ell^- \bar{\nu}_{\ell}$ \unboldmath}

\begin{center}
\begin{figure}[H]
\begin{center}
\includegraphics[width=0.6\textwidth]{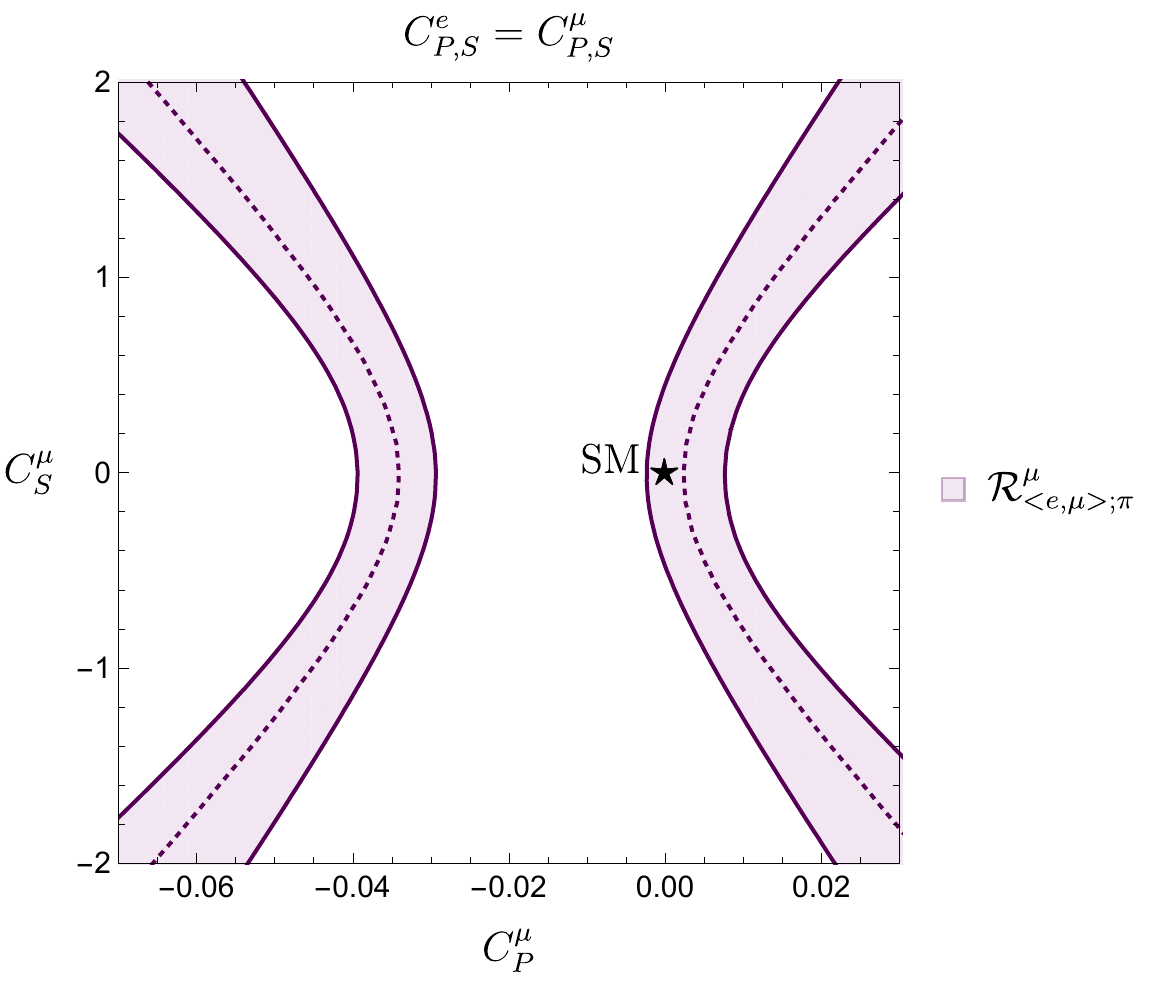} \qquad
\caption{Constraints in the  $C^{\mu}_P$--$C^{\mu}_S$ plane obtained from the leptonic over semileptonic ratio $\mathcal{R}^{\mu}_{\Braket{e, \mu};\pi}$.}
\label{fig:lepsempi}
\end{center}
\end{figure}
\end{center}

Thanks to the observable in Eq.~(\ref{eq:leptonicoversemileptonicpi}), we may now obtain stronger bounds for $C^{\ell}_P$ and  $C^{\ell}_S$. If we make the assumptions
\begin{eqnarray}
C_{P}^e &=& C_{P}^\mu,\nonumber\\
C_{S}^e &=& C_{S}^\mu,
\end{eqnarray}
we obtain the situation shown in Fig.~\ref{fig:lepsempi}, where we notice that the SM point is included in the allowed region.

\subsection{Combining leptonic and semileptonic constraints}

We now proceed with the combination of all constraints from the different leptonic and semileptonic channels. By combining the branching fractions for the decays $\bar{B}\rightarrow \rho \ell^- \bar{\nu}_{\ell}$ and $\bar{B}\rightarrow \pi \ell^- \bar{\nu}_{\ell}$, we can introduce the following extra observable:
\begin{eqnarray}
\mathcal{R}^{\Braket{e, \mu};\rho~[q^2_{\rm{min}} \leq q^2 \leq q^2_{\rm{max}}]}_{\Braket{e, \mu};\pi}
&=&
\Braket{ \mathcal{B}(\bar{B}\rightarrow \rho \ell^- \bar{\nu}_{\ell})}_{[\ell=e, \mu]}\Bigl|^{q^2_{\rm{max}}}_{q^2_{\rm{min}}}/
\Braket{\mathcal{B}(\bar{B}\rightarrow \pi \ell^- \bar{\nu}_{\ell})}_{[\ell=e, \mu]},
\label{eq:semirhosemipi}
\end{eqnarray}
where the numerator is calculated by integrating the differential expression in Eq.~(\ref{eq:difrho}) over the interval $q^2_{\rm{min}} \leq q^2 \leq q^2_{\rm{max}}$. We start by evaluating the ratio in Eq.~(\ref{eq:semirhosemipi}) in the low-$q^2$ regime, i.e.\ within the interval 
$q^2 \leq 12~\rm{GeV}^2$. Therefore, using the results in Eqs.~(\ref{eq:Brhoaverage}) and (\ref{eq:piexp}), we obtain
\begin{eqnarray}\label{eq:semileptonicsemileptonic}
\mathcal{R}^{\Braket{e, \mu};\rho~[0 \leq q^2 \leq 12]~\rm{GeV}^2}_{\Braket{e, \mu};\pi}
&=&
1.29 \pm 0.09.
\end{eqnarray}

\begin{center}
\begin{figure}[H]
\begin{center}
\includegraphics[width=0.65\textwidth]{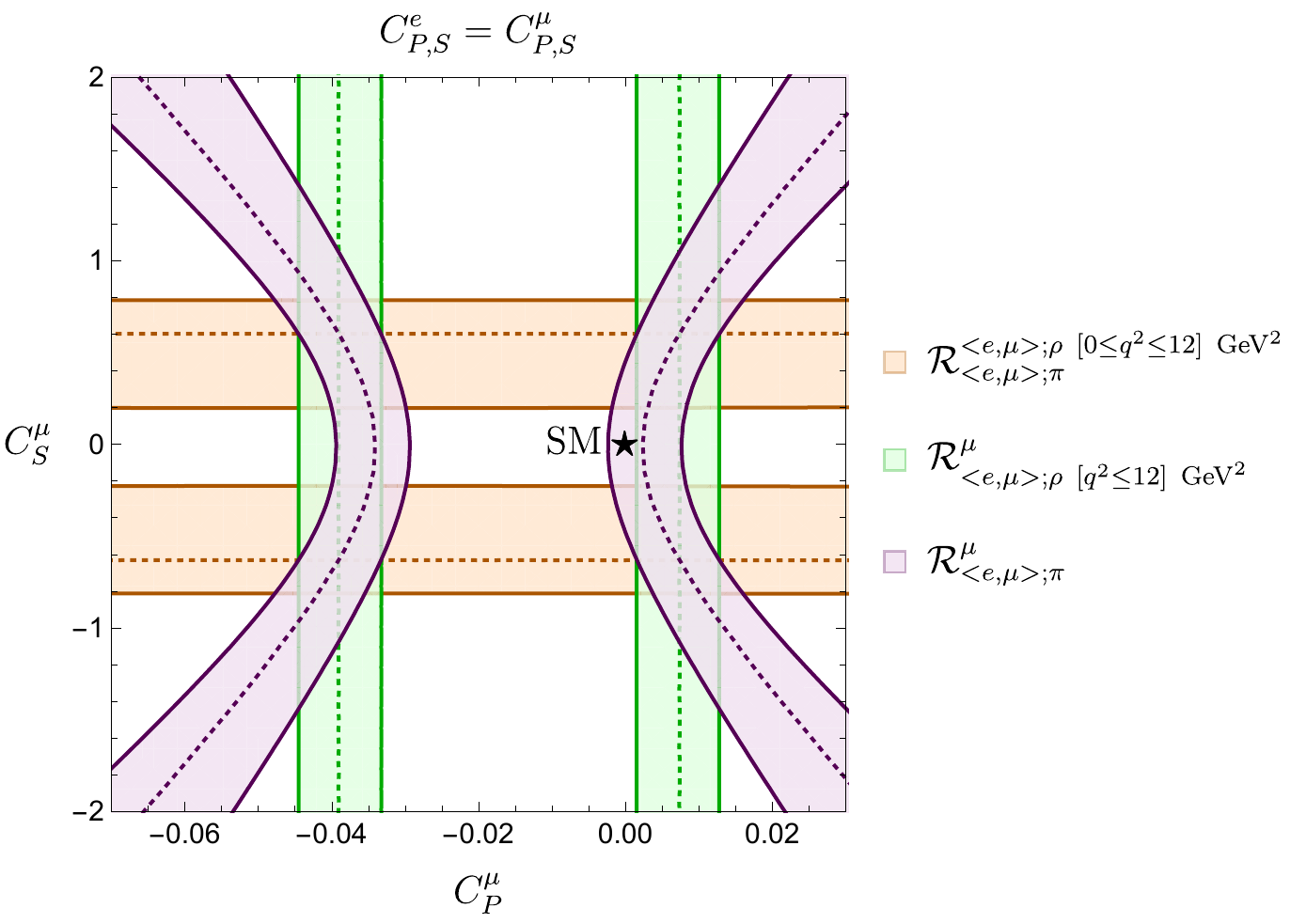} \qquad
\caption{Allowed regions in the  $C^{\mu}_P$--$C^{\mu}_S$ plane considering the observables
$\mathcal{R}^{\Braket{e, \mu};\rho~[0 \leq q^2 \leq 12]~{\rm GeV^2}}_{\Braket{e, \mu};\pi}$, $\mathcal{R}^{\mu}_{\Braket{e, \mu};\rho~[q^2 \leq 12]~{\rm GeV^2}}$ and $\mathcal{R}^{\mu}_{\Braket{e, \mu};\pi}$.}
\label{fig:semilepoversemilep}
\end{center}
\end{figure}
\end{center}

Making the assumption $C^e_P=C^{\mu}_P$, we use the ratio in Eq.~(\ref{eq:semileptonicsemileptonic}) to obtain stronger constraints on $C^{\mu}_P$ and $C^{\mu}_S$. The combination of observables
\[
\mathcal{R}^{\mu}_{\braket{e,\mu};\rho~[q^2\leq 12]~\rm{GeV}^2}, \quad \mathcal{R}^{\mu}_{\braket{e, \mu};\pi} \quad \text{and} \quad \mathcal{R}^{\Braket{e, \mu};\rho~[0 \leq q^2 \leq 12]~\rm{GeV}^2}_{\Braket{e, \mu};\pi}
\] 
in Eqs.~(\ref{eq:leptonicoversemileptonicrho-theo}), (\ref{eq:leptonicoversemileptonicpi}) and (\ref{eq:semirhosemipi}), respectively, leads to the regions shown in Fig.~\ref{fig:semilepoversemilep}. Interestingly, the semileptonic over semileptonic ratio defines two horizontal bands that exclude the SM point by $(1$--$2) \sigma$.

\begin{center}
\begin{figure}[H]
\begin{center}
\includegraphics[width=0.49\textwidth]{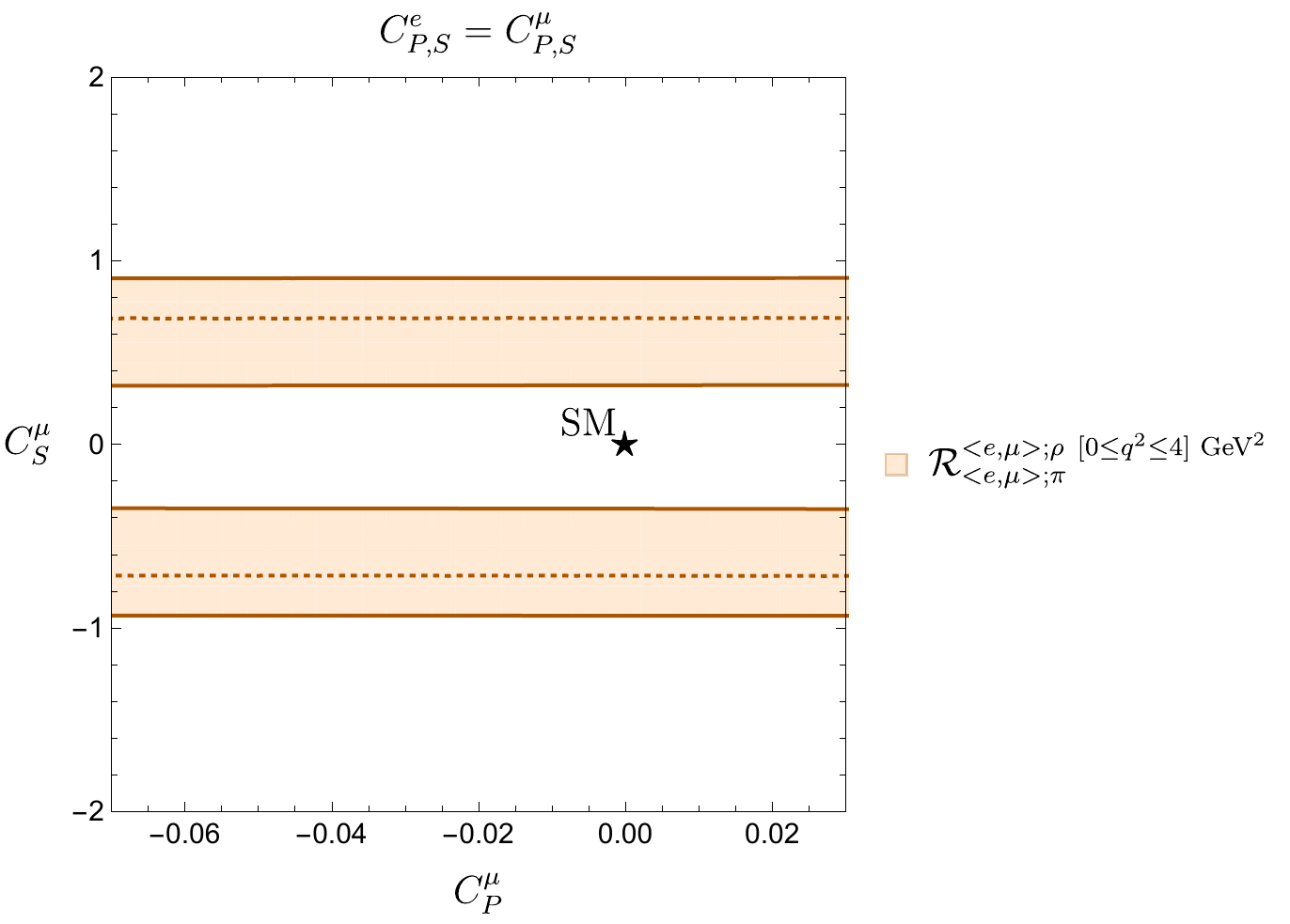} 
\includegraphics[width=0.49\textwidth]{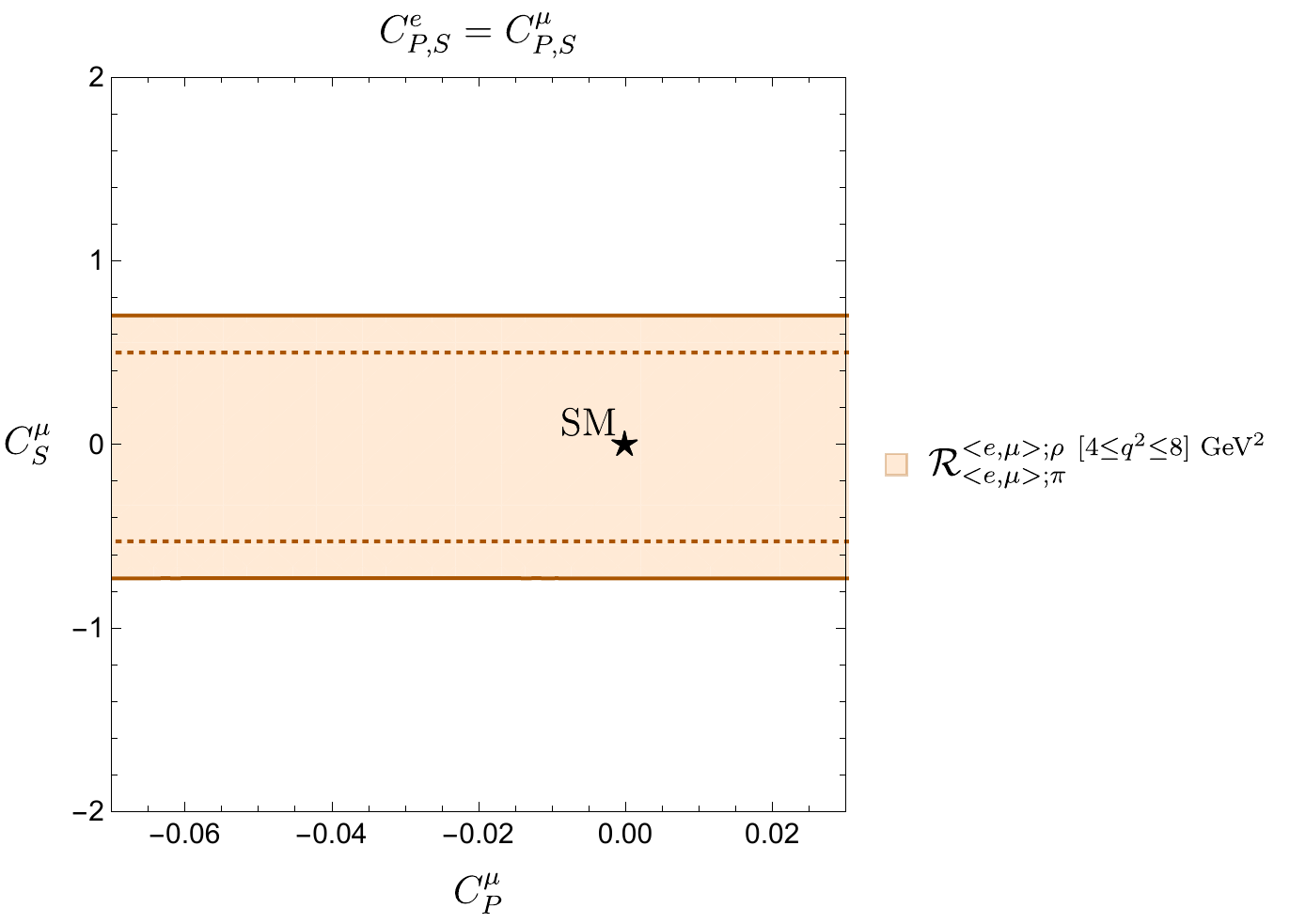}
\includegraphics[width=0.49\textwidth]{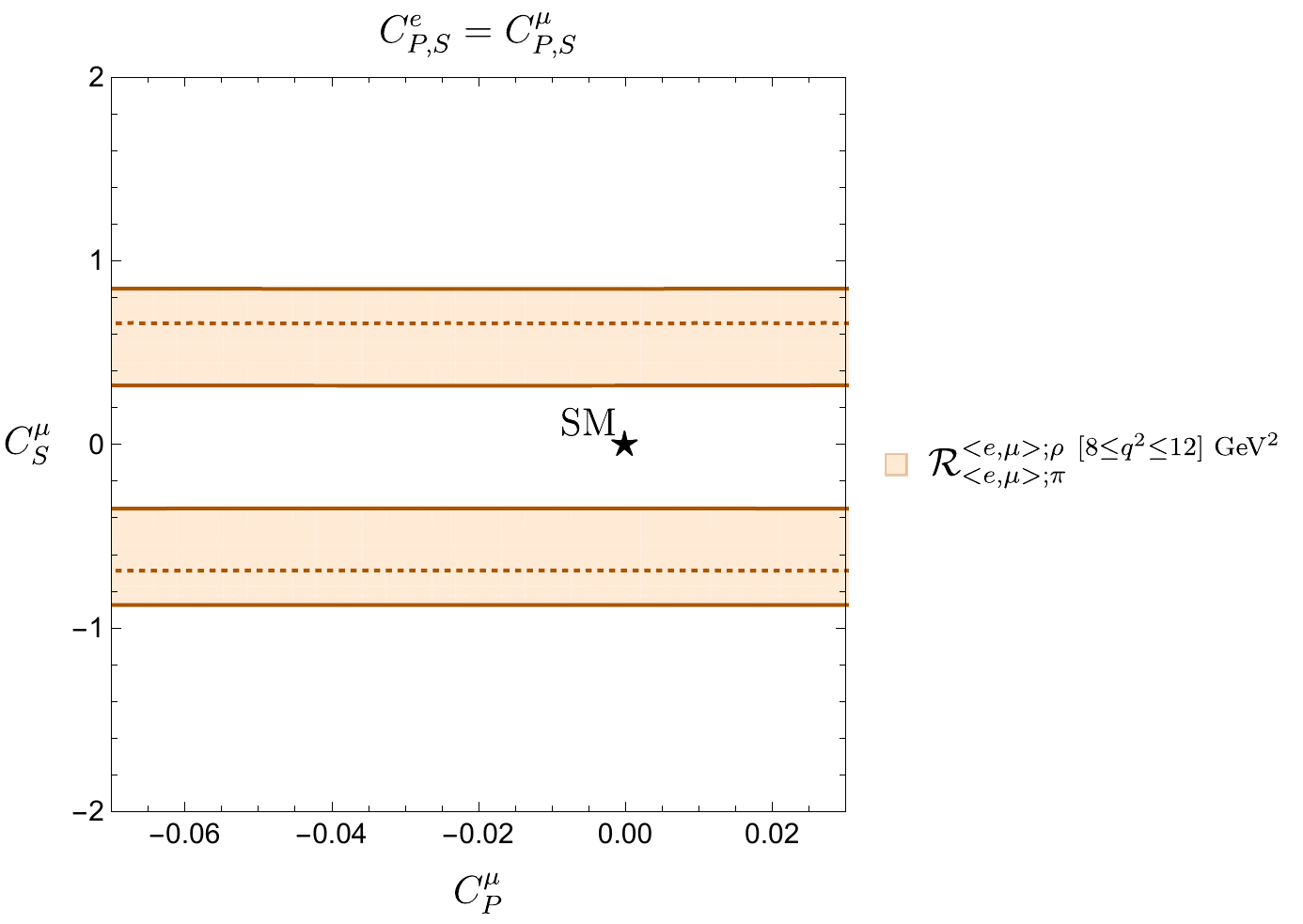}
\caption{Constraints from $\mathcal{R}^{\Braket{e, \mu};\rho}_{\Braket{e, \mu};\pi}$ in the $C_P^\mu$--$C_S^\mu$ plane for different intervals of $q^2$.}
\label{fig:semilepoversemilepbinbybin}
\end{center}
\end{figure}
\end{center}

The tension with the SM found in Fig.~\ref{fig:semilepoversemilep} is an interesting effect that we proceed to investigate in more detail. To this end, we consider the partition of the interval $0\leq q^2\leq 12~\rm{GeV}^2$ given in Table~\ref{tab:binbybin}. Calculating the observable 
$\mathcal{R}^{\Braket{e, \mu};\rho}_{\Braket{e, \mu};\pi}$ in each subinterval yields
\begin{eqnarray}\label{eq:semilepoversemilepbinbybin}
\mathcal{R}^{\Braket{e, \mu}; \rho~[0\leq q^2 \leq 4 ]~\rm{GeV}^2}_{\Braket{e, \mu};\pi}&=&0.31 \pm 0.05,\nonumber\\
\mathcal{R}^{\Braket{e, \mu}; \rho~[4\leq q^2 \leq 8 ]~\rm{GeV}^2}_{\Braket{e, \mu};\pi}&=&0.50 \pm 0.05,\nonumber\\
\mathcal{R}^{\Braket{e, \mu}; \rho~[8\leq q^2 \leq 12 ]~\rm{GeV}^2}_{\Braket{e, \mu};\pi}&=&0.47 \pm 0.05.
\end{eqnarray}
We present the constraints from these observables in Fig.~\ref{fig:semilepoversemilepbinbybin}. We observe that the SM point is excluded within the sub-intervals $[0,4]~\rm{GeV}^2$ and $[8,12]~\rm{GeV}^2$. However, it is contained within $[4,8]~\rm{GeV}^2$. Thus, we can now identify the source of the tension with the SM point found in 
Fig.~\ref{fig:semilepoversemilep}.

\begin{center}
\begin{table}
\begin{tabular}{ |c|c|c|c| }
\hline
             &&&\\
$\Delta q^2$ $(\rm{GeV}^2)$ & $2\braket{\mathcal{B}(B^-\rightarrow \rho^0 \ell^- \bar{\nu}_{\ell} )}_{[\ell=e, \mu]}$ 
& $\braket{\mathcal{B}(\bar{B}^0\rightarrow \rho^+ \ell^- \bar{\nu}_{\ell} )}_{[\ell=e, \mu]}$ & 
$\braket{\mathcal{B}(\bar{B}\rightarrow \rho \ell^- \bar{\nu}_{\ell} )}_{[\ell=e, \mu]}$ \\
             &&&\\
\hline
$[0,~4]$  & $(5.54 \pm 0.92)\times 10^{-5}$ & $(3.73 \pm 1.06) \times 10^{-5}$ & $(4.76 \pm 0.69)\times 10^{-5}$ \\
$[4,~8]$  & $(7.92 \pm 0.96)\times 10^{-5}$ & $(7.18 \pm 1.16) \times 10^{-5}$ & $(7.62 \pm 0.74)\times 10^{-5}$ \\
$[8,~12]$ & $(6.84 \pm 0.89)\times 10^{-5}$ & $(8.06 \pm 1.23) \times 10^{-5}$ & $(7.26 \pm 0.72)\times 10^{-5}$ \\ 
\hline
\end{tabular}
\caption{Experimental values of $\mathcal{B}(\bar{B} \to \rho \ell^- \bar{\nu}_\ell)$ in different $q^2$ intervals \cite{Sibidanov:2013rkk}. The fourth column gives the isospin averages of the values in the second and third columns.}\label{tab:binbybin}
\end{table}
\end{center}

In view of the tension with the SM found in Figs.~\ref{fig:semilepoversemilep} 
and \ref{fig:semilepoversemilepbinbybin}, we investigate whether this effect persists for $12~{\rm{GeV}}^2\leq q^2$. For high $q^2$ values, the theoretical determination of the form factors is done through lattice calculations. To the best of our knowledge, the most recent determination of the $\bar{B}\rightarrow \rho$ form factors available is discussed in Ref.~\cite{Bowler:2004zb}, where the range
\begin{eqnarray}
12.7~{\rm{GeV}}^2 \leq q^2\leq 18.2~{\rm{GeV}}^2
\end{eqnarray}
is considered. It should be noted that this reference is more than $14$ years old. Moreover, there is not an analytical parameterization of the form factors similar to the one for the low-$q^2$ regime presented in Appendix~\ref{sec:HFF}. Consequently, we extract the required information directly from the distributions presented in Ref.~\cite{Bowler:2004zb} that have large errors. In the absence of analytical expressions for the form factors, we run the risk of over estimating the uncertainties associated with the branching fraction $\mathcal{B}(\bar{B}\rightarrow \rho \ell^- \bar{\nu}_{\ell})$. We can avoid this problem by using the differential branching ratio $d\mathcal{B}(B^-\rightarrow \rho^0 \ell^{-}\bar{\nu}_\ell)/dq^2$  at specific values of $q^2$ presented in Eq.~(\ref{eq:difrho}). For this part of the analysis, we cannot use isospin-averaged quantities because the experimental partition for $\mathcal{B}(\bar{B}_d^0 \to \rho^+ \ell^- \bar{\nu}_\ell)$ cannot be compared against the corresponding theoretical range given by the form factors. Therefore we restrict ourselves to the decay channel $B^-\rightarrow \rho^0 \ell^- \bar{\nu}_\ell$ and consider the following observable:
\begin{eqnarray}\label{eq:dR}
d\mathcal{R}^{\Braket{e, \mu};\rho}_{\Braket{e, \mu};\pi}&=&
\frac{2\Braket{d\mathcal{B}(B^-\rightarrow \rho^0 \ell^{-}\bar{\nu}_\ell)/dq^2}_{[\ell=e, \mu]}}{\Braket{\mathcal{B}(\bar{B}\rightarrow \pi \ell^- \bar{\nu}_{\ell})}_{[\ell=e, \mu]}}.
\end{eqnarray}

\begin{center}
\begin{figure}[H]
\begin{center}
\includegraphics[width=0.45\textwidth]{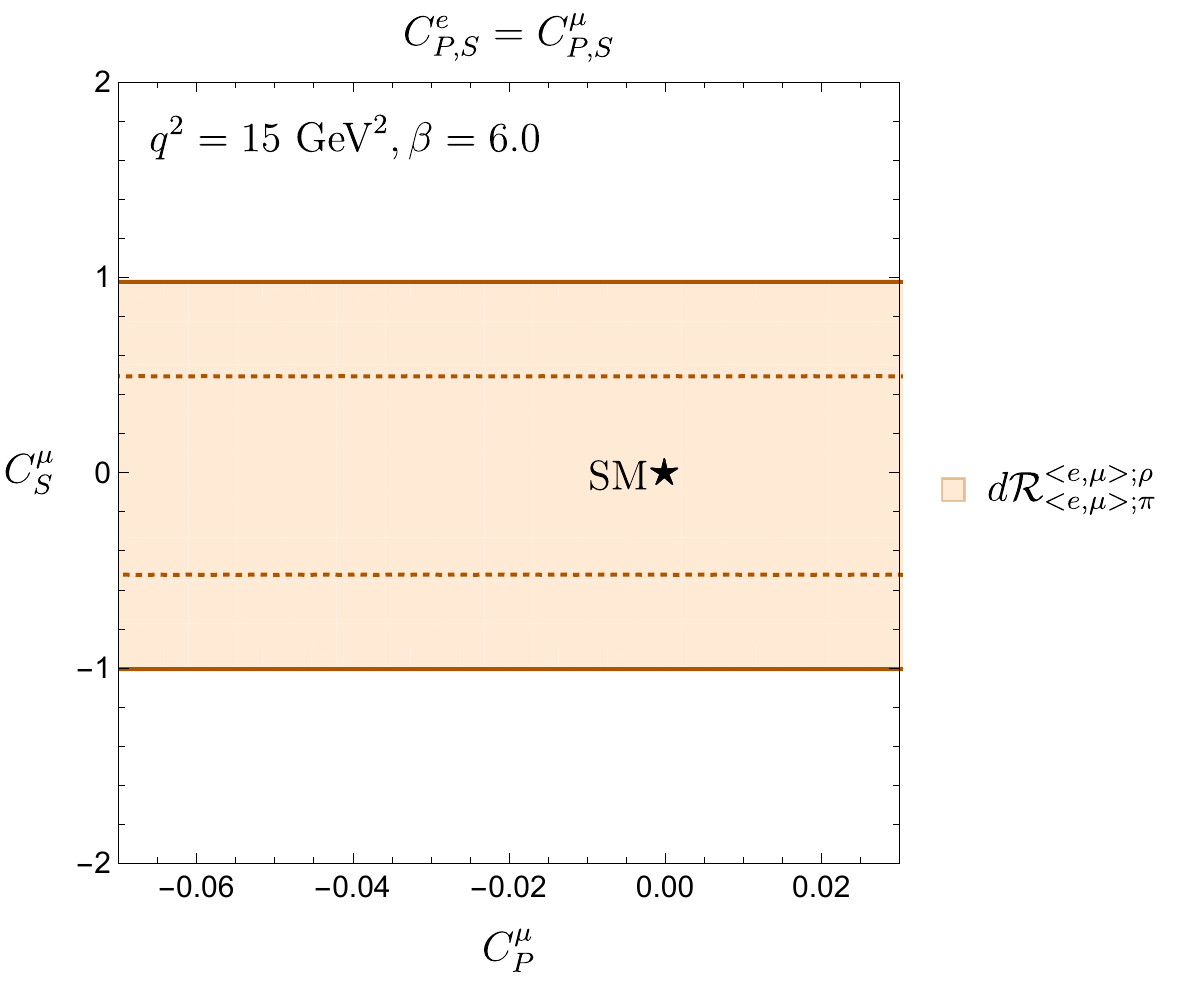}
\includegraphics[width=0.45\textwidth]{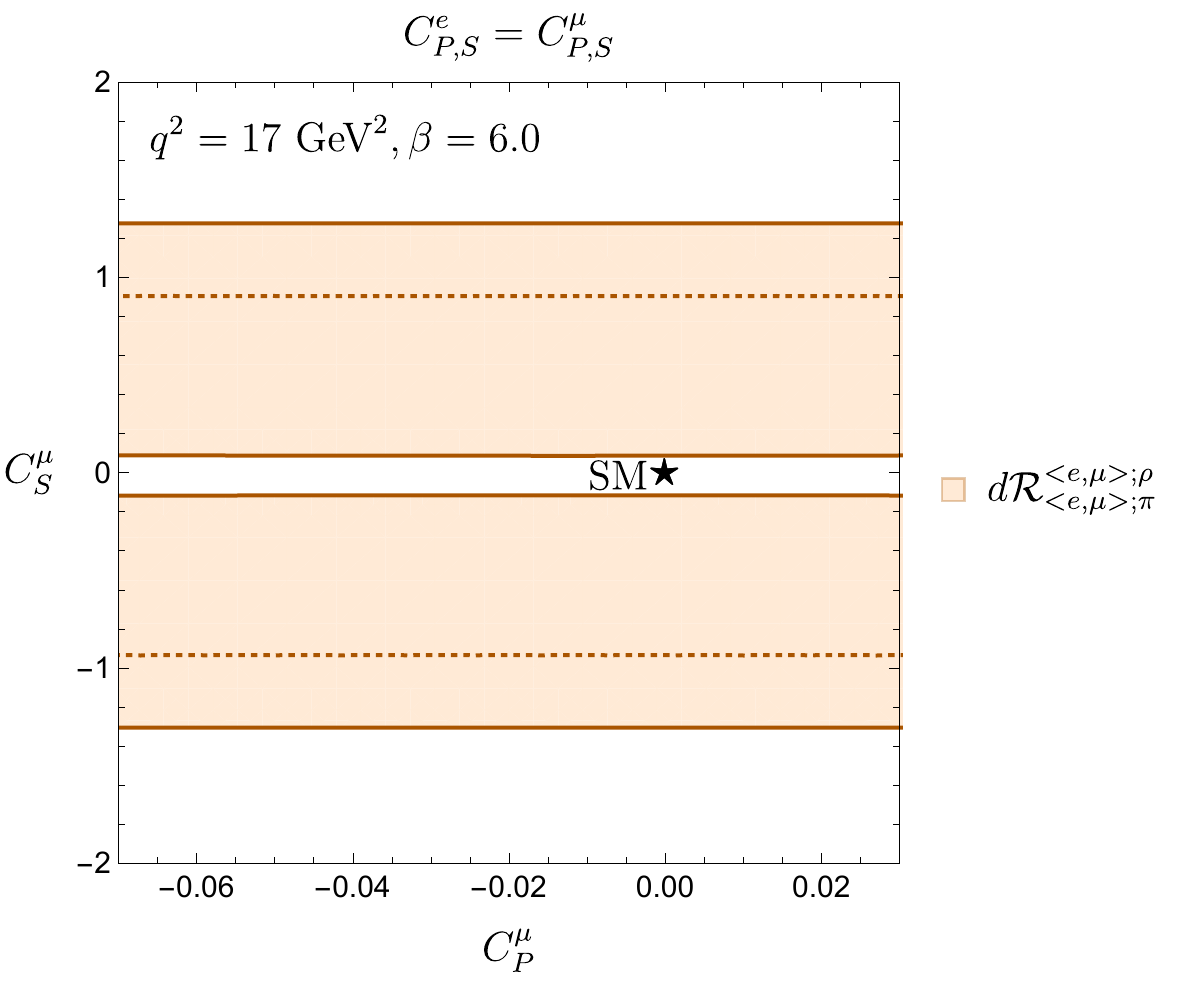}
\caption{Regions in the $C^{\mu}_P$--$C^{\mu}_S$ plane allowed by the observable $d\mathcal{R}^{\Braket{e, \mu};\rho}_{\Braket{e, \mu};\pi}$ 
for $q^2=15~\rm{GeV}^2$ (left) and $q^2=17~\rm{GeV}^2$ (right), considering $\beta=6.0$.}
\label{fig:dsemilepoversemilepbinbybinbeta6}
\end{center}
\end{figure}
\end{center}

\begin{center}
\begin{figure}[H]
\begin{center}
\includegraphics[width=0.45\textwidth]{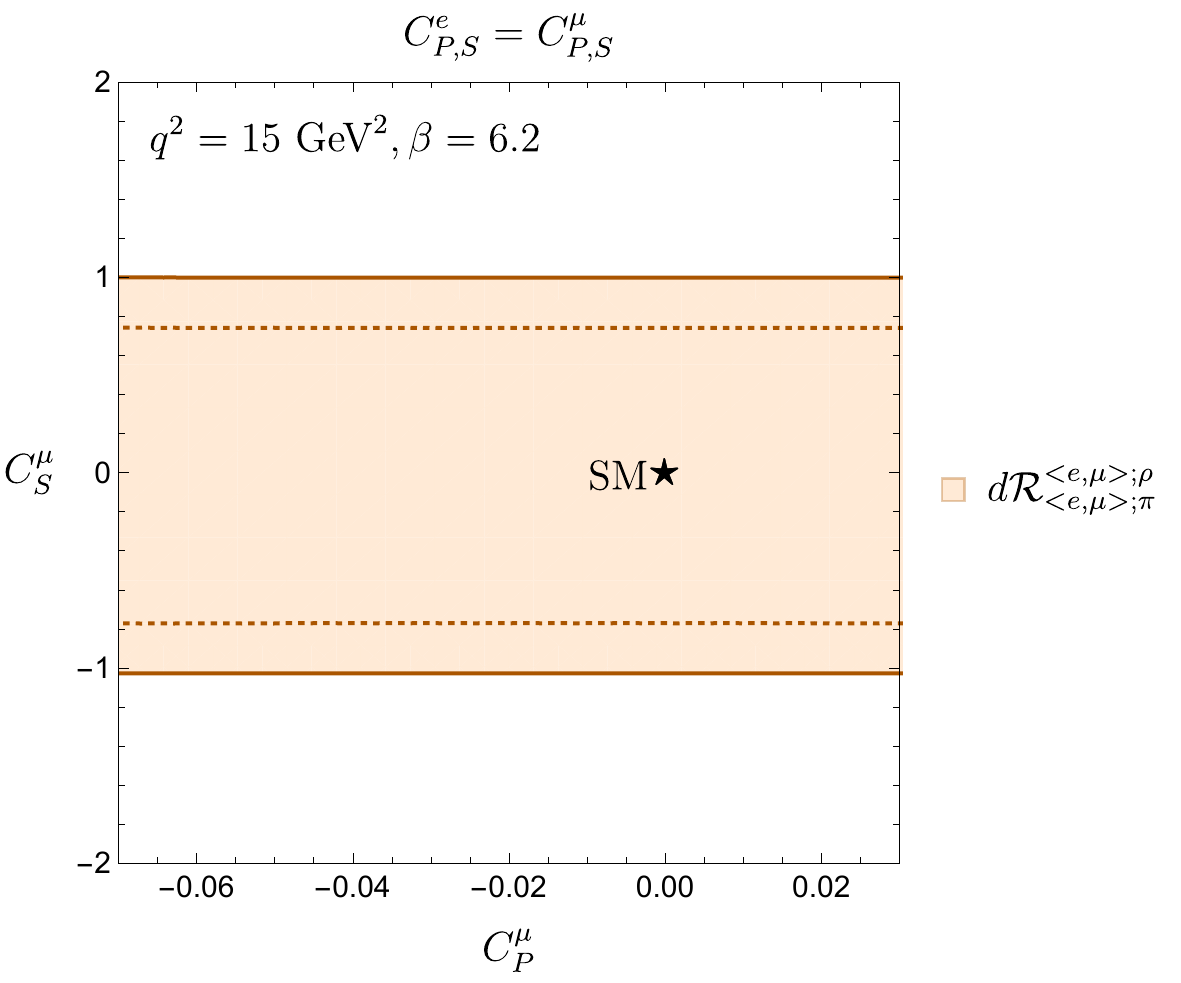}
\includegraphics[width=0.45\textwidth]{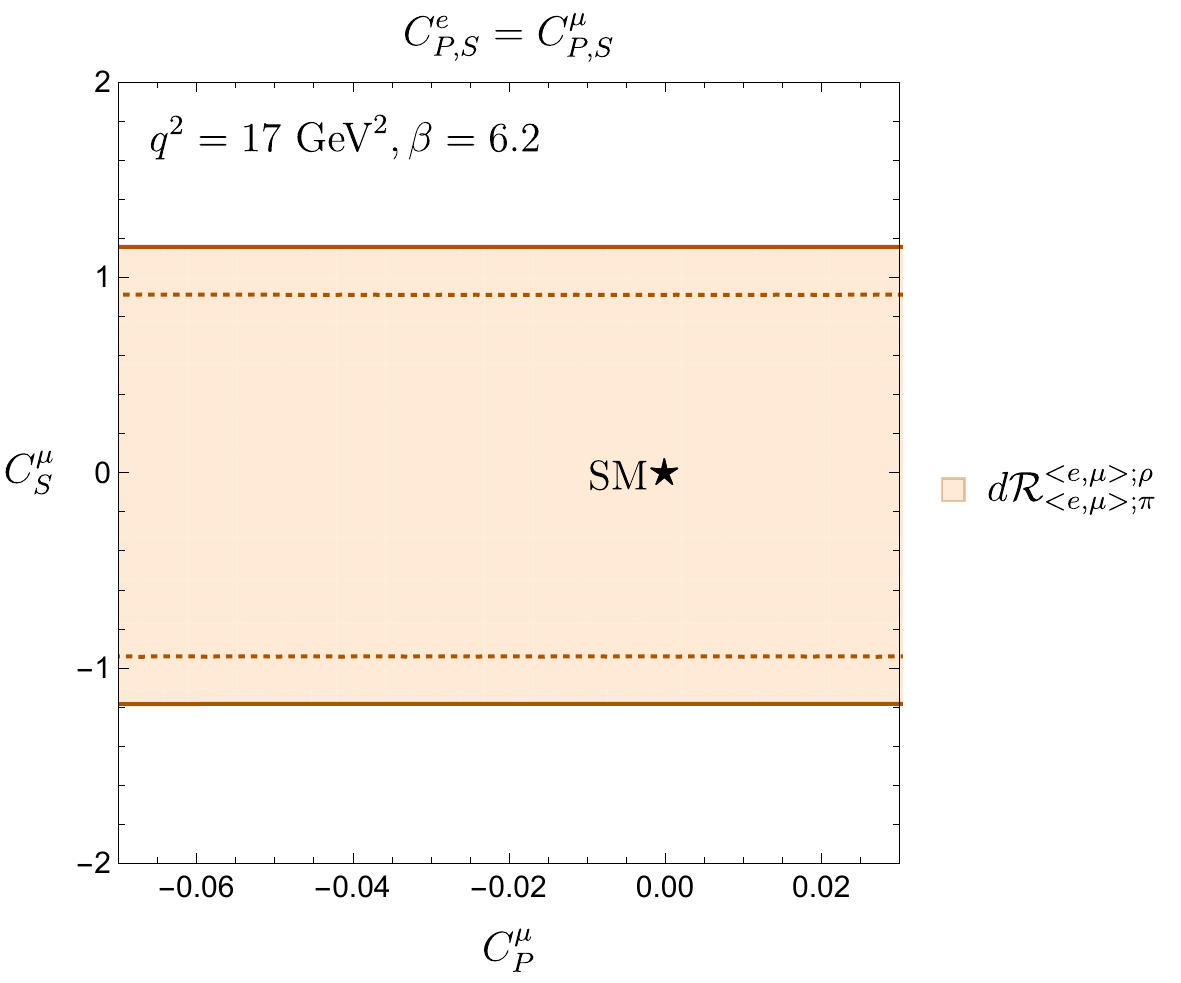}
\caption{Regions in the $C^{\mu}_P$--$C^{\mu}_S$ plane allowed by the observable $d\mathcal{R}^{\Braket{e, \mu};\rho}_{\Braket{e, \mu};\pi}$ 
for $q^2=15~\rm{GeV}^2$ (left) and $q^2=17~\rm{GeV}^2$ (right), considering $\beta=6.2$.}
\label{fig:dsemilepoversemilepbinbybinbeta62}
\end{center}
\end{figure}
\end{center}

In Ref.~\cite{Bowler:2004zb}, two different determinations of the form factors are available depending on the value of the coupling constant $\beta=6/g^2_0$. In particular, we have $\beta=6.0$ and $\beta=6.2$. Moreover, the available experimental data allow us to evaluate the numerator in Eq.~(\ref{eq:dR}) at $q^2=15~\rm{GeV}^2$ and $q^2=17~\rm{GeV}^2$, yielding
\begin{equation}
d\mathcal{R}^{\Braket{e, \mu};\rho}_{\Braket{e, \mu};\pi}\Bigl|_{q^2=15~\rm{GeV}^2} = 0.14 \pm 0.02, \qquad d\mathcal{R}^{\Braket{e, \mu};\rho}_{\Braket{e, \mu};\pi}\Bigl|_{q^2=17~\rm{GeV}^2} = 0.11 \pm 0.02.
\end{equation}
The corresponding plots are shown in Fig.~\ref{fig:dsemilepoversemilepbinbybinbeta6} for $\beta=6.0$
and in Fig.~\ref{fig:dsemilepoversemilepbinbybinbeta62} for $\beta=6.2$.

Just as for the low-$q^2$ region, a small tension with the SM appears in the case of $\beta=6.0$ with $q^2=17~\rm{GeV}^2$. However, a more precise determination of the form factors in the high-$q^2$ regime is required in order to understand the origin of this discrepancy: it can certainly be triggered by the theoretical precision of the  non-perturbative contributions. Indeed, the study presented in \cite{Bowler:2004zb} was
performed when  the lattice calculations technology was in its early stages of development and an underestimation of the uncertainties cannot be discarded. A very interesting prospect would be the presence of NP; this possibility is quite exciting and is in principle allowed by the theoretical and experimental information available at the moment. 
 In addition, an interpolation between the low- and high-$q^2$ regimes for the $B\rightarrow \rho$ transitions will allow a full use of the experimental determinations.

\subsection{CP Violation}\label{sec:CPV}
Finally, we would like to study the implications of CP-violating phases once we combine the different leptonic and semileptonic constraints described at the beginning of this Section and in  Sec.~\ref{sec:lept}. Since the direct CP asymmetries defined in Eq.~(\ref{aCP-dir}) would take essentially vanishing values for the (semi)leptonic decays, we follow the approach introduced in Sec.~\ref{sec:CPlept} for leptonic processes and explore the implications of new CP-violating phases in the short distance contributions, i.e. complex Wilson coefficients. Specifically, we analyse correlations between the norms and phases of the short-distance contributions, as well as between norms of different coefficients.

\begin{center}
	\begin{figure}[H]
		\begin{center}
			\includegraphics[width=0.6\textwidth]{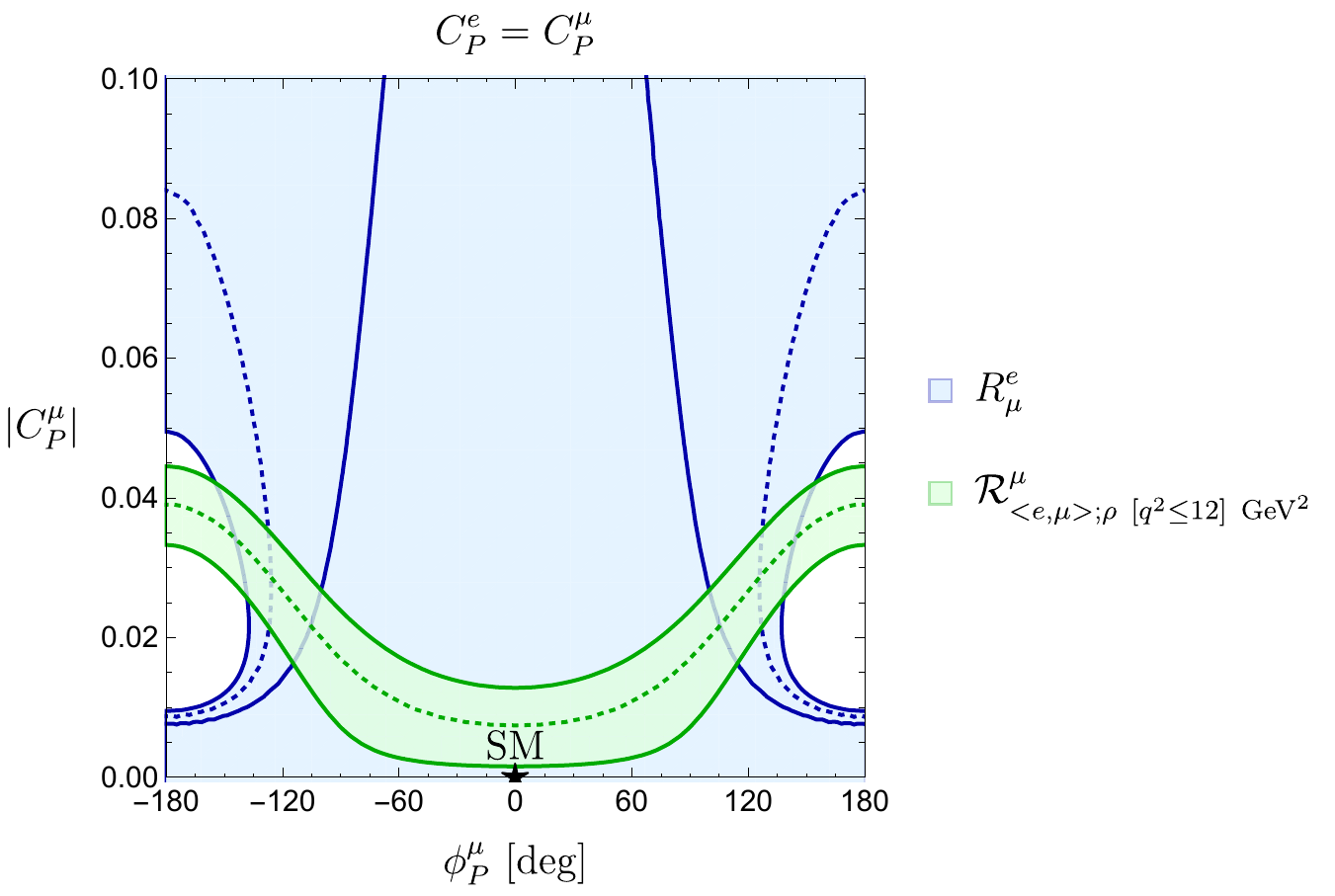}
			\caption{Regions obtained in the $\phi_P^\mu$--$|C_P^\mu|$ plane from $R^{e}_{\mu}$ and $\mathcal{R}^{\mu}_{\Braket{e,\mu};\rho~[q^2 \leq 12]~{\rm GeV}^2}$, assuming universality for the light lepton flavours.}
			\label{fig:phiPCPmu}
		\end{center}
	\end{figure}
\end{center}

To begin with, we consider the constraints in the $\phi^{\mu}_P$--$|C^{\mu}_P|$ plane shown in Fig.~\ref{fig:d}. This analysis was performed under the assumption $C_P^e = C_P^\mu$ using only the observable $R^{e}_{\mu}$. We complement this study by including the ratio $\mathcal{R}^{\mu}_{\Braket{e,\mu};\rho~[q^2 \leq 12]~{\rm GeV}^2}$, introduced in Eq.~(\ref{eq:leptonicoversemileptonicrho-theo}). The new regions are shown in Fig.~\ref{fig:phiPCPmu}. We notice that the SM point falls within the allowed regions. Additionally, the norm of the pseudo-scalar Wilson coefficient is bounded, at the one sigma level this bound reads
\begin{eqnarray}
|C_P^\mu| \leq 0.042.
\end{eqnarray}

\begin{center}
	\begin{figure}[H]
		\includegraphics[width=\textwidth]{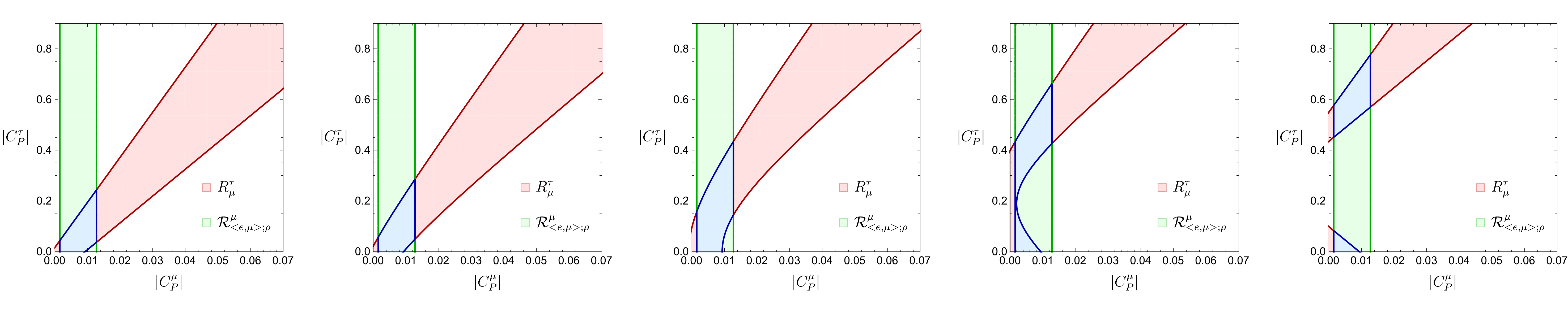}
		\caption{Evolution of the $|C^{\mu}_P|$--$|C^{\tau}_P|$ plane for $\phi^{\mu}_P=0^\circ$ and different values of $\phi^{\tau}_P$. The first and the last plots correspond to $\phi^{\tau}_P=0^\circ$ and $\phi^{\tau}_P=180^\circ$, respectively, whereas the plots in between show increasing values of $\phi_P^\tau$ in steps of $45^\circ$ from left to right.}
		\label{fig:evol}
	\end{figure}
\end{center}

We continue by adding the observable $\mathcal{R}^{\mu}_{\Braket{e,\mu};\rho~[q^2 \leq 12]~{\rm GeV}^2}$ to the analysis shown in Fig.~\ref{fig:aa}; to incorporate this observable we assume $C_P^e = C_P^\mu$. We explore the correlations between $|C^{\mu}_P|$--$|C_P^\tau|$ considering different values for the phases $\phi^{\mu}_P$ and $\phi^{\tau}_P$. We first fix $\phi^{\mu}_P=0^{\circ}$ and allow $\phi^{\tau}_P$ to change in steps of $45^{\circ}$ up to the value $\phi^{\tau}_P=180^{\circ}$. The resulting patterns are shown in Fig.~\ref{fig:evol}, where the overlapping region of the two constraints is indicated in blue. We can see how the regions evolve along the vertical direction. Once the value $\phi^{\tau}_P=180^{\circ}$ is reached, the behaviour is cyclic and the resulting patterns come back into themselves.

\begin{center}
	\begin{figure}[H]
		\begin{center}
			\includegraphics[width=0.6\textwidth]{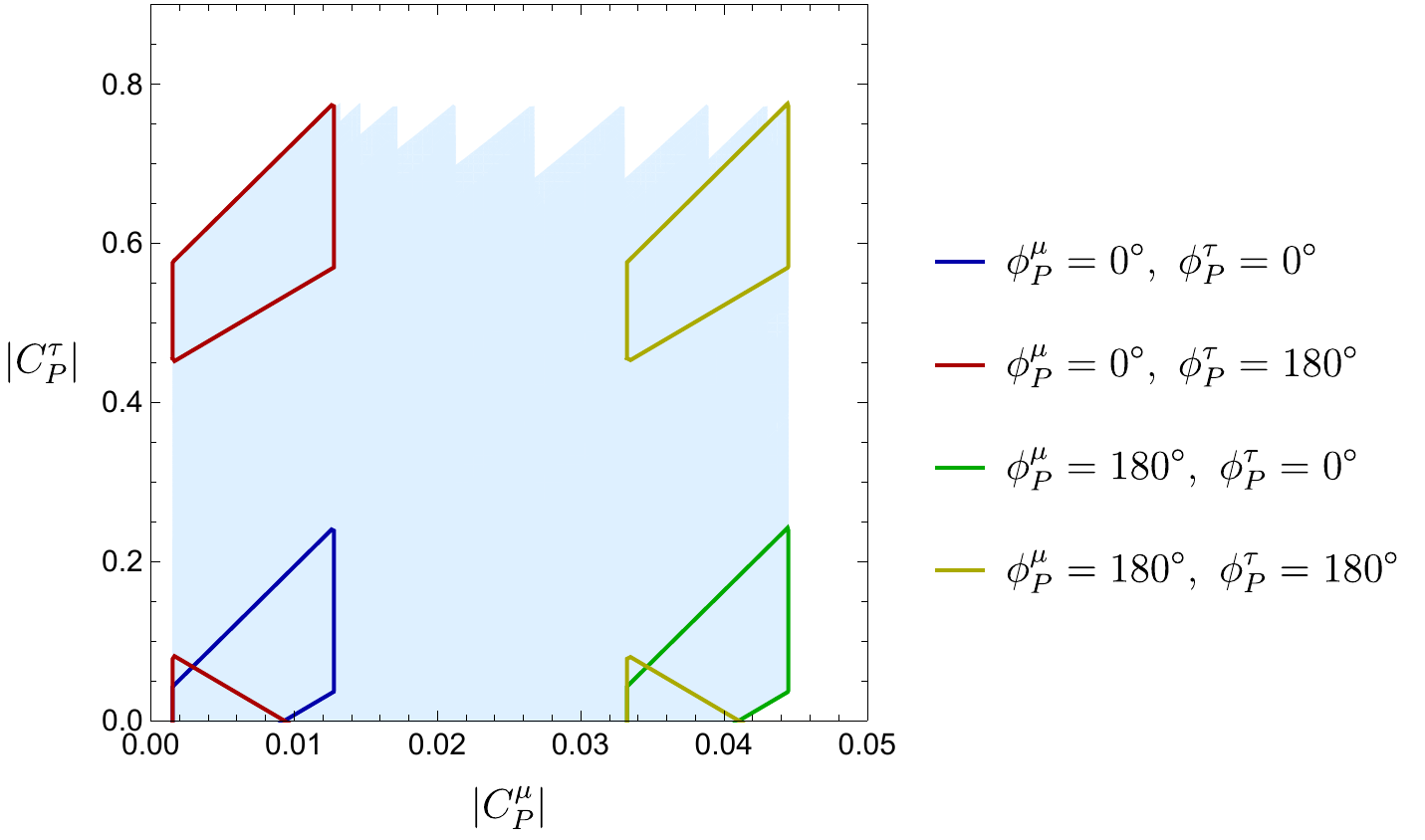}
			\caption{Region in the $|C_P^\mu|$--$|C_P^\tau|$ plane obtained from the overlap between the contours following from $R^{\tau}_{\mu}$ and $\mathcal{R}^{\mu}_{\Braket{e,\mu};\rho~[q^2 \leq 12]~{\rm GeV}^2}$. We vary the phases $\phi_P^\mu$ and $\phi_P^\tau$ independently in the interval $[0^{\circ}, 180^{\circ}]$, giving the blue background region. The jagged upper boundary is due to varying $\phi_P^\mu$ in steps of $20^\circ$. The coloured regions indicate solutions obtained for $\phi_P^\mu,\phi_P^\tau \in \{0^\circ,180^\circ\}$.}
			\label{fig:smear}
		\end{center}
	\end{figure}
\end{center}

Finally, we allow $\phi^{\mu}_P$ to change as well. Unlike the previous case, the evolution is now along the horizontal direction. By scanning $\phi^{\mu}_P$ and $\phi^{\tau}_P$ within the interval $[0^{\circ}, 180^{\circ}]$ we generate the smeared plot shown in Fig.~\ref{fig:smear}. We have highlighted the steps corresponding to: $(\phi^{\mu}_P=0^{\circ}, \phi^{\tau}_P=0^{\circ})$, $(\phi^{\mu}_P=0^{\circ}, \phi^{\tau}_P=180^{\circ})$, $(\phi^{\mu}_P=180^{\circ}, \phi^{\tau}_P=0^{\circ})$ and $(\phi^{\mu}_P=180^{\circ}, \phi^{\tau}_P=180^{\circ})$.

\section{Determination of \boldmath$|V_{ub}|$\unboldmath}
\label{sec:Vub}
The extraction of $|V_{ub}|$ from semileptonic decays is usually done under the assumption of the SM, although NP contributions may also have an impact \cite{Crivellin:2014zpa}. For instance, the effect of a new right-handed vector current on the determination of $|V_{ub}|$ has been discussed in Ref.~\cite{Bernlochner:2014ova}, where also new ways to search for such NP effects using $\bar{B} \to \rho \ell^- \bar{\nu}_\ell$ decays are presented. Here we provide a general strategy that allows us to determine $|V_{ub}|$ in the presence of new scalar and pseudoscalar contributions. We remind ourselves that the branching fractions of the leptonic decays and the semileptonic $\bar{B} \to \rho \ell^- \bar{\nu}_\ell$ transitions are only sensitive to the pseudoscalar NP operator. On the other hand, $\mathcal{B}(\bar{B}\rightarrow \pi \ell^{-}\nu)$ depends exclusively on the scalar Wilson coefficient. Throughout this section and Sec.~\ref{sec:predictions}, we consider only the range $0 \leq q^2 \leq 12$ GeV$^2$ for the $\bar{B} \to \rho \ell^- \bar{\nu}_\ell$ transition and therefore we omit this information from the labels of the ratios. Consequently, unless stated otherwise, we take
\begin{eqnarray}
\mathcal{R}^{\mu}_{\Braket{e, \mu};\rho} \equiv \mathcal{R}^{\mu}_{\Braket{e, \mu};\rho~[q^2\leq 12]~\rm{GeV}^2}.
\end{eqnarray}

We start our discussion by focussing our attention on observables containing only the Wilson coefficient $C^{\ell}_P$. Moreover, we will assume universal scalar and pseudoscalar interactions for  light leptons, i.e.\ $C^e_S=C^{\mu}_S$, $C^e_P=C^\mu_P$. There are then two key steps to obtain $|V_{ub}|$ that can be summarized as follows: 
\begin{enumerate}
	\item Perform a  $|V_{ub}|$-independent extraction of $C^{\ell}_P$.  This can be achieved using the ratios introduced in Secs.~\ref{sec:lept} and \ref{sec:semilept}.
	\item Substitute the ranges for $C^{\ell}_P$ in any of the leptonic or semileptonic branching ratios available, i.e. $\mathcal{B}(B^-\rightarrow \mu^-\bar{\nu}_{\mu})$ or $\mathcal{B}(\bar{B}\rightarrow \rho \ell^-\bar{\nu}_{\ell})$, and then solve for $|V_{ub}|$.
\end{enumerate}

\begin{center}
	\begin{figure}[H]
		\begin{center}
			\includegraphics[width=\textwidth]{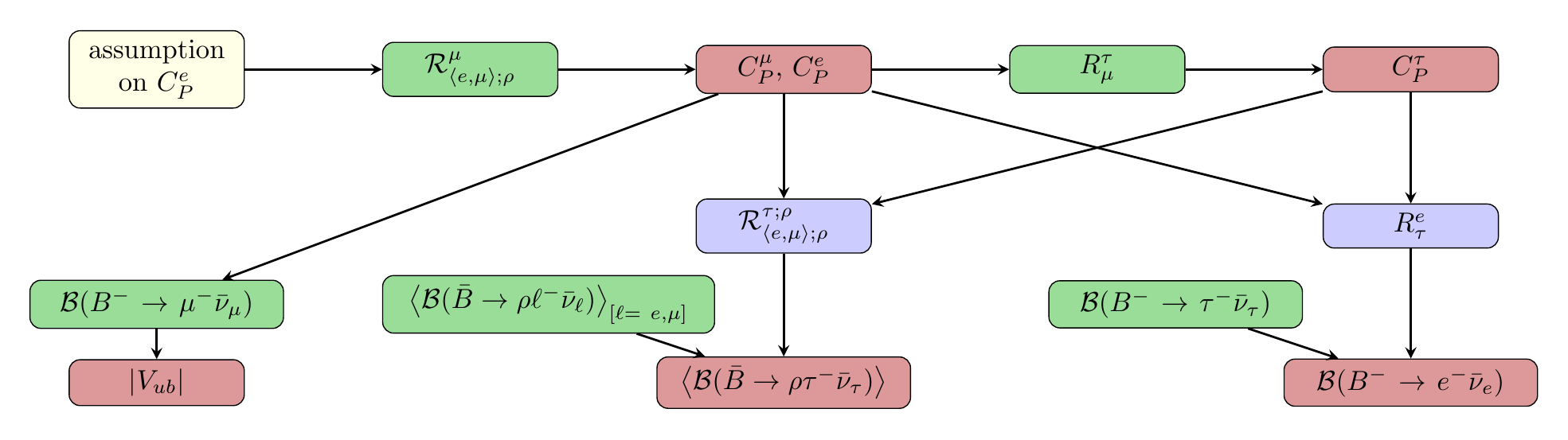}
			\caption{Flowchart illustrating the strategies for the determination of $|V_{ub}|$ and the predictions of $\Braket{\mathcal{B}(\bar{B} \to \rho \tau^- \bar{\nu}_\tau)}$ and $\mathcal{B}(B^- \to e^- \bar{\nu}_e)$.}
			\label{fig:flowchart}
		\end{center}
	\end{figure}
\end{center}

This procedure can be implemented in different ways employing the constraints discussed in the previous sections. For instance, we can use the bounds for the pseudoscalar NP short-distance contributions derived in Secs.~\ref{sec:lept}, \ref{sec:semilept} and presented in Fig.~\ref{fig:lepsemmutau}. One of the problems with this approach is that possible correlations between the observables are not taken into account. Let us now elaborate on an alternative strategy which avoids this issue:
\begin{itemize}
	\item Using the expressions for $\mathcal{R}^{\mu}_{\Braket{e, \mu};\rho}$ introduced in Eq.~(\ref{eq:leptonicoversemileptonicrho-theo}), we solve for $C^{\mu}_P$. Since we are assuming universal NP contributions for electrons and muons, this ratio depends only on one single NP coefficient.
	\item The previous step leads to the function $C^{\mu}_P(\mathcal{R}^{\mu}_{\Braket{e, \mu};\rho})$. There are two solutions satisfying independently $C^{\mu}_P<0$ and $0 < C^{\mu}_P$. Looking at Fig.~\ref{fig:lepsemmutau}, we see that only $0 < C^{\mu}_P$ is consistent with all the available constraints.
	\item Finally, we evaluate any of the individual branching fractions $\mathcal{B}(B^-\rightarrow \mu^-\bar{\nu}_{\mu})$ or $\Braket{\mathcal{B}(\bar{B}\rightarrow \rho \ell^-\bar{\nu}_{\ell})}_{[\ell=e, \mu]}$ in the interval for $C^{\mu}_P$ obtained above. From the resulting expression, we can determine the only unknown left: the value of $|V_{ub}|$.
\end{itemize}
This strategy has been summarized in the flowchart in Fig.~\ref{fig:flowchart}.

Up to now we have shown how it is possible to extract $|V_{ub}|$ from observables involving $C^{\mu}_P$. We can, however, incorporate also the constraints for $C^{\tau}_P$. With this in mind, we consider $R^{\tau}_{\mu}$ defined in Eq.~(\ref{eq:Ru}), which depends both on $C^{\mu}_P$ and on $C^{\tau}_P$. We reduce the number of independent parameters by substituting $C^{\mu}_P(\mathcal{R}^{\mu}_{\Braket{e, \mu};\rho})$ in $R^{\tau}_{\mu}$. The resulting expression will depend only on $C^{\tau}_P$ and can be inverted to obtain this coefficient as a function of  $R^{\tau}_{\mu}$ and $\mathcal{R}^{\mu}_{\Braket{e, \mu};\rho}$, which can then be inserted into $\mathcal{B}(B^-\rightarrow \tau^-\bar{\nu}_{\tau})$ to extract $|V_{ub}|$.

Following any of the two methods described above leads to consistent results. This is actually not surprising since by adding $\mathcal{B}(B^- \to \tau^- \bar{\nu}_\tau)$ to our set of observables we are also including an additional coefficient $C_P^\tau$. The result will be the same if we consider ratios 
containing $\Braket{\mathcal{B}(\bar{B} \to \pi \ell^- \bar{\nu}_\ell)}_{[\ell=e, \mu]}$, which bring $C_S^\mu$ as an extra parameter into the analysis.

Following any of the procedures described above, we find for the universal scenario
\begin{equation} \label{eq:vubSolUniversal}
|V_{ub}| = (3.31 \pm 0.32) \times 10^{-3}.
\end{equation}
This result is in agreement with the CKMFitter value in Eq.~(\ref{eq:VubCKMFitter}) but the latter is three times more precise. However, our aim is to illustrate how to account properly for NP effects during the determination of $|V_{ub}|$.

We may also relax the universality condition for the light leptons. However, in order to use the experimental result in Eq.~(\ref{eq:Brhoaverage}), we have to make an assumption on the correlation between $C_P^e$ and $C_P^\mu$. Here we consider four scenarios:
\begin{enumerate}
	\item $C^{e}_P\ll C^{\mu}_P$; in particular, we explore
	\begin{equation}
	C^{e}_P=(1/10)C^{\mu}_P.
	\end{equation}
	\item $C^{\mu}_P \ll C^{e}_P$; we focus on
	\begin{equation}
	C^{e}_P= 10C^{\mu}_P.
	\end{equation}
	\item The 2HDM, where according to Eq.~(\ref{eq:C2HDM}), we have
	\begin{eqnarray}\label{eq:WCTHDM}
	C^e_P=\frac{m_e}{m_{\mu}}C^{\mu}_P, \quad \quad \quad C^{\tau}_P=\frac{m_{\tau}}{m_{\mu}}C^{\mu}_P.
	\end{eqnarray}
	\item NP entering only through the 3rd generation:
	\begin{eqnarray}
	C^{\tau}_P\neq 0, \quad \quad \quad C^{e}_P=C^{\mu}_P=0.
	\end{eqnarray}
\end{enumerate}

\begin{center}
	\begin{figure}[H]
		\begin{center}
			\includegraphics[width=0.45\textwidth]{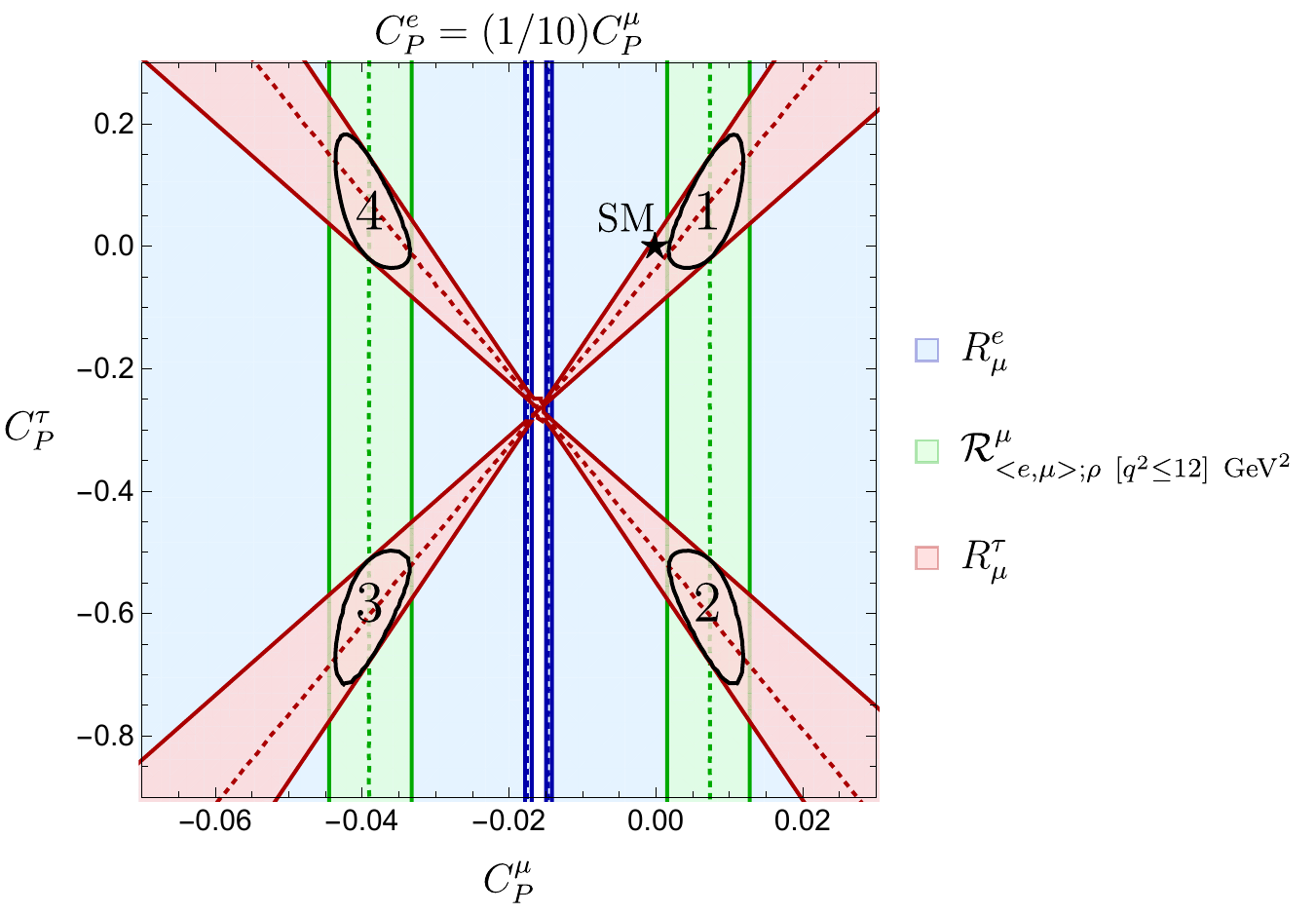}
			\includegraphics[width=0.45\textwidth]{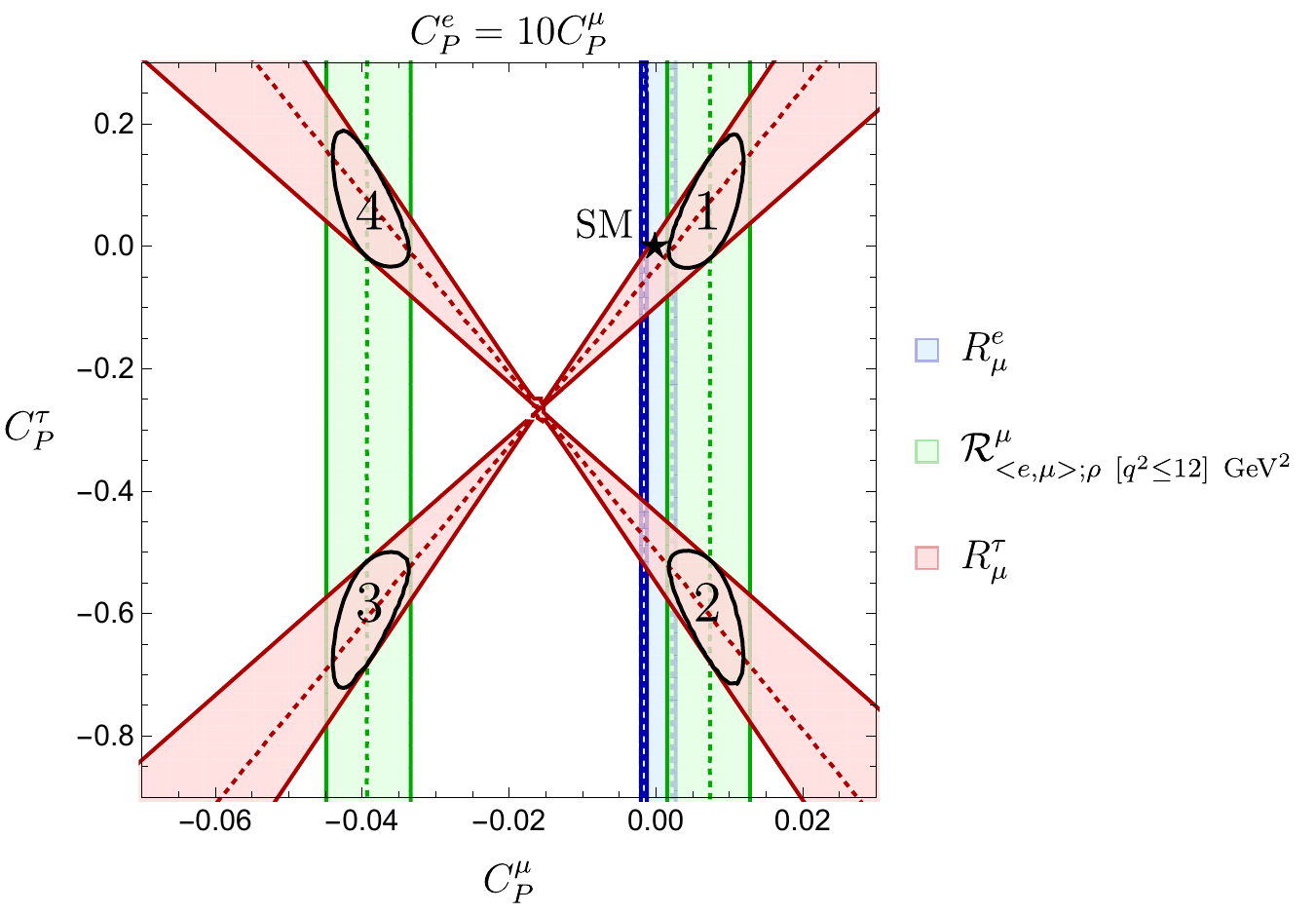}
			\caption{Regions allowed for $C_P^\mu$ and $C_P^\tau$ for the scenarios discussed in the text.}
			\label{fig:Vubalternative}
		\end{center}
	\end{figure}
\end{center}

Let us consider first the cases $C^{e}_P= (1/10)C^{\mu}_P$ and $C^{e}_P= 10C^{\mu}_P$.
After imposing the relevant leptonic and semileptonic constraints, we obtain the plots shown in Fig.~\ref{fig:Vubalternative}. From the left plot, we see how for $C^e_P=(1/10)C^{\mu}_P$ the four regions lying in the intersections of the observables $\mathcal{R}^{\mu}_{\Braket{e, \mu};\rho}$ and 
$R^{\tau}_{\mu}$ are allowed. They are enclosed by four ellipses shown in the plot and numbered clockwise starting with the one in the upper-right corner. We obtain $|V_{ub}|$ by applying the methods described at the beginning of this section, and summarize our results in the second column of Table \ref{tab:summaryResults}.

If we consider the correlation $C^e_P=10 C^{\mu}_P$, we obtain the right plot in Fig.~\ref{fig:Vubalternative}, where the observable $R^e_{\mu}$ selects two narrow vertical sections inside the two ellipses on the right. As for $C^e_P=(1/10)C^{\mu}_P$, the numerical results are summarized in Table \ref{tab:summaryResults}.

For the 2HDM the only relevant constraint is given by $\mathcal{R}^{\mu}_{\Braket{e, \mu};\rho}$. According to Eq.~(\ref{eq:WCTHDM}), all the Wilson coefficients depend only on $C^{\mu}_P$. Using the corresponding experimental information, we may solve for this coefficient, yielding
\begin{eqnarray} \label{eq:CPmu2HDM}
C^{\mu}_P = -0.0391 \pm 0.0055 \quad   \vee  \quad  C^{\mu}_P= 0.0074 \pm 0.0054.
\end{eqnarray}

Finally, in our 4th scenario, NP enters exclusively though $C^{\tau}_P$. Therefore the only useful constraint is given by $R^{\tau}_{\mu}$. Using the experimental determination in Eq.~(\ref{eq:RTaumu}) leads to the following two solutions:
\begin{eqnarray}
C^{\tau}_P= -0.499 \pm 0.056 \quad   \vee  \quad  C^{\tau}_P= -0.034 \pm 0.055.
\end{eqnarray}
The resulting values for $|V_{ub}|$ are summarized in Table~\ref{tab:summaryResults}.

For most of these studies, the values of $|V_{ub}|$ coincide with one another at the level of the significant digits. However, in the scenario where the NP enters only through the third generation, our numerical result for $|V_{ub}|$ is higher in comparison with the other cases. In this respect it agrees with the inclusive $|V_{ub}|$ determinations. This is certainly an interesting observation, although the uncertainty is still too large to draw any further conclusions.

\begin{table}[H]
	\begin{center}
		\begin{tabular}{|c c|c|c|c|}
			\hline
			Scenario & &  $|V_{ub}|$ & $\mathcal{B}(B^-\rightarrow e^- \bar{\nu}_e)$ & $\Braket{{\mathcal B}(\bar{B}\rightarrow \rho \tau^- \bar{\nu}_{\tau})}$\\
			\hline\hline
			\multirow{4}{*}{$C_P^e = C_P^\mu$} & 1 & \multirow{2}{*}{$(3.31 \pm 0.32) \times 10^{-3}$} & \multirow{2}{*}{$(6.7_{-6.7}^{+9.3})\times 10^{-8}$} & $(7.81\pm0.66)\times 10^{-5}$\\ 
			& 2 & & &$(6.30\pm0.45)\times 10^{-5}$\\
			& 3 & \multirow{2}{*}{-}& \multirow{2}{*}{-} & \multirow{2}{*}{-} \\
			& 4 & & & \\
			\hline
			\multirow{4}{*}{$C_P^e = (1/10)C_P^\mu$} & 1 & \multirow{2}{*}{$(3.31 \pm 0.32) \times 10^{-3}$} & \multirow{2}{*}{$(8.0_{-8.0}^{+10.1})\times 10^{-10}$} & $(7.81\pm0.66)\times 10^{-5}$ \\ 
			& 2 & & &$(6.30\pm0.45)\times 10^{-5}$ \\
			& 3 & \multirow{2}{*}{$(3.31 \pm 0.32) \times 10^{-3}$} & \multirow{2}{*}{$(1.76\pm0.47)\times 10^{-8}$} & $(6.30\pm0.45)\times 10^{-5}$  \\
			& 4 & & & $(7.82\pm0.66)\times 10^{-5}$ \\
			\hline
			\multirow{4}{*}{$C_P^e = 10 C_P^\mu$} & 1 & \multirow{2}{*}{$(3.31 \pm 0.32) \times 10^{-3}$} & \multirow{2}{*}{$(6.6_{-6.6}^{+9.2})\times 10^{-6}$} & $(7.81\pm0.66)\times 10^{-5}$\\ 
			& 2 & & &$(6.29\pm0.45)\times 10^{-5}$\\
			& 3 & \multirow{2}{*}{-} & \multirow{2}{*}{-} & \multirow{2}{*}{-} \\
			& 4 & & & \\
			\hline\hline
			\multirow{2}{*}{2HDM} & 1 & $(3.31 \pm 0.32) \times 10^{-3}$ & $(1.15\pm0.25)\times 10^{-11}$ & $(6.26\pm0.45) \times 10^{-5}$\\ 
			& 2 & $(3.31 \pm 0.32) \times 10^{-3}$ & $(1.15\pm0.25)\times 10^{-11}$ &$(8.00\pm0.74)\times 10^{-5}$\\
			\hline
			\multirow{2}{*}{$C_P^e = C_P^\mu=0$} & 1 & $(4.85 \pm 1.03) \times 10^{-3}$ & \multirow{2}{*}{$(1.51\pm0.64)\times 10^{-11}$} &$(6.42\pm0.45)\times 10^{-5}$ \\ 
			& 2 & $(4.85 \pm 1.03)\times 10^{-3}$ & &$(7.45\pm0.66)\times 10^{-5}$\\
			\hline
		\end{tabular}
		\caption{Summary of the determination of $|V_{ub}|$ and the predictions for $\mathcal{B}(B^-\rightarrow e^- \bar{\nu}_e)$ and $\Braket{{\mathcal B}(\bar B\rightarrow \rho \tau^- \bar{\nu}_{\tau})}$ in the different scenarios discussed in the text.}\label{tab:summaryResults}
	\end{center}
\end{table}

\section{Predictions of Branching Ratios} \label{sec:predictions}
Here we provide predictions for branching ratios which have not yet been measured:
\begin{equation}
\mathcal{B}(B^-\rightarrow e^- \bar{\nu}_{e}),\quad
\mathcal{B}(\bar B\rightarrow \rho \tau^- \bar{\nu}_{\tau}),\quad
\mathcal{B}(\bar B\rightarrow \pi \tau^- \bar{\nu}_{\tau}).
\end{equation}
We will again consider scalar and pseudoscalar NP contributions and shall follow the studies discussed in Secs.~\ref{sec:semilept} and \ref{sec:Vub}.

We begin by having a closer look at $\mathcal{B}(B^-\rightarrow e^- \bar{\nu}_{e})$. As discussed in Secs.~\ref{sec:intro} and \ref{sec:lept}, within the SM, this branching fraction  is helicity suppressed due to the tiny value of the mass of the electron. However, the presence of the pseudoscalar NP contribution $C^{\ell}_P$ can potentially lift the helicity suppression. In Ref.~\cite{FJTX-1}, we have explored an analogous mechanism that may enhance the  branching fraction for the leptonic rare decays $B_{s,d}\rightarrow e^+e^-$. We now describe the main steps of our procedure using as an example the universal NP scenario:
\begin{enumerate}
	\item With the values of $C_P^\mu$ and $C_P^\tau$ calculated as in Sec.~\ref{sec:Vub}, we determine $R^{e}_{\tau}$.  In the case of universal Wilson coefficients for the light leptons, we obtain
	\begin{eqnarray}\label{eq:Retau}
	R^{e}_{\tau} = (5.8_{-5.8}^{+8.2})\times 10^3.
	\end{eqnarray}
	\item In order to obtain $\mathcal{B}(B^-\rightarrow e^- \bar{\nu}_{e})$, we multiply the theoretical determination of $R^{e}_{\tau}$ with the experimental value of $\mathcal{B}(B^-\rightarrow \tau^- \bar{\nu}_{\tau})$ and the relevant mass factors (see Eq.~(\ref{eq:Ru})). We employ 
	the experimental value in Eq.~(\ref{br-btaunu-avg}) which yields
	\begin{eqnarray}\label{eq:predBtoenu}
	\mathcal{B}(B^-\rightarrow e^- \bar{\nu}_e) = (6.7_{-6.7}^{+9.3})\times 10^{-8}.
	\end{eqnarray}
	Consequently, the branching ratio for the process $B^- \to e^- \bar{\nu}_e$ could be enhanced by up to four orders of magnitude with respect to the SM value given in Eq.~(\ref{eq:SMleptBr}). Interestingly, our determination in Eq.~(\ref{eq:predBtoenu}) is only one order of magnitude below the current experimental bound in Eq.~(\ref{Belle-enu-limit}).
\end{enumerate}

\begin{center}
	\begin{figure}[H]
		\begin{center}
			\includegraphics[width=0.7\textwidth]{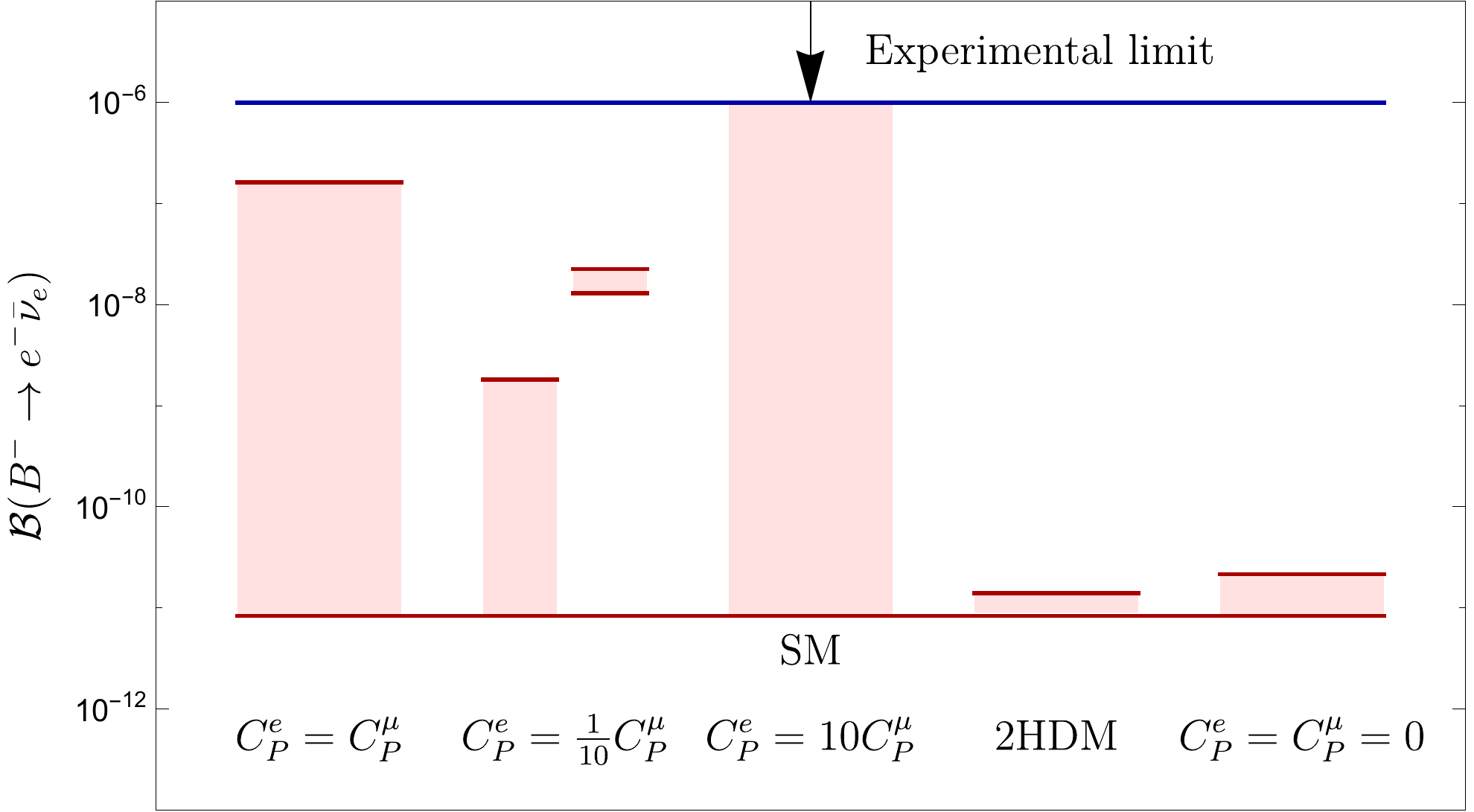}
			\caption{Illustration of the possible enhancement of $\mathcal{B}(B^-\rightarrow e^- \bar{\nu}_{e})$ for the scenarios discussed in the text. The blue line gives the current experimental upper bound on $\mathcal{B}(B^-\rightarrow e^- \bar{\nu}_{e})$, whereas the red horizontal line on the bottom represents the SM value. The red regions indicate the values of the branching ratio that may be obtained.}
			\label{fig:prop-plot-e}
		\end{center}
	\end{figure}
\end{center}

For completeness, we evaluate also the observable $\mathcal{B}(B^-\rightarrow e^- \bar{\nu}_e)$ within the four scenarios introduced in Sec.~\ref{sec:Vub}. The corresponding predictions are summarized in Table \ref{tab:summaryResults}. We illustrate graphically how our predictions for the branching fractions compare with the SM value in Fig.~\ref{fig:prop-plot-e}.

\newpage

We proceed in an analogous way in order to determine ${\mathcal B}(\bar B\rightarrow \rho \tau^- \bar{\nu}_{\tau})$. The steps are as follows:
\begin{enumerate}
	\item Substitute the results for  $C^{\mu}_P(\mathcal{R}^{\mu}_{\Braket{e, \mu}; \rho}, $) and $C^{\tau}_P(\mathcal{R}^{\mu}_{\Braket{e, \mu}; \rho}, R^{\tau}_{\mu})$ obtained in Sec.~\ref{sec:Vub} inside the ratio
	\begin{eqnarray}
	\mathcal{R}^{\tau;\rho~[m^2_{\tau} \leq q^2 \leq 12]~\rm{GeV}^2}_{\Braket{e, \mu};\rho}&\equiv&
	\Braket{ \mathcal{B}(\bar{B}\rightarrow \rho \tau^- \bar{\nu}_{\tau})}\Bigl|^{12~\rm{GeV}^2}_{m^2_{\tau}}/
	\Braket{\mathcal{B}(\bar{B}\rightarrow \rho \ell^- \bar{\nu}_{\ell})}_{[\ell=e, \mu]},
	\end{eqnarray}
	constructed in analogy with $\mathcal{R}^{\Braket{e, \mu};\rho~[0 \leq q^2 \leq 12]~\rm{GeV}^2}_{\Braket{e, \mu};\pi}$ as given by Eq.~(\ref{eq:semirhosemipi}). In the case of universality for the light leptons our theoretical predictions are
	\begin{equation}
    \mathcal{R}^{\tau;\rho~[m^2_{\tau} \leq q^2 \leq 12]~\rm{GeV}^2}_{\Braket{e, \mu};\rho}= 0.395 \pm 0.025, \quad \mathcal{R}^{\tau;\rho~[m^2_{\tau} \leq q^2 \leq 12]~\rm{GeV}^2}_{\Braket{e, \mu};\rho}= 0.318 \pm 0.011
	\end{equation}
	Note that we have two solutions, corresponding to the two allowed regions in Fig.~\ref{fig:lepsemmutau}, which have the same $C_P^\mu$ but different values of $C_P^\tau$.
	\item Multiply the theoretical determination of $\mathcal{R}^{\tau;\rho~[m^2_{\tau} \leq q^2 \leq 12]~\rm{GeV}^2}_{\Braket{e, \mu};\rho}$ by the experimental value of the branching fraction $\Braket{\mathcal{B}(\bar{B}\rightarrow \rho \ell^- \bar{\nu}_{\ell})}_{[\ell=e, \mu]}$. The resulting value is precisely $\Braket{ \mathcal{B}(\bar{B}\rightarrow \rho \tau^- \bar{\nu}_{\tau})}\Bigl|^{12~\rm{GeV}^2}_{m^2_{\tau}}$. In the universal case for light leptons, we obtain
\begin{eqnarray}
\Braket{ \mathcal{B}(\bar{B}\rightarrow \rho \tau^- \bar{\nu}_{\tau})}\Bigl|^{12~\rm{GeV}^2}_{m^2_{\tau}} &=& (7.81\pm0.66)\times 10^{-5}, \nonumber \\
\Braket{ \mathcal{B}(\bar{B}\rightarrow \rho \tau^- \bar{\nu}_{\tau})}\Bigl|^{12~\rm{GeV}^2}_{m^2_{\tau}} &=& (6.30\pm0.45)\times 10^{-5}.
\end{eqnarray}
\end{enumerate}		

\begin{center}
	\begin{figure}[H]
		\begin{center}
			\includegraphics[width=0.7\textwidth]{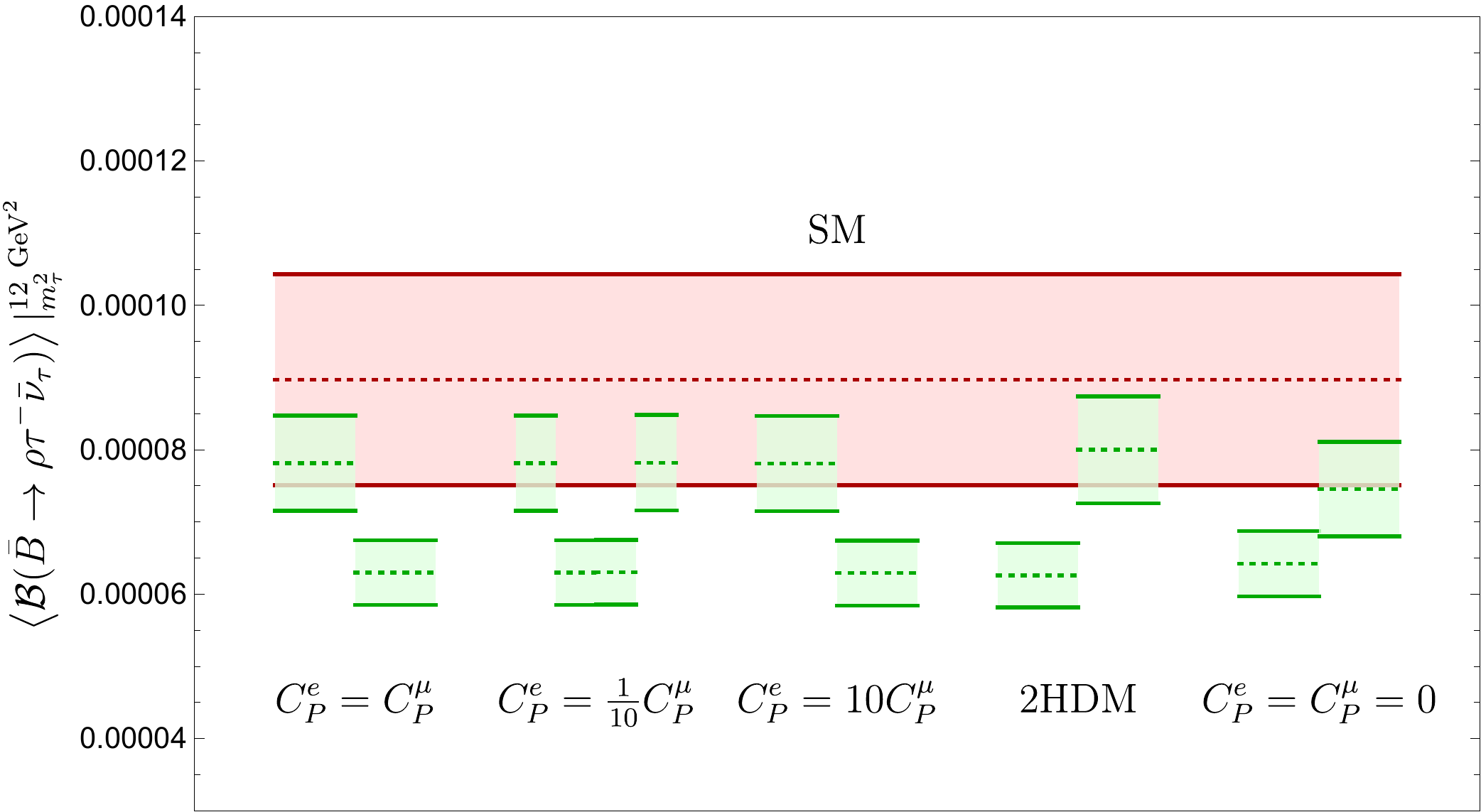}
			\caption{Illustration of the possible values that $\mathcal{B}(\bar B\rightarrow \rho \tau^- \bar{\nu}_{\tau})$ may take for the scenarios discussed in the text. The red horizontal band gives the SM value, whereas the green regions indicate the values of the branching ratio that may be obtained.}
			\label{fig:prop-plot-tau}
		\end{center}
	\end{figure}
\end{center}

We have also estimated $\Braket{ \mathcal{B}(\bar{B}\rightarrow \rho \tau^- \bar{\nu}_{\tau})}\Bigl|^{12~\rm{GeV}^2}_{m^2_{\tau}}$ for the different models introduced in Sec.~\ref{sec:Vub}. Our results are presented in fourth column of Table \ref{tab:summaryResults} and are illustrated in Fig.~\ref{fig:prop-plot-tau}. We observe that the predictions are very stable with respect to the model under consideration. However, a measurement of this observable can be used to distinguish between the two different solutions for
$C_P^\tau$. The strategies described in this section are schematically presented in Fig.~\ref{fig:flowchart}.

In order to predict $\Braket{ \mathcal{B}(\bar{B}\rightarrow \pi \tau^- \bar{\nu}_{\tau})}$, we require sensitivity on $C^{\tau}_S$. Unfortunately,  none of the observables considered in our analysis give us direct access to this coefficient. Therefore, in order to make predictions for this observable, we need to make extra assumptions that in general will be model dependent.  For instance, the 2HDM gives us access to $C^{\tau}_S$ through Eq.~(\ref{eq:C2HDM}), yielding
\begin{equation} \label{eq:2HDMPSrelation}
C^{\tau}_S=\frac{m_{\tau}}{m_{\mu}}C^{\mu}_S = \frac{m_{\tau}}{m_{\mu}}C^{\mu}_P.
\end{equation}

To finally extract $\Braket{ \mathcal{B}(\bar{B}\rightarrow \pi \tau^- \bar{\nu}_{\tau})}$,  we consider the observable
\begin{eqnarray}\label{eq:tausemilept}
\mathcal{R}^{\tau;\pi}_{\Braket{e, \mu};\pi}\equiv \Braket{\mathcal{B}(\bar{B}\rightarrow \pi \tau^- \bar{\nu}_{\tau})}/
\Braket{\mathcal{B}(\bar{B}\rightarrow \pi \ell^- \bar{\nu}_{\ell})}_{[\ell=e, \mu]},
\end{eqnarray}
which can then be multiplied by the experimental value of $\Braket{\mathcal{B}(\bar{B}\rightarrow \pi \ell^- \bar{\nu}_{\ell})}_{[\ell=e, \mu]}$ to obtain $\Braket{\mathcal{B}(\bar{B}\rightarrow \pi \tau^- \bar{\nu}_{\tau})}$. 
 Employing the relations in Eq.~(\ref{eq:2HDMPSrelation}) and the experimental value in Eq.~(\ref{eq:piexp}) we get
\begin{eqnarray}
\Braket{\mathcal{B}(\bar{B}\rightarrow \pi \tau^- \bar{\nu}_{\tau})} &=& (0.91\pm0.17)\times 10^{-4}, \nonumber \\ \Braket{\mathcal{B}(\bar{B}\rightarrow \pi \tau^- \bar{\nu}_{\tau})} &=& (1.27\pm0.31)\times 10^{-4}.
\end{eqnarray}

\boldmath
\section{Conclusions}\label{sec:concl}
\unboldmath
We have presented a detailed analysis of leptonic $B^-\to \ell^- \bar\nu_\ell$ decays and their semileptonic counterparts 
$\bar B \to \rho \ell^- \bar\nu_\ell$ and $\bar B \to \pi \ell^- \bar\nu_\ell$, aiming at tests of lepton flavour universality in processes
caused by $b\to u \ell^- \bar\nu_\ell$ transitions. A key requirement to constrain the short-distance coefficients of NP operators is 
to consider only quantities which do not depend on $|V_{ub}|$. The point is that the values of this CKM parameter extracted from 
semileptonic decays assume the SM while we allow for NP contributions to these processes. Since the leptonic decays, which 
exhibit helicity suppression in the SM, play a key role in this endeavour, we focused on new (pseudo)-scalar operators which 
may lift the helicity suppression, thereby having a potentially dramatic impact on these modes. 

The $B^-\to \ell^- \bar\nu_\ell$ 
decays involve actually the pseudoscalar coefficient $C_P^\ell$. Using a recent Belle result for the $B^-\to \mu^- \bar\nu_\mu$ 
branching ratio in combination with the measured ${\mathcal B}(B^-\to \tau^- \bar\nu_\tau)$, we obtained theoretically clean 
constraints in the $C_P^\mu$--$C_P^\tau$ plane. One branch of the solutions is consistent with the SM picture within the current
uncertainties. Thanks to the lift of the helicity suppression, we obtain a remarkably constrained picture despite the significant 
experimental uncertainty for the $B^-\to \mu^- \bar\nu_\mu$ mode. 

In order to further constrain the pseudoscalar NP coefficients, we employ the semileptonic $\bar B \to \rho \ell^- \bar\nu_\ell$ modes
which involve $C_P^\ell$ as well. While the leptonic decays depend on the $B^-$ decay constant as the only non-perturbative 
parameter, the semileptonic decay requires a variety of hadronic form factors which can be determined by means of QCD
sum rule and lattice calculations. Using results available in the literature, we have performed a comprehensive study of the available 
data. Interestingly, to the best of our knowledge, measurements of differential decay rates of $\bar B \to \rho \ell^- \bar\nu_\ell$
for $\ell=\mu$ and $\ell=e$ are not available. It would be important for probing violations of lepton flavour universality if
experimental collaborations would report such analyses. We obtain a picture which is consistent with the SM at the $1\,\sigma$ 
level, taking both experimental and theoretical uncertainties into account.

The general low-energy effective Hamiltonian including NP effects has also a scalar operator which does not contribute
to the $B^-\to \ell^- \bar\nu_\ell$ and $\bar B \to \rho \ell^- \bar\nu_\ell$ modes but has an impact on the semileptonic 
$\bar B \to \pi \ell^- \bar\nu_\ell$ channels. A comment similar to the one for the $\bar B \to \rho \ell^- \bar\nu_\ell$ modes
applies also in this
case, i.e.\ it would be very useful to have experimental results for electrons and muons in the final states. Making a simultaneous
analysis of the leptonic $B^-\to \mu^- \bar\nu_\mu$ and semileptonic  $\langle\bar B \to \pi  \ell^- \bar\nu_\ell\rangle_{[\ell=e,\mu]}$ 
decays, we derived a constraint in the $C_P^\mu$--$C_S^\mu$ plane, showing one solution in agreement with the SM. Yet another
constraint follows from the ratio of the differential  $\langle\bar B \to \rho  \ell^- \bar\nu_\ell\rangle_{[\ell=e,\mu]}$  and
$\langle\bar B \to \pi  \ell^- \bar\nu_\ell\rangle_{[\ell=e,\mu]}$ rates, which we discussed for various values of the momentum transfer
$q^2$. Interestingly, for certain values, we obtain tension with the SM at the $1\sigma$ level which will be interesting to monitor
in the future. It would be very desirable to have more sophisticated non-perturbative analyses of the form factors available, in 
particular for the semileptonic $\bar B \to \rho$ transitions. In our study, we have also explored the impact of CP-violating 
phases of the NP coefficients.

Using the NP constraints, we could make corresponding predictions for decay observables which have not yet been 
measured. In particular, we find a potentially huge enhancement of the  $B^-\to e^- \bar\nu_e$ branching ratio, lifting it up
to the regime of the experimental upper bound.  Moreover, we determined the CKM element $|V_{ub}|$, obtaining
values in agreement with other analyses in the literature although having larger uncertainties. 

The method which we proposed and explored for decays caused by $b\to u  \ell^- \bar\nu_\ell$ quark-level processes can 
actually also be implemented for exclusive $\bar B$ decays originating from $b\to c \ell^- \bar\nu_\ell$ modes. In this case, 
the leptonic decay $B_c^-\to \ell^- \bar\nu_\ell$ channels are key ingredients. Unfortunately, these decays are very challenging 
from an experimental point of view and no measurements are currently available, despite the fact that many $B_c$ mesons are 
produced at the LHC. Hopefully, in the future, innovative ways will be found to get a handle on the leptonic $B_c$ modes. 

It will be very interesting to apply the strategy presented in this paper in the future high-precision era of $B$ physics, thereby
shedding more light on contributions of new (pseudo)-scalar operators and probing lepton flavour universality in yet another
territory of the flavour physics landscape.


\section*{Acknowledgements}
This research has been supported by the Netherlands Foundation for Fundamental Research of Matter (FOM) 
programme 156, ``Higgs as Probe and Portal'', and by the National Organisation for Scientific Research (NWO).
G.B. acknowledges the support through a Volkert van der Willigen grant of the University of Amsterdam.


\newpage

\appendix

\section{Hadronic Form Factors}\label{sec:HFF}
In order to constrain the NP coefficients $C^\ell_S$ and $C^\ell_P$
through observables involving  the semileptonic ratios
${\mathcal B}(\bar{B}\rightarrow \rho \ell^- \bar{\nu}_{\ell})$ and ${\mathcal B}(\bar{B}\rightarrow \pi \ell^- \bar{\nu}_{\ell})$, we need to 
estimate the non-perturbative contributions to the different branching fractions. Depending on the value of $q^2$, two approaches can be followed:
\begin{itemize}
	\item QCD sum rules analyses, which  apply to $0 \leq q^2 \leq q^2_{max}$, with $q^2_{max}$ inside the interval
	$[12,~16]~\rm{GeV}^2$.
	\item Lattice QCD calculations, which provide results for $q^2$ close to the maximal leptonic momentum transfer. For $B\rightarrow \rho$ transitions we consider the interval
	\begin{equation}
	q^2_{max} \leq q^2\leq (M_{B^0}-M_{\rho})^2=20.29~\rm{GeV}^2,
	\end{equation}
	on the other hand, for $B\rightarrow \pi$ processes we use
	\begin{equation}
	q^2_{max} \leq q^2\leq (M_{B^0}-M_{\pi})^2=26.42~\rm{GeV}^2.
	\end{equation}
\end{itemize}

\boldmath
\subsection{Form factors for ${\mathcal B}(\bar{B}\rightarrow \rho \ell^- \bar{\nu}_{\ell})$}
\unboldmath

In the helicity basis, the hadronic form factors are given by \cite{STTW, Straub:2015ica}:
\begin{eqnarray}
H^{\rho}_{V, \pm}&=&(m_B + m_{\rho}) A_1(q^2) \mp \frac{2M_B|\vec{p}_{\rho}|}{m_B +m_{\rho}}V(q^2),\label{eq:Hrhopm}\\
H^{\rho}_{V,0}&=&-\frac{8 m_B m_{\rho}}{\sqrt{q^2}}A_{12}(q^2),\label{eq:Hrho0}\\
H^{\rho}_{V,t}(q^2)&=&-\frac{2 M_B |\vec{p}_{\rho}|}{\sqrt{q^2}}A_{0}(q^2),\label{eq:Hvt}\\
H^{\rho}_{S}(q^2)&=&-\frac{2 M_B  |\vec{p}_{\rho}|}{m_b + m_u}A_{0}(q^2).\label{eq:HS}
\end{eqnarray}
Following Ref.~\cite{Straub:2015ica}, the parametrization of the form factors for the range $0\leq q^2\leq 14~\rm{GeV}^2 $ obeys the generic expression
\begin{eqnarray}\label{eq:ffactors}
F_i(q^2)&=&\frac{k_{b\to u}^\rho}{(1-q^2/m^2_{R,i})}\sum^3_{k=1}\alpha^i_k \Bigl[ z_{\rho}(q^2, t_0) - z_{\rho}(0, t_0) \Bigl]^k,
\end{eqnarray}
for $F_i=A_0, A_1,  A_{12}, V$. The required coefficients $\alpha^i_k$ can be found in Table~\ref{parametersform} and the mass factors $m_{R,i}$ in Eq.~(\ref{eq:ffactors}) are evaluated according to the scheme presented in Table \ref{massfactors}. The function $z_{\rho}(q^2,t^{\rho}_0)$ is calculated using 
\begin{eqnarray}\label{eq:zrho}
z_{\rho}(q^2,t^{\rho}_0)&=&\frac{\sqrt{(M_B+m_{\rho})^2-q^2}-\sqrt{(M_B+m_{\rho})^2-t^{\rho}_0}}{\sqrt{(M_B+m_{\rho})^2-q^2}+
	\sqrt{(M_B+m_{\rho})^2-t^{\rho}_0}},\nonumber\\
t^{\rho}_0&=&(M_B+m_{\rho})\Bigl[(M_B+m_{\rho})-2\sqrt{M_B m_{\rho}}\Bigl].
\end{eqnarray}
The parameter $k_{b\to u}^\rho$ in Eq.~(\ref{eq:ffactors}) accounts for the fact that the form factors $F_i$ above refer to $b \to u$ transitions as discussed in Ref.~\cite{Straub:2015ica}. It is defined as
\begin{equation}
k_{b\to u}^\rho \equiv \frac{f_{\rho^0}^{(u)}}{\bar{f}_\rho^{\rho^I}},
\end{equation}
where
\begin{equation}
f_{\rho^0}^{(u)} = (221.5 \pm 0.3) \times 10^{-3}~{\rm GeV}, \qquad \bar{f}_\rho^{\rho^I} = (213 \pm 5) \times 10^{-3}~{\rm GeV}.
\end{equation}

For the high $q^2$ regime our only source is the lattice study in \cite{Bowler:2004zb},
where the interval $12.7~{\rm GeV}^2\leq q^2\leq 18.2~{\rm GeV}^2$ is taken into account. Unfortunately  no analytical parametrization of the form factors in this region is provided. Therefore we use directly the distributions provided in \cite{Bowler:2004zb}. These results are relatively old and potential underestimations of the uncertainties are possible. However, to the best of our knowledge this is the only study available to investigate the high $q^2$ region. Updated lattice calculations are crucial to obtain a more precise assessment of the effects on the $B\rightarrow \rho$ channels discussed in this work.

The form factors obtained through the parametrization in Eq.~(\ref{eq:ffactors}) correspond to the transition $\bar{B}^0\rightarrow \rho^+ \ell^- \bar{\nu}_{\ell}$. The expressions for $\bar{B}^-\rightarrow \rho^0 \ell^- \bar{\nu}_{\ell}$ can then be calculated using isospin symmetry. Hence when evaluating the branching ratio for the process $\bar{B}^-\rightarrow \rho^0 \ell^- \bar{\nu}_{\ell}$  the non-perturbative contributions $F_i$ discussed in this section should
include the correction factor $1/\sqrt{2}$.

\begin{table}
	\begin{center}
		\begin{tabular}{|c|c|c|c|c|} 
			\hline
			& $\alpha^k_0$ & $\alpha^k_1$ & $\alpha^k_2$ \\ [0.5ex] 
			\hline\hline
			$A_0$ & $0.356 \pm 0.042$ & $-0.833\pm 0.204$ & $1.331\pm 1.050$ \\ 
			\hline
			$A_1$ & $0.262\pm 0.026$ & ~$0.394\pm 0.139$ & $0.163\pm0.408$  \\
			\hline
			$A_{12}$ & $0.297\pm 0.035$  &~$0.759\pm 0.197$  & $0.465\pm 0.756$ \\
			\hline
			$V$ & $0.327\pm 0.031$  & $-0.860\pm 0.183$  & $1.802\pm 0.965$ \\
			\hline
		\end{tabular}
		\caption{Parameters used for the determination of the $\bar{B}\rightarrow \rho \ell^- \bar{\nu}_{\ell}$ form factors as given in Ref.~\cite{Straub:2015ica}.}\label{parametersform}
	\end{center}
\end{table}

\begin{table}
	\begin{center}
		\begin{tabular}{|c|c|} 
			\hline
			$F_i$ & $m_{R_i}/\rm{GeV}$  \\ [0.5ex] 
			\hline\hline
			$A_0$ & $5.279$\\
			\hline 
			$V$ & $5.325$\\
			\hline
			$A_1$, $A_{12}$& 5.724\\
			\hline		
		\end{tabular}
		\caption{Mass terms required for the evaluation of the different $\bar{B}\rightarrow \rho$ form factors as given in Ref.~\cite{Straub:2015ica}.}\label{massfactors}
	\end{center}
\end{table}

\boldmath
\subsection{Form factors for ${\mathcal B}(\bar{B}\rightarrow \pi \ell^- \bar{\nu}_{\ell})$}\label{sec:pionform}
\unboldmath

The form factors for the process $\bar{B}\rightarrow \pi \ell^- \bar{\nu}_{\ell}$ in the helicity basis are \cite{STTW,BNP}
\begin{equation}
H^{\pi}_{V,0}=\frac{2 M_B |\vec{p}_{\pi}|}{\sqrt{q^2}}f_{+}(q^2), 
\quad\quad H^{\pi}_{V,t}=\frac{M^2_B-m^2_{\pi}}{\sqrt{q^2}}f_{0}(q^2),
\quad\quad
H^{\pi}_{S}=\frac{M^2_B-m^2_{\pi}}{m_b-m_u}f_{0}(q^2).
\end{equation}
To obtain $f_+(q^2)$ and $f_0(q^2)$ we use the Bourrely-Caprini-Lellouch (BCL) \cite{Bourrely:2008za} parametrization:
\begin{eqnarray}\label{eq:fpf0}
f_0(q^2)&=&\sum^{3}_{n=0}b^0_n z_{\pi}(q^2,t^{\pi}_0)^n,\nonumber\\
f_+(q^2)&=&\frac{1}{1-q^2/M^2_{B^*}}\sum^3_{n=0}  b^+_n \Bigl[z_{\pi}(q^2,t^{\pi}_0)^n - (-1)^{n-4}\frac{n}{4}z_{\pi}(q^2,t^{\pi}_0)^{4}\Bigl].
\end{eqnarray}
The corresponding numerical coefficients $b^0_n$ and $b^+_n$ are shown in Table \ref{parametersformpi} and were presented originally in Ref. \cite{Lattice:2015tia}. They are the result of a fit to lattice data extrapolated to the full kinematical range in $q^2$.

\begin{table}
	\begin{center}
		\begin{tabular}{|c|c|c|c|c|c|} 
			\hline
			& $b^k_0$ & $b^k_1$ & $b^k_2$ & $b^k_3$\\ [0.5ex] 
			\hline\hline
			$f_+$ & $0.407 \pm 0.015$ & $-0.65\pm 0.16$ & $-0.46\pm 0.88$ &$0.4 \pm 1.3$\\
			\hline
			$f_0$ & $0.507 \pm 0.022$ & $-1.77\pm 0.18$ & $1.27\pm 0.81$ &$4.2 \pm 1.4$\\			 
			\hline			
		\end{tabular}
		\caption{Parameters used for the determination of the $\bar{B}\rightarrow \pi \ell^- \bar{\nu}_{\ell}$ form factors as given in Ref.~\cite{Lattice:2015tia}.}\label{parametersformpi}
	\end{center}
\end{table}

The function $z_{\pi}(q^2,t^{\pi}_0)$ is analogous to $z_{\rho}(q^2,t^{\rho}_0)$ presented in Eq.~(\ref{eq:zrho}) and is  given by
\begin{eqnarray}
z_{\pi}(q^2,t^{\pi}_0)&=&\frac{\sqrt{(M_B+m_{\pi})^2-q^2}-\sqrt{(M_B+m_{\pi})^2-t^{\pi}_0}}{\sqrt{(M_B+m_{\pi})^2-q^2}+
	\sqrt{(M_B+m_{\pi})^2-t^{\pi}_0}},\nonumber\\
t^{\pi}_0&=&(M_B+m_{\pi})\Bigl[(M_B+m_{\pi})-2\sqrt{M_B m_{\pi}}\Bigl].
\end{eqnarray}

The form factors in Eq.~(\ref{eq:fpf0})  correspond to $\bar{B}^0\rightarrow \pi^+\ell^-\bar{\nu}_{\ell}$. Those for $B^-\rightarrow \pi^0\ell^-\bar{\nu}_{\ell}$ should be estimated including an extra multiplicative factor of $1/\sqrt{2}$.


\begin{thebibliography}{99}
%
%
%
\bibitem{Aoki:2016frl}
  S.~Aoki {\it et al.},
  Eur.\ Phys.\ J.\ C {\bf 77} (2017) 112
  doi:10.1140/epjc/s10052-016-4509-7
  [arXiv:1607.00299 [hep-lat]].

\bibitem{Charles:2004jd}
  J.~Charles {\it et al.} [CKMfitter Group],
  Eur.\ Phys.\ J.\ C {\bf 41} (2005) 1
  doi:10.1140/epjc/s2005-02169-1
  [hep-ph/0406184],
  {\it updated results and plots available at: http://ckmfitter.in2p3.fr}

\bibitem{PDG}
C. Patrignani {\it et al.} [Particle Data Group], Chin.\ Phys.\ C {\bf 40} (2016) 100001 and 2017 update.

\bibitem{Belle-munu}
  A.~Sibidanov {\it et al.} [Belle Collaboration],
  arXiv:1712.04123 [hep-ex].
  
\bibitem{Belle-enu}
  N.~Satoyama {\it et al.} [Belle Collaboration],
  Phys.\ Lett.\ B {\bf 647} (2007) 67
  doi:10.1016/j.physletb.2007.01.068
  [hep-ex/0611045].
  
\bibitem{Lees:2012xj}
  J.~P.~Lees {\it et al.} [BaBar Collaboration],
  Phys.\ Rev.\ Lett.\  {\bf 109} (2012) 101802
  doi:10.1103/PhysRevLett.109.101802
  [arXiv:1205.5442 [hep-ex]].

\bibitem{Lees:2013uzd}
  J.~P.~Lees {\it et al.} [BaBar Collaboration],
  Phys.\ Rev.\ D {\bf 88} (2013) 072012
  doi:10.1103/PhysRevD.88.072012
  [arXiv:1303.0571 [hep-ex]].

\bibitem{Huschle:2015rga}
  M.~Huschle {\it et~al.} [Belle Collaboration],
  Phys. \ Rev. \ D {\bf 92}  (2015) 072014
  doi:10.1103/PhysRevD.92.072014
  [arXiv:1507.03233 [hep-ex]]
  
\bibitem{BFNT}
  D.~Becirevic, S.~Fajfer, I.~Nisandzic and A.~Tayduganov,
  arXiv:1602.03030 [hep-ph].

\bibitem{Sato:2016svk}
  Y.~Sato {\it et~al.} [Belle Collaboration],
  Phys. \ Rev. \ D {\bf 94}  (2016) 072007
  doi:10.1103/PhysRevD.94.072007
  [arXiv:1607.07923 [hep-ex]]

\bibitem{Hirose:2016wfn}
  S.~Hirose {\it et~al.} [Belle Collaboration],
  Phys. \ Rev. \ Lett.  {\bf 118}  (2017) 211801
  doi:10.1103/PhysRevD.94.072007
  [arXiv:1612.00529 [hep-ex]]

\bibitem{Aaij:2015yra}
  R.~Aaij {\it et~al.} [LHCb Collaboration],
  Phys. \ Rev. \ Lett. {\bf 115} (2015) 111803
  doi:10.1103/PhysRevLett.115.159901
  [arXiv:1506.08614 [hep-ex]]

\bibitem{ABDS}
  W.~Altmannshofer, P.~S.~Bhupal Dev and A.~Soni,
  Phys.\ Rev.\ D {\bf 96} (2017) 095010
  doi:10.1103/PhysRevD.96.095010
  [arXiv:1704.06659 [hep-ph]].
  
\bibitem{WNSS}
  W.~Altmannshofer, C.~Niehoff, P.~Stangl and D.~M.~Straub,
  Eur.\ Phys.\ J.\ C {\bf 77} (2017) 377
  doi:10.1140/epjc/s10052-017-4952-0
  [arXiv:1703.09189 [hep-ph]].
  
\bibitem{HN}
  G.~Hiller and I.~Nisandzic,
  Phys.\ Rev.\ D {\bf 96} (2017) 035003
  doi:10.1103/PhysRevD.96.035003
  [arXiv:1704.05444 [hep-ph]].
  
\bibitem{FJTX-1}
  R.~Fleischer, R.~Jaarsma and G.~Tetlalmatzi-Xolocotzi,
  JHEP {\bf 1705} (2017) 156
  doi:10.1007/JHEP05(2017)156
  [arXiv:1703.10160 [hep-ph]].
  
\bibitem{RF-rev}
  R.~Fleischer,
  Phys.\ Rept.\  {\bf 370} (2002) 537
  doi:10.1016/S0370-1573(02)00274-0
  [hep-ph/0207108].
  
\bibitem{BSEGR}
  S.~Bar-Shalom, G.~Eilam, M.~Gronau and J.~L.~Rosner,
  Phys.\ Lett.\ B {\bf 694} (2011) 374
  doi:10.1016/j.physletb.2010.10.025
  [arXiv:1008.4354 [hep-ph]].

\bibitem{JZ}
  J.~Zupan,
  arXiv:1101.0134 [hep-ph].
  
\bibitem{BZ}
  J.~Brod and J.~Zupan,
  JHEP {\bf 1401} (2014) 051
  doi:10.1007/JHEP01(2014)051
  [arXiv:1308.5663 [hep-ph]].
  
\bibitem{FV}
  R.~Fleischer and K.~K.~Vos,
  Phys.\ Lett.\ B {\bf 770} (2017) 319
  doi:10.1016/j.physletb.2017.04.056
  [arXiv:1606.06042 [hep-ph]].

\bibitem{CPV-paper}
  R.~Fleischer, D.~Gal\'arraga Espinosa, R.~Jaarsma and G.~Tetlalmatzi-Xolocotzi,
  Eur.\ Phys.\ J.\ C {\bf 78} (2018) 1
  doi:10.1140/epjc/s10052-017-5488-z
  [arXiv:1709.04735 [hep-ph]].
  
\bibitem{Khodjamirian:2011}
  A.~Khodjamirian, T.~Mannel, N.~Offen and Y.-M.~Wang,
  Phys.\ Rev.\ D {\bf 83} (2011) 094031
  doi:10.1103/PhysRevD.83.094031
  [arXiv:1103.2655 [hep-ph]].
  
\bibitem{Bernlochner:2014ova}
  F.~U.~Bernlochner, Z.~Ligeti and S.~Turczyk,
  Phys.\ Rev.\ D {\bf 90} (2014) 094003
  doi:10.1103/PhysRevD.90.094003
  [arXiv:1408.2516 [hep-ph]].
  
\bibitem{Crivellin:2014zpa}
  A.~Crivellin and S.~Pokorski,
  Phys.\ Rev.\ Lett.\  {\bf 114} (2015) 011802
  doi:10.1103/PhysRevLett.114.011802
  [arXiv:1407.1320 [hep-ph]].
  
\bibitem{Tanaka:2016ijq}
  M.~Tanaka and R.~Watanabe,
  PTEP {\bf 2017} (2017) no.1,  013B05
  doi:10.1093/ptep/ptw175
  [arXiv:1608.05207 [hep-ph]].
  
\bibitem{Dutta:2016eml}
  R.~Dutta and A.~Bhol,
  Phys.\ Rev.\ D {\bf 96} (2017) no.3,  036012
  doi:10.1103/PhysRevD.96.036012
  [arXiv:1611.00231 [hep-ph]].
  
\bibitem{Ivanov:2017hun}
  M.~A.~Ivanov, J.~G.~K\"orner and C.~T.~Tran,
  Phys.\ Part.\ Nucl.\ Lett.\  {\bf 14} (2017) 669.
  doi:10.1134/S1547477117050053

\bibitem{Hou:1992sy}
  W. S. Hou,
  Phys.\  Rev. \ D {\bf 48} (1993) 2342
  doi:10.1103/PhysRevD.48.2342 
  
\bibitem{BGPS}
  A.~Biswas, D.~K.~Ghosh, S.~K.~Patra and A.~Shaw,
  arXiv:1801.03375 [hep-ph].
  
\bibitem{Kamenik:2008tj}
  J.~F.~Kamenik and F. Mescia,
  Phys.\ Rev.\ D {\bf 78} (2008) 014003
  doi:10.1103/PhysRevD.78.014003
  [arXiv:0802.3790 [hep-ph]].
  
\bibitem{Lattice:2015tia}
  J.~A.~Bailey  {\it et al.},
  Phys. \ Rev. \ D {\bf 92} (2015) 014024
  doi:10.1103/PhysRevD.92.014024
  [arXiv:1503.07839 [hep-lat]]
  
\bibitem{Bowler:2004zb}
  K.~C.~Bowler {\it et al.} [UKQCD Collaboration],
  JHEP {\bf 0405} (2004) 035
  doi:10.1088/1126-6708/2004/05/035
  [hep-lat/0402023].
  
\bibitem{STTW}
  Y.~Sakaki, M.~Tanaka, A.~Tayduganov and R.~Watanabe,
  Phys.\ Rev.\ D {\bf 88} (2013) 094012
  doi:10.1103/PhysRevD.88.094012
  [arXiv:1309.0301 [hep-ph]].
  
\bibitem{Altmannshofer:2008dz}
  W.~Altmannshofer, P.~Ball, A.~Bharucha, A.~J.~Buras, D.~M.~Straub and M.~Wick,
  JHEP {\bf 0901} (2009) 019
  doi:10.1088/1126-6708/2009/01/019
  [arXiv:0811.1214 [hep-ph]].
  
\bibitem{Sibidanov:2013rkk}
  A.~Sibidanov {\it et al.} [Belle Collaboration],
  Phys. \ Rev. \ D {\bf 88} (2013) 032005
  doi:10.1103/PhysRevD.88.032005
  [arXiv:1306.2781 [hep-ex]]  
  
\bibitem{Straub:2015ica}
  A.~Bharucha, D.~M.~Straub and R.~Zwicky,
  JHEP {\bf 1608} (2016) 098
  doi:10.1007/JHEP08(2016)098
  [arXiv:1503.05534 [hep-ph]].
  
\bibitem{BNP}
  S.~Bhattacharya, S.~Nandi and S.~K.~Patra,
  Phys.\ Rev.\ D {\bf 95} (2017) 075012
  doi:10.1103/PhysRevD.95.075012
  [arXiv:1611.04605 [hep-ph]].
  
\bibitem{Bourrely:2008za}
  C.~Bourrely, I.~Caprini and L.~Lellouch,
  Phys.\ Rev.\ D {\bf 79} (2009) 013008
   Erratum: [Phys.\ Rev.\ D {\bf 82} (2010) 099902]
  doi:10.1103/PhysRevD.82.099902, 10.1103/PhysRevD.79.013008
  [arXiv:0807.2722 [hep-ph]].
 

%
%
%
\end{thebibliography}
\end{document}